\documentclass[3p,number,sort&compress,times,fleqn,uniquelist=false]{elsarticle} 

\makeatletter
\def\ps@pprintTitle{%
 \let\@oddhead\@empty
 \let\@evenhead\@empty
 \def\@oddfoot{\centerline{\thepage}}%
 \let\@evenfoot\@oddfoot}
\makeatother

\usepackage{graphicx}
\usepackage{color}
\usepackage{mathtools}
\usepackage{amsmath}
\usepackage{amssymb}
\usepackage[only,llbracket,rrbracket]{stmaryrd}
\usepackage[misc]{ifsym}
\usepackage{bm}                    
\usepackage{setspace}
\usepackage{xspace}
\usepackage{hyphenat}
\usepackage{comment}
\usepackage{array} 
\usepackage{float} 
\floatstyle{ruled}
\newfloat{Fbox}{thp}{lop}
\floatname{Fbox}{Box}
\usepackage[flushmargin]{footmisc} 
\usepackage[colorlinks,pdfpagelabels,bookmarksopen,bookmarksnumbered,citecolor=blue,urlcolor=blue,linkcolor=blue]{hyperref}
\usepackage[title]{appendix}
\usepackage{tcolorbox}
\usepackage[utf8x]{inputenc} 
\usepackage{natbib}

\newtheorem{theorem}{Theorem}
\newtheorem{lemma}{Lemma}
\newtheorem{remark}{Remark}
\newtheorem{proof}{Proof}
\newtheorem{definition}{Definition}
\newtheorem{corollary}{Corollary}[theorem]

\newcommand{\bs}[1]{\mathbf{#1}}
\newcommand{\bms}[1]{\bm{#1}}

\usepackage[font=normalsize]{caption}
\usepackage[font=normalsize]{subcaption}

\graphicspath{{./Figures/}}
\usepackage{float}

\newcommand{\newplaceholder}[0]{{\cdot}}

\begin{document}
\title{Variational approach to relaxed topological optimization: closed form solutions for structural problems in a sequential pseudo-time framework}

\author[cimne,etseccpb]{J. Oliver\corref{cor}}
\ead{oliver@cimne.upc.edu}
\author[cimne,eseiaat]{D. Yago}
\author[cimne,eseiaat]{J. Cante}
\author[cimne,etseccpb]{O. Lloberas-Valls}

\cortext[cor]{Corresponding author}

\address[cimne]{
	Centre Internacional de M\`{e}todes Num\`{e}rics en Enginyeria (CIMNE)\\ 
	Campus Nord UPC, M\`{o}dul C-1 101, c/ Jordi Girona 1-3, 08034 Barcelona, Spain
}
\address[eseiaat]{
	Escola Superior d'Enginyeries Industrial, Aeroespacial i Audiovisual de Terrassa (ESEIAAT)\\
	Technical University of Catalonia (UPC/Barcelona Tech), Campus Terrassa UPC, c/ Colom 11, 08222 Terrassa, Spain
}
\address[etseccpb]{
	E.T.S d'Enginyers de Camins, Canals i Ports de Barcelona (ETSECCPB)\\
	Technical University of Catalonia (UPC/Barcelona Tech), Campus Nord UPC, M\`{o}dul C-1, c/ Jordi Girona 1-3, 08034 Barcelona, Spain
}

\begin{abstract}

The work explores a specific scenario for structural computational optimization based on the following elements: (a) a \textit{relaxed optimization setting} considering the ersatz (bi-material) approximation, (b) a treatment based on a \textit{non-smoothed characteristic function field} as a topological design variable, (c) the consistent derivation of a \textit{relaxed topological derivative} whose determination is simple, general and efficient, (d) formulation of the \textit{overall increasing cost function topological sensitivity} as a suitable optimality criterion, and (e) consideration of a pseudo-time framework for the problem solution, ruled by the problem constraint evolution. 

In this setting, it is shown that the optimization problem can be analytically solved in a  variational framework, leading to, nonlinear, \textit{closed-form algebraic solutions for the characteristic function}, which are then solved, in every time-step, via fixed point methods based on a \textit{pseudo-energy cutting algorithm} combined with the \textit{exact fulfillment of the constraint}, at every iteration of the non-linear algorithm, via a bisection method. The issue of the ill-posedness (mesh dependency) of the topological solution, is then easily solved via a Laplacian smoothing of that pseudo-energy. 

In the aforementioned  context, a number of (3D) topological structural optimization benchmarks are solved, and the solutions obtained with the explored \textit{closed-form solution method}, are analyzed, and compared, with their solution through an alternative level set method. Although the obtained results, in terms of the cost function and topology designs, are very similar in both methods, the associated computational cost is about five times smaller in the closed-form solution method this possibly being one of its advantages. Some comments, about the possible application of the method to other topological optimization problems, as well as envisaged modifications of the explored method to improve its performance close the work. 
\end{abstract}


\maketitle
\section{Motivation}
\label{Sec_intro}
Computational mechanics tools for solving topological optimization problems raise a large number of challenges, both from the mathematical and computational points of view. The problem is originally formulated in the context of a design domain, $\Omega\subset{\mathbb R}^3$, in which a \textit{solid} domain, $\Omega^{+}\subset\Omega$, whose topology is going to be designed, is embedded. Then, the \textit{voids domain} is defined as $\Omega^{-}=\Omega\backslash\Omega^+$, and a cost function (typically the structural compliance or a related structural measure) is aimed at being  optimized (minimized) in terms of some \textit{design variables}, defining the material distribution (topology) in $\Omega$, and subject to some constraints\footnote{Typically, but not necessarily, the volume occupied by the solid.}. This setting poses a number of difficulties, among which we will mention the following:
\begin{itemize} 
	\item [(a)]\textit {Discrete character of the design variables}.
	 The distribution of the material at every point, ${\bf{x}}\in\Omega$, is defined in terms of the characteristic function, $\chi:\Omega\rightarrow\{0,1\}$ ; $\chi({\bf x})=1\ \forall {\bf x}\in\Omega^+$ (material)  and $\chi{\bf(x)}=0\ \forall ({\bf x})\in\Omega^-$ (voids). Therefore, the design variables $\chi({\bf x})$ map into the \textit{discrete image set} $\{0,1\}$ and the optimization problem is highly non smooth.
	\item [(b)]\textit {Number of design variables}. The design variable $\chi({\bf x})$ is valued at all points ${\bf x}\in\Omega$. Therefore, unlike in standard discrete optimization problems \cite{Bertsekas1996}, there are, in principle, an \textit{infinite number} of design variables, $\chi({\bf x})\ \forall {\bf x}\in\Omega$, to be determined.
	\item [(c)]\textit {The unknown characteristic function, $\chi({\bf x})$, is non-Lipschitz and non-differentiable}. This translates into the fact that optimality conditions, based on standard differentiation of the cost functional, are not, in principle, applicable. 
\end{itemize}

Along the last decades, a large number of approaches to computational topological optimization have been proposed. With no aim of exhaustiveness, two large \textit{families} of methods to overcome those challenges have been proposed (see, for instance, \cite{eschenauer2001topology,Rozvany2008,Sigmund2013} for exhaustive reviews on topological optimization methods):
\begin{itemize}
	\item[(I)] \textit{Regularize the characteristic function}. This is the choice of the most popular, and successful, of the current approaches: the \textit{Solid Isotropic Material with Penalization} (SIMP) \cite{MartinPh.Bendsoe2003}, in which the discrete characteristic function, $\chi$,  is replaced by a regularized (continuous) function, (\textit{density-like function}), $\rho:\Omega\rightarrow[0,1]$. The regularization domain, $\Omega^{reg}\subset\Omega$, (the set of points of $\Omega$ where $\rho({\bf x})\in(0,1)$) is not necessarily small, and it requires to be covered by a, relatively large, number of \textit{discretization domains} (finite elements or Voxels). The approach succeeds in breaking the challenge of continuity, and standard finite-dimensional optimization  procedures\footnote{Typically based on efficient optimization techniques, like the Method for Moving Asymptotes (MMA).} can be applied to the resulting optimization problem, now formulated in terms of a finite (but large) number of design variables\footnote{The value of $\rho({\bf x}_i)$ at the sampling points ${\bf x}_i\in\Omega$.}, which are, then, supplemented by additional constraints imposing  $0\le\rho({\bf x}_i)\le 1$. The success of the method brings also some drawbacks, typically:
	\begin{itemize}
		\item[$\bullet$] Intrinsic to the existence of the regularization domain, $\Omega^{reg}\subset\Omega$, is the appearance of 
		\textit{gray zones} (zones with \textit{diffuse} material/voids coexistence, $0<\rho<1$) \cite{Bourdin2001,MartinPh.Bendsoe2003}, 
		 where the kind of material provided by the optimization is unclear. This requires to, subsequently, resort to \textit{filtering} techniques to eliminate them.
		 \item[$\bullet$] The \textit{mesh-dependence} \cite{Sigmund1998} or \textit{lack of mesh-size objectivity} of the results. Again these requires resorting to filtering techniques \cite{Yamasaki2010}.		
		  \item[$\bullet$] The appearance of \textit{checkerboards} when using finite element discretizations. This type of problems is identified as a numerical instability (not fulfillment of the Ladyzhenskaya--Babuska--Brezzi stability conditions) and it should be solved by resorting to specific finite elements in modeling the problem \cite{Jang2003}.   
		\end{itemize}
\item[(II)] \textit{Change the design variable}. The so called \textit{level-set methods} for topological optimization \cite{Dijk2013}, overcome the difficulties inherent to the discrete character of the characteristic function, $\chi({\bf x})$, by replacing it by a different, smoother, function  (the level-set function $\phi\in H^1(\Omega):\Omega\rightarrow{\mathbb R}$) related to each other through
 $\chi({\bf x})={\cal H}\left(\phi({\bf x})\right)$, where ${\cal H}:{\mathbb R}\rightarrow\{0,1\}$ stands for the Heaviside function. The values of $\phi({\bf x}_i)$ at the sampling points are the new design variables. The optimization problem is no longer solved through standard optimization methods, but by means of the solution of a Hamilton-Jacobi (pseudo-time evolutionary) equation \cite{Osher1988,Wang2003}. This family of methods have proven to be very robust in solving different types of topological optimization methods, albeit they suffer, also, from some drawbacks, e.g.: 
 	\begin{itemize}
 		\item[$\bullet$] They also require regularization techniques, for instance Tikhonov-regularization based methods \cite{Yamada2010}, to obtain a well posed (mesh-size objective) optimization problem. 
 		\item[$\bullet$] Their robustness, and rate of convergence, relies on the size of the pseudo time increment adopted, in turn depending on heuristic considerations. In general this translates into a high number of iterations and, thus, into a high computational cost. In some cases, these methods use initially the signed distance to the solid boundaries as level set function, and they may require re-initialization techniques during the analysis \cite{Yamasaki2010}.
 		\item[$\bullet$]A fundamental issue in this method is the selection of an appropriated \textit{sensitivity} of the cost function to changes on the design variable (the level-set function) which was initially related to the \textit{shape derivative concept} \cite{Allaire2005a}. More recently, connections with the \textit{topological derivative concept} were introduced (see for instance \cite{Allaire2005a,Burger2004,Yamada2010}). 
 		
 		The use of an appropriate sensitivity of the cost function to topological changes in the design domain, in front of the introduction of holes in the solid material, plays a fundamental role in the performance of this family of methods \cite{Norato2004}. Indeed, derivation of analytical expressions for the topological derivative, in linear elastic problems,  becomes crucial, and much work has been done in this sense (\cite{Novotny2003,Novotny2013,Giusti2008,Giusti2016,Amstutz2010}). In these cases, the topological sensitivity is mathematically derived in terms of the perturbation of the cost function, after material removal (introduction of a hole) in a certain point of $\Omega$, ruled by the elastic state problem in the perturbed solid. This analytical derivation, which specifically depends on the \textit{type of optimization problem} and the \textit{considered elastic material} (isotropic, orthotropic, anisotropic etc.), has to be previously derived via specific, and sometimes heavy, analytical methods, i.e. limit and asymptotic analysis to a null size of the hole. This could be also understood as a specific burden of the method.    
	\end{itemize}
 \end{itemize}
On the light of the, so far, depicted  situation, this work aims at exploring the impact and consequences of an alternative option for computational topological optimization: facing, as much as possible, the topological optimization problem in its original setting, i.e. considering the following scenario:
\begin{itemize}
	\item[$\bullet$]\textit{A non-smoothed characteristic function}, $\chi:\Omega\rightarrow\{\beta,1\}\ ;\ \beta\in(0,1)$, as design variable.
	\item[$\bullet$]\textit{An infinite-dimensional design space}, considering the values of the characteristic function $\chi{\bf(x)}$ at all points of $\Omega$, as design variables of the optimization problem.
	\item[$\bullet$]\textit{A relaxed optimization setting}, which facilitates the obtainment of the topological derivative, not requiring a complex mathematical derivation.
	\item[$\bullet$]\textit{Deriving} the corresponding \textit{closed-form solutions}, for such a non-smooth optimization problem, \textit{in a number of structural problems}\footnote{As precursors of solutions, in the chosen context, for broader families of topological optimization problems.}. 
\end{itemize}   

Then, an algorithmic setting for solving the obtained (highly non-linear)  closed-form solutions is devised, consisting of:
\begin{itemize}
	\item[(1)] \textit{A relaxed, bi-material, topological optimization, setting}. The original \textit{embedded-solid setting}, where a hard material (solid) domain $\Omega^+$ is embedded into the design space, $\Omega$, is relaxed to a \textit{bi-material} (hard-phase + soft-phase, ${\mathfrak M}^+\slash{\mathfrak M}^-$) setting. This relaxed scenario is commonly found in current topological optimization approaches and it is, sometimes referenced as \textit{ersatz material approximation} \cite{Allaire1997,Dambrine2009}. Its main advantage is that it makes the optimization problem simpler than the alternatives, e.g. the \textit{immersed boundary methods} \cite{Allaire2012,Sethian2000,Norato2007}. In this way, \textit {at all points of the design domain $\Omega$, there exists (some type of)  material}, and the classical concept of topological perturbation, initially understood as \textit{solid material removal or addition}, is replaced by that of \textit{material switching} or \textit{material exchange} (${\mathfrak M}^{+}({\bf x})\leftrightarrow{\mathfrak M}^{-}({\bf x})$). A number of, non-trivial, consequences derive from this, apparently minor, modification: (a) the standard (solid + voids) setting is asymptotically approximated by assigning to the soft material, ${\mathfrak M}^-$, very weak elastic properties (characterized by the classical \textit{contrast factor} $\alpha\rightarrow 0$), and (b) the sensitivity of the cost function to the topological perturbation at a given point ${\bf x}\in\Omega$, now understood as a \textit{local material exchange} at the point.
	
	\item[(2)]\textit{A consistent relaxed topological derivative (RTD)}.  The aforementioned relaxed bi-material setting is used, not only for solving the \textit{static elastic problem}, but also for the purposes of derivation of the topological sensitivity, this providing additional \textit{consistence} to the approach. Derivation of this \textit{consistent} topological sensitivity can be done, in a non-smooth variational setting, and computed, independently of the considered elastic material family, as a directional derivative of the cost functional. This derivation of RTD is extremely simple in comparison with the \textit{exact topological derivative}  classically used in the literature.
	
	\item[(3)]\textit{Sequential pseudo-time framework}. In this framework, a pseudo-time parametrization is introduced. Solutions of the topological optimization problem are obtained for increasing values of the restriction (i.e. the soft-phase volume $\vert\Omega^-\vert$), that plays the role of a \textit{pseudo-time}. In this context, \textit{the time evolution} of the topology of the optimal solution (and the associate cost function values) is obtained at no additional computational cost. In this way the optimization process provides information, not only on the optimal topology at every pseudo-time-step and, thus, at the corresponding soft-phase volume level, but also on the \textit{evolution of the minima} for a chosen set of times (volumes). In some cases this is a very useful information for the designer, who obtains, in a single run, additional information about, for instance, \textit{the minimum of the minima} in the computed designs.
	\item[(4)]\textit{Exact fulfillment of the restriction at every iteration of the algorithm}. This concept is borrowed from numerical techniques frequently used in non-linear analysis of structures: the so-called \textit{load control, or arc-length} methods \cite{Crisfield1983}. This substantially contributes to the robustness, and to the computational cost diminution, in the resolution of the resulting non-linear problem.
	\item[(5)]\textit{Regularization of the obtained closed-form topological solutions} via Laplacian smoothing techniques. In order to prevent the aforementioned \textit{ill-posedness of the problem}, leading to the mesh-size dependence, a Laplace-type smoothing is done, which removes from the solution the small (noisy) wavelengths below a predefined threshold.
	Distinctly from other approaches, this regularization is not done by introducing an additional contribution to the cost function \cite{Yamada2010}, but by \textit{smoothing the closed-form solution obtained from the original optimization problem}. Additionally, this allows: a) controlling the minimum attainable width/section of the \textit{filaments of the solid material} in the obtained solutions topology and, b) obtaining  $2\frac{1}{2}D$ (\textit{extruded to the third dimension}) topological designs.
	\end{itemize}
In next sections the theoretical and algorithmic aspects of the considered formulation are, first, presented in detail and, then, assessed by application to a set of representative examples. Comparisons of the obtained results, with those obtained with level-set (\textit{Hamilton-Jacobi type}) techniques using the derived \textit{relaxed topological derivative}, are done in order to check the performance of the analyzed approach, both in terms of the  
cost function values, and also in terms of the required computational cost.
\section{Problem set up}
\begin{figure}[h]
	\centering
	\includegraphics[width=12cm, height=4.5cm]{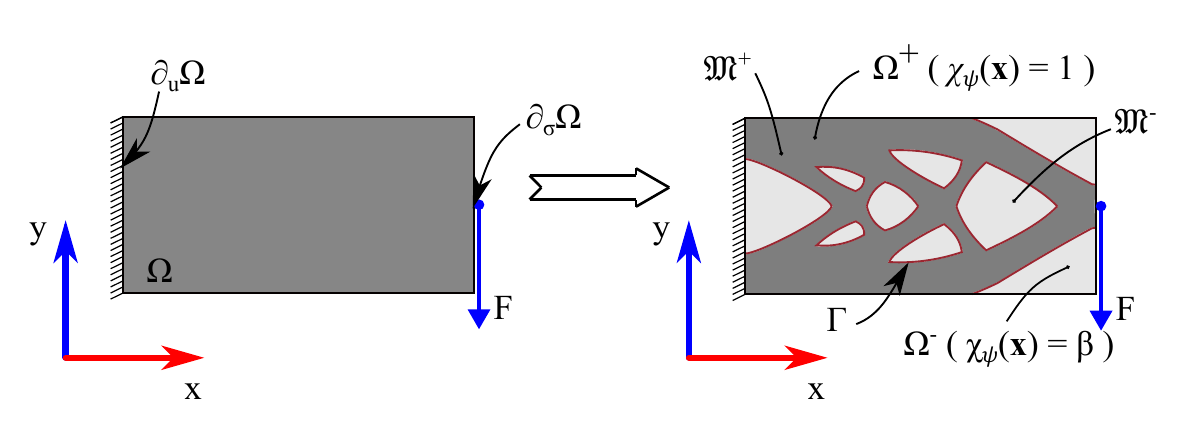}
	\caption{$\beta$-relaxed setting: bi-material design domain.}
	\label{fig_design_domain}
\end{figure}
\subsection{Relaxed characteristic function}
\label{sec_Relaxed_characteristic_function}
\begin{figure}[h]
	\centering
	\includegraphics[width=12cm, height=6cm]{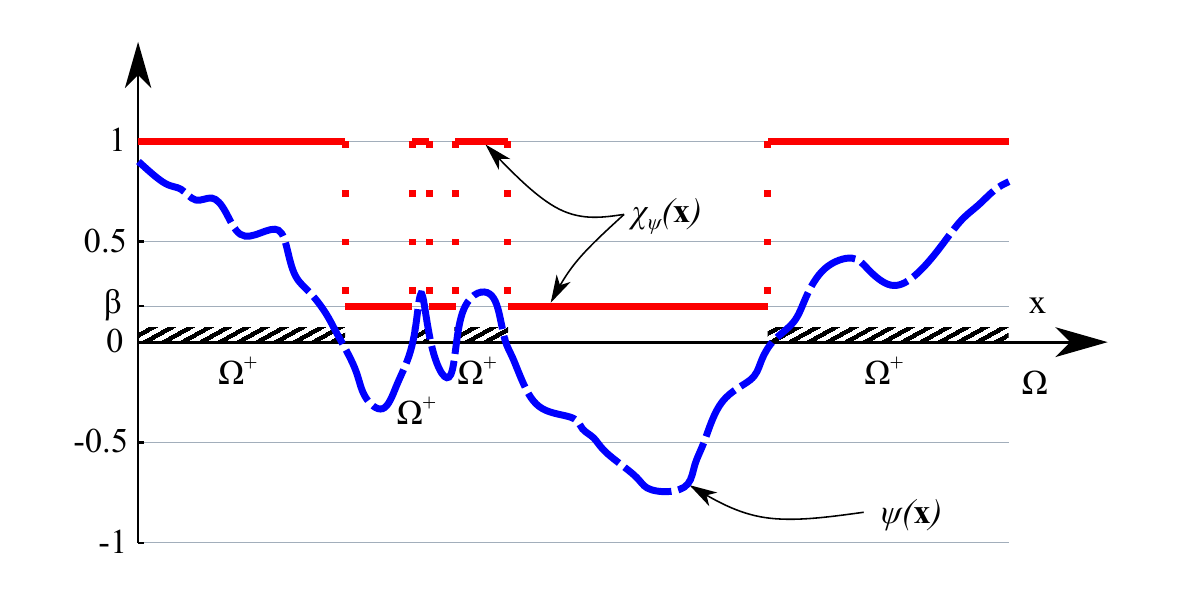}
	\caption{Discrimination function, $\psi$ and the corresponding relaxed characteristic function, $\chi_{\psi}$. }
	\label{fig_characteristic_function}
\end{figure}

Let us consider the \textit{design domain}, $\Omega\subset\mathbb{R}^n\   ; \  n\in\{2,3\}$, and the distribution over this domain  of two different material phases: a) \textit{hard-phase} ($ {\mathfrak M}^{+} $)  and b) \textit{soft-phase} ($ {\mathfrak M}^{-} $), the type of phase at every point,  ${\bf x} \in\Omega$, being defined by the {\em characteristic function} $\chi$ as:
\begin{equation}
\label{eq_chi_definition}
\begin{split}
&\chi : \Omega \rightarrow \{\beta,1\}\coloneqq
\left\{
\begin{split}
&\chi(\bs{x})=1  \quad \textit{if material phase at $\mathbf x$ is} \ {\mathfrak M}^{+} \\
&\chi(\bs{x})=\beta \quad 	\textit{if material phase at $\mathbf x$ is} \ {\mathfrak M}^{-}
\end{split}
\right.&(a)\\
&0<\beta<1\Rightarrow(1-\beta)>0&(b)
\end{split}
\end{equation}  
In a second step we shall consider a \textit{relaxed} version of the characteristic function, denoted as $\chi_{\psi}$, in terms of the, here termed, \textit{discrimination function} 
 $\psi \in H^1({\Omega})$ 
which is assumed single-valued and 
defined as (see figure \ref{fig_characteristic_function})
\begin{equation}
\label{eq_relaxed_chi}
\chi_{\psi}(\bs{x})={\cal H}_\beta(\psi(\bs{x}))
\end{equation} 
where ${\cal H}_{\beta}(\newplaceholder)$ stands for a \textit{relaxed} Heaviside function\footnote{Which recovers to the classical Heaviside function as $\beta\rightarrow 0$.}, defined as
\begin{equation}
\label{eq_heaviside}
\begin{split}
&{\cal H}_{\beta}(\psi)=1\; &\textit{for} \quad{\psi}>0\\
&{\cal H}_{\beta}(\psi)=\beta\; &\textit{for} \quad{\psi}<0
\end{split}
\end{equation}
 Therefore, function ${\psi}(\bs{x})$ defines, through its value, two possible values of $\chi_{\psi}$, i.e.
\begin{equation}
\label{eq_discriminating}
\begin{split}
\chi_{\psi}(\bs{x})={\cal H}_{\beta}({\psi}(\bs{x}))=1 \quad \textit{for} \quad{\psi}(\bs{x})>0   
  \\
\chi_{\psi}(\bs{x})={\cal H}_{\beta}({\psi}(\bs{x}))=\beta \quad \textit{for} \quad{\psi}(\bs{x})<0
\end{split}
\end{equation}
and discriminates points  $\mathbf{x}\in\Omega$ belonging either to the \textit{hard-phase domain}, $\Omega^{+}$, or to the \textit{soft-phase domain}  $\Omega^{-}$:
\begin{equation}
\label{eq_domain_splitting}
\left\{
\begin{split}
&\Omega^{+}({\chi_\psi})\coloneqq\{\mathbf{x}\in \Omega \;;\quad \psi\mathbf{(x)}>0\} \Rightarrow\chi_\psi\mathbf{(x)}={1} &\quad(a)\\
&\Omega^{-}({\chi_\psi})\coloneqq\{\mathbf{x}\in \Omega \;;\quad \psi\mathbf{(x)}<0\} \Rightarrow\chi_\psi\mathbf{(x)}={\beta} &\quad(b)\\
\end{split}
\right.\\
\end{equation}
Equations (\ref{eq_domain_splitting}) define the \textit{topology} of $\Omega$.
\begin{remark}
Notice that, since only two different phases are considered, ${\Omega^{+}}\cup {\Omega^{-}}=\Omega$, the description of the topology in equation (\ref{eq_domain_splitting}) can be  rephrased only in terms of the positive counterpart of the discrimination function, $\psi({\bf x})$, i.e.
\begin{equation}
\label{eq_domain_splitting_rephrased}
\left\{
\begin{split}
&\Omega^{+}(\chi_\psi)\coloneqq\{\mathbf{x}\in \Omega \; ;\quad \psi\mathbf{(x)}>0 \} &\Leftrightarrow\chi_\psi\mathbf{(x)}={1} &\quad(a)\\
&\Omega^{-}(\chi_\psi)\coloneqq\{\mathbf{x}\in \Omega \; ;\quad  \mathbf{x}\notin {\Omega^+}\}  &\Leftrightarrow\chi_\psi\mathbf{(x)}={\beta} &\quad(b)\\
\end{split}
\right.
\end{equation}
or, alternatively, in terms of the negative counterpart of $\psi({\bf x})$ i.e.
\begin{equation}
\label{eq_domain_splitting_rephrased_1}
\left\{
\begin{split}
&\Omega^{-}(\chi_\psi)\coloneqq\{\mathbf{x}\in \Omega \; ;\quad \psi\mathbf{(x)}<0 \} &\Rightarrow\chi_\psi\mathbf{(x)}={\beta} &\quad(a)\\
&\Omega^{+}(\chi_\psi)\coloneqq\{\mathbf{x}\in \Omega \; ;\quad  \mathbf{x}\notin {\Omega^-}\}  &\Rightarrow\chi_\psi\mathbf{(x)}={1} &\quad(b)\\
\end{split}
\right.
\end{equation}
\end{remark}
From equations (\ref{eq_domain_splitting_rephrased}) or (\ref{eq_domain_splitting_rephrased_1}) the following equalities can be stated:	
\begin{equation}
\label{eq_domain_measures}
\left\{
\begin{split}
	&\vert\Omega\vert=\vert\Omega^{+}(\chi_{\psi})\vert+\vert\Omega^{-}(\chi_{\psi})\vert&\quad(a)\\
	&\vert\Omega^{+}(\chi_{\psi})\vert=\int_{\Omega}\dfrac{\chi_{\psi}({\bf x})-\beta}{1-\beta} d\Omega &\quad\quad(b)  \\
	&\vert\Omega^{-}(\chi_{\psi})\vert=\int_{\Omega}\dfrac{1-\chi_{\psi}({\bf x})}{1-\beta} d\Omega
&\quad\quad(c)\end{split}
\right.
\end{equation}
where $\lvert\Omega\rvert$, $\lvert\Omega^{+}\rvert$  and $\lvert\Omega^{-}\rvert$ stand, respectively, for the measure (surface/volume) of the design domain, $\Omega$, the \textit{hard-phase} domain, $\Omega^{+}$, and the \textit{soft-phase} domain, $\Omega^{-}$.   	

\subsection{Relaxed, bi-material, linear elastic problem}
\label{sec_relaxed_bimaterial_linear_elastic_problem}
Let us now consider a \textit{bi-material elastic solid} occupying the whole domain $\Omega$. Then, the material distribution in $\Omega$ can be defined in terms of the relaxed characteristic function\footnote{From now on, subscript $(\newplaceholder)_\psi$ will be omitted, unless it is strictly necessary.}, ${\chi(\bf x)}$, in equations (\ref{eq_chi_definition}) to (\ref{eq_domain_splitting}). The corresponding elastic problem reads
\begin{equation} \label{eq_elastic_problem_2}
\begin{split}
&\bm{\nabla}\cdot\bm{\sigma}_{\chi}(\mathbf{x})+\mathbf{b}(\mathbf{x})=\mathbf{0}\,;\quad
\left\{
\begin{split}
&\bm{\sigma}_{\chi}(\mathbf{x})\equiv\bm{\sigma}(\mathbf{u}_{\chi}(\mathbf{x}))=
\underbrace{\chi^{m}{\mathbb C}}_{\textstyle {\mathbb C}_\chi({\bf x})}:\bm{\varepsilon}(\mathbf{u}_{\chi})
={{\mathbb C}_\chi(\mathbf{x})}:\bm{\varepsilon}(\mathbf{u}_{\chi}) \\
&\bm{\varepsilon}(\mathbf{u}_{\chi})\equiv\left( \bm\nabla\otimes\mathbf{u}_{\chi}(\mathbf{x})\right)^{sym}
\end{split}
\right. &\forall\mathbf{x}\in\Omega\quad&(a)\\
&\bm{\sigma}_{\chi}(\mathbf{x})\cdot\mathbf{n}=\mathbf{t}^*(\mathbf{x})&\forall\mathbf{x}\in\partial_{\sigma}\Omega\quad&(b) \\
&\mathbf{u}_{\chi}(\mathbf{x})=\mathbf{0}&\forall\mathbf{x}\in\partial_{u}\Omega\quad&(c) \\
\end{split}
\end{equation}
where $\mathbf{u}_{\chi}(\mathbf{x})$ is the displacement vector field, parametrized by the current topology,  $\bm\sigma_{\chi}(\mathbf{x})$ stands for the second order stress tensor field, $\mathbf{b}(\mathbf{x})$, stands for the density of the body forces, ${\mathbf t}^*(\mathbf{x})$ stands for the boundary tractions and ${\bm\varepsilon}(\mathbf{\mathbf{u}_{\chi}})$ is the (infinitesimal) strain tensor. In equation (\ref{eq_elastic_problem_2})-(a), $\mathbb{C}$ is the, fourth order, elastic constitutive tensor of the hard-phase (assumed constant), and the exponent $m$ is typically taken $m>1$. In addition, $\partial_u\Omega\subset\partial\Omega$ and $\partial_\sigma\Omega\subset\partial\Omega$, (where
$\partial_u\Omega\cap\partial_{\sigma}\Omega=\emptyset$ and $\partial_u\Omega\cup\partial_{\sigma}\Omega=\partial\Omega$) are, respectively, those parts of the boundary, $\partial\Omega$, where displacements, ${\bf u}_{\chi}$, and tractions, ${\bf t}={\bms \sigma}_{\chi}\cdot\bf{n}$ or forces, are, prescribed (see figure \ref{fig_design_domain}).
In addition $\chi(\mathbf{x})$ is recalled, from equations (\ref{eq_chi_definition}) as:
\begin{equation}
\label{eq_chi}
\chi(\mathbf{x})
=\left\{
\begin{split}
&1\quad&\forall\mathbf{x}\in\Omega^{+} \\
&\beta&\forall\mathbf{x}\in\Omega^{-}
\end{split}
\right.
\end{equation}
where ${0}<\beta<1$ will be, from now on, termed \textit{the relaxation factor}. 
Notice that, according to equations (\ref{eq_elastic_problem_2})-(a) and (\ref{eq_chi}) the  elastic constitutive tensor, ${\mathbb C}_\chi({\bf x})$ can be specified for all the domain points as 
\begin{equation}\label{eq_contrasted_constitutive_tensor}
\begin{split} 
& {\mathbb C}_{\chi}({\bf x})\equiv\chi^m(\mathbf{x})\mathbb{C}
=\left\{
\begin{split}
&\mathbb{C}\quad&\quad\forall\mathbf{x}\in\Omega^{+}(\chi) \\
&\beta^{m}\mathbb{C}=\alpha\mathbb{C}\approx\mathbf{0}\quad&\quad\forall\mathbf{x}\in\Omega^{-}(\chi)
\end{split}
\right. \quad&(a)\\
&\alpha=\beta^m\in(0,1)  \;;\qquad \beta=\alpha^{\frac{1}{m}}\in(0,1)\quad&(b)\\
&\dfrac{\partial{\mathbb{C}_{\chi}}}{\partial \chi}=m\chi^{m-1}\mathbb{C}=
\left\{
\begin{split}
&m\mathbb{C}\quad&\forall\mathbf{x}\in\Omega^{+}(\chi) \\
&m\beta^{(m-1)}\mathbb{C}=m\alpha^{\frac{m-1}{m}}\mathbb{C}&\forall\mathbf{x}\in\Omega^{-}(\chi)
\end{split}
\right. \quad&(c)
\end{split}
\end{equation}
where the factor $\alpha$ and the exponent $m$ will be, respectively, termed the \textit{contrast stiffness factor} and the \textit{contrast stiffness exponent}. For the standard isotropic case the elasticity tensor can be written \cite{Oliver2016a}
\begin{equation}
\label{eq_isotropic_constitutive_tensor}
 \left\{ 
   \begin{split}
	&\bm{\mathbb{C}}=\bar{\lambda} {\bf 1} \otimes {\bf 1} + 2\mu {\mathbb I}=\dfrac{\nu E}{\left(1+\nu \right)\left(1-2\nu\right)}  {\bf 1}+ \dfrac{E}{\left(1+\nu \right)} {\mathbb I}\\ 
	&\mathbb{C}_{ijkl} =\bar\lambda \delta_{ij} \delta_{kl} +\mu \left[\delta_{ik} \delta_{jl} + \delta_{il} \delta_{jk} \right] \; ;\quad  i,j,k,l \in \{ 1,2,3\}
	\end{split}
 \right.
\end{equation}
where ${\bar{\lambda}}$ and $\mu$ are the Lamé coefficients and $E$ and $\nu$ stand, respectively, for the Young modulus and the Poisson ratio. In addition, $\bf 1$ and ${\mathbb I}$ are, respectively, the second order unit tensor and the fourth order (symmetric) unit tensor.
The problem in equations (\ref{eq_elastic_problem_2}) can be alternatively written in a variational form as
\begin{equation} \label{eq_bilinear_form_problem}
\begin{split}
&\textit{\bf{RELAXED BI-MATERIAL ELASTIC PROBLEM}}\\
&GIVEN: \chi(\mathbf{x}), \ \mathbf{b}(\mathbf{x}), \ \mathbf{t}^*(\mathbf{x}),\quad &(a) \\
&\begin{split}	
&{\cal V}_{\bm \eta}\coloneqq\left\lbrace \bm{\eta}\in {\bf H}^{(1)}(\Omega)\; ;\quad\bm{\eta}=
\mathbf{0} \quad\text{on}\quad \partial_{u}\Omega \right\rbrace \\ 
\end{split} \quad&(b)\\
&FIND: \quad {\mathbf{u}}(\chi,\mathbf{x})\equiv{\mathbf{u}}_{\chi}(\mathbf{x}), \quad \bf{u}_\chi\in {\cal V}_{\bm \eta}\quad &(c)\\
&FULFILLING: \quad a_{\chi}(\mathbf{u}_{\chi},\mathbf{w}) =l(\mathbf{w}) 
\quad \forall \mathbf{w}\in{\cal V}_{\bm \eta} \quad &(d)\\ 
&\hspace{1.5cm}\left\{
\begin{split}
&a_{\chi}(\mathbf{u}_{\chi},\mathbf{w})\equiv\int_{\Omega}
\bm{\varepsilon}(\mathbf{u}_{\chi}):
{\mathbb C}_{\chi}:
\bm{\varepsilon}(\mathbf{w})
d\Omega\ \; ;\quad \bm{\varepsilon}\left( \bm{\eta}\right) =\left( \bm{\nabla}\otimes\bm{\eta}\right)^{sym}\\
&l(\mathbf{w})=\int_{\Omega} \mathbf{b}\cdot\mathbf{w}d\Omega+
\int_{\partial_{\sigma}\Omega}
{\mathbf{t}}^{*}\cdot\mathbf{w}
d\Gamma
\end{split}
\right.\quad &(e)
\end{split}
\end{equation}
where, $a_{\chi}(\mathbf{u}_{\chi},\mathbf{w})$ is a symmetric bilinear form ($a_{\chi}(\mathbf{u}_{\chi},\mathbf{w})=a_{\chi}( \mathbf{w},\mathbf{u}_{\chi})$).
\begin{remark}
	Notice that, unlike in the strong form of the relaxed elastic problem, in equations (\ref{eq_elastic_problem_2}), no derivative of the, spatially discontinuous, constitutive tensor ${\mathbb C}_{\chi}$ appears in the variational form of problem in equations  (\ref{eq_bilinear_form_problem}). The  problem description in these equations will be the one considered, from now on, and termed the (relaxed) \textnormal{state problem (or equation)}. With the boundary conditions appropriately precluding the rigid body motions, it has a unique solution. 
\end{remark}	
\begin{remark}
Additionally, it is assumed that, as the relaxation factor $\beta\rightarrow 0$ (and, thus, $\alpha\rightarrow 0$), the relaxed elastic problem, in equations (\ref{eq_elastic_problem_2}) to (\ref{eq_bilinear_form_problem}), asymptotically converges, to the solution of the \textit{single material} (${\mathfrak M}^+\equiv solid$ and ${\mathfrak M}^-\equiv voids$) elastic problem defined as:
\begin{equation} \label{eq_bilinear_form__singel_material_problem}
\begin{split}
&\bf{\textbf{SINGLE-MATERIAL ELASTIC PROBLEM}}\\
&GIVEN: \Omega, \Omega^+, \mathbf{b}(\mathbf{x}), \mathbf{t}^*(\mathbf{x}),\quad &(a) \\
&\begin{split}	
&{\cal V_{\bm \eta}}\coloneqq\left\lbrace \bm{\eta}\in {\bf H}^{(1)}(\Omega^+)\; ;\quad\bm{\eta}=
\mathbf{0} \quad\text{on}\quad \partial_{u}\Omega\cap\partial\Omega^+ \right\rbrace \\ 
\end{split} \quad&(b)\\
&FIND: \quad {\mathbf{u}}(\mathbf{x})\in {\cal V}_{\bm \eta}  \quad &(c)\\
&FULFILLING: \quad a(\mathbf{u},\mathbf{w}) =l(\mathbf{w}) 
\quad \forall \mathbf{w}\in{\cal V}_{\bm \eta} \quad &(d)\\ 
&\hspace{1.5cm}\left\{
\begin{split}
&a(\mathbf{u},\mathbf{w})\equiv\int_{\Omega^{+}}
\bm{\varepsilon}(\mathbf{u}):
{\mathbb C}:
\bm{\varepsilon}(\mathbf{w})
d\Omega\ \; ;\quad \bm{\varepsilon}\left( \bm{\eta}\right) =\left( \bm{\nabla}\otimes\bm{\eta}(\mathbf{x})\right)^{sym}\\
&l(\mathbf{w})=\int_{\Omega^{+}} \mathbf{b}\cdot\mathbf{w}d\Omega+
\int_{\partial_{\sigma}\Omega^+}
{\mathbf{t}}^{*}\cdot\mathbf{w}
d\Gamma
\end{split}
\right.\quad &(e)
\end{split}
\end{equation}
i.e. the linear elastic solution of the \textnormal{(non-relaxed or \textit{embedded}) solid/voids problem}.
\end{remark}
The fact that the relaxed bi-material version (in equations (\ref{eq_bilinear_form_problem})) is considered here as the target \textit{state problem} for the subsequent developments, is emphasized.

\subsection{Finite element discretization}
\label{sec_finite_element_discretization}
So far, no mention has been made of the numerical method used to solve the elastic state problem in equation  (\ref{eq_bilinear_form_problem}). If the finite element method is chosen, the state-equation (\ref{eq_bilinear_form_problem})-(d) yields\footnote{In the following, matrix (Voigt's) notation is used for discrete finite element formulations.}, after discretization,
\begin{equation} \label{eq_equilibrium}
\begin{split}
&{\mathbb K}_{\chi}\mathbf{d}_{\chi}=\mathbf{f}
\; ;\quad 
{\mathbb K}_{\chi}=\int_{\Omega}\mathbf{B}^T(\mathbf{x})\ {\mathbb D}_{\chi}({\bf x})\ \mathbf{B}(\mathbf{x})d{\Omega}
\; ;\quad {\mathbb D}_{\chi}({\bf x})=\left(\chi(\mathbf{x})\right)^m{\mathbb D}
&\quad(a)\\
&\bm{\sigma}_{\chi}(\mathbf{x})={\mathbb D}_{\chi}{\bm\varepsilon}_{\chi}(\mathbf{x}) \; ;\quad
{\bm\varepsilon}_{\chi}(\mathbf{x})=\mathbf{B}(\mathbf{x})\mathbf{d}_\chi
\quad ; \quad 
\mathbf{u}_{\chi}(\mathbf{x})=\mathbf{N}_{u}(\mathbf{x})\mathbf{d}_\chi 
&\quad(b)
\end{split}
\end{equation}
where $\mathbb{K}_{\chi}$ stands for the stiffness matrix, $\mathbf{f}$ for the external forces vector and ${\mathbf d}_{\chi}$ is the nodal displacement vector. In addition, $\bm{\sigma}_{\chi}(\mathbf{x})$ and $\bm{\varepsilon}_{\chi}(\mathbf{x})$ stand, respectively, for the stress and strain vectors, $\mathbf{B}({\bf x})$ is the strain matrix, ${\mathbb D}$ is the matrix version of the hard-phase constitutive tensor $\mathbb{C}$, and $\mathbf{N}_{u}({\bf x})$ is the, displacement, interpolation (shape-function) matrix.
\subsection{Relaxed Topological Derivative. Definition.}
Let us consider the function space of solutions, ${\cal V}_{\bf {u}_{\chi}}$, of the relaxed state problem in equation (\ref{eq_bilinear_form_problem}) for all possible topologies $\chi({\bf x})$, i.e.
\begin{equation}
\label{eq_space_of_displacements}
\left\{
\begin{split}
&{\cal V}_{\chi}\coloneqq\left\{\chi : \Omega \rightarrow \{\beta,1\}\ \right\}\\
&{\cal V}_{\bf {u}_{\chi}}\coloneqq\left\{{\bf u}_{\chi}\;/\;{\bf u}_{\chi}:\Omega\rightarrow{\mathbb R}^n ,\ n\in\{2,3\}\; ;\quad \chi\in{\cal V}_{\chi}\; ;\quad a_{\chi}(\mathbf{u}_{\chi},\mathbf{w}) =l(\mathbf{w}) \quad \forall \mathbf{w}\in{\cal V}_{\bm \eta}\right\}
\end{split}
\right.
\end{equation}
and the function space, ${\cal V}_{\Omega}$, of $L^2$-integrable mappings $G\left(\chi({\bf x}),{\bf u}_{\chi}({\bf x}),{\bf x}\right)$, in $\mathbb R$, and the corresponding functional  ${\cal J(\chi)}$ i.e.
\begin{equation}
\label{eq_functional}
\left\{
\begin{split}
&{\cal V}_{\Omega}\coloneqq\left\{G\;/\;G:{\cal V}_{\chi}\times{\cal V}_{\bf {u}_{\chi}}\times\Omega\rightarrow{\mathbb R}\right\}\\
&{\cal J}(\chi)\equiv{\cal J}_{\chi}:{\cal V}_\Omega\rightarrow {\mathbb R}\; ;\quad{\cal J}_{\chi}=
\int_{\Omega}\underbrace{G\left({\chi,\bf u}_{\chi}({\bf x}),{\bf x}\right)}_{\displaystyle{F(\chi,{\bf x})}}d\Omega=\int_{\Omega}F(\chi,{\bf x})d \Omega\\
\end{split}
\right.
\end{equation}

\begin{figure}[h]
	\centering
	\includegraphics[width=14cm, height=9.5cm]{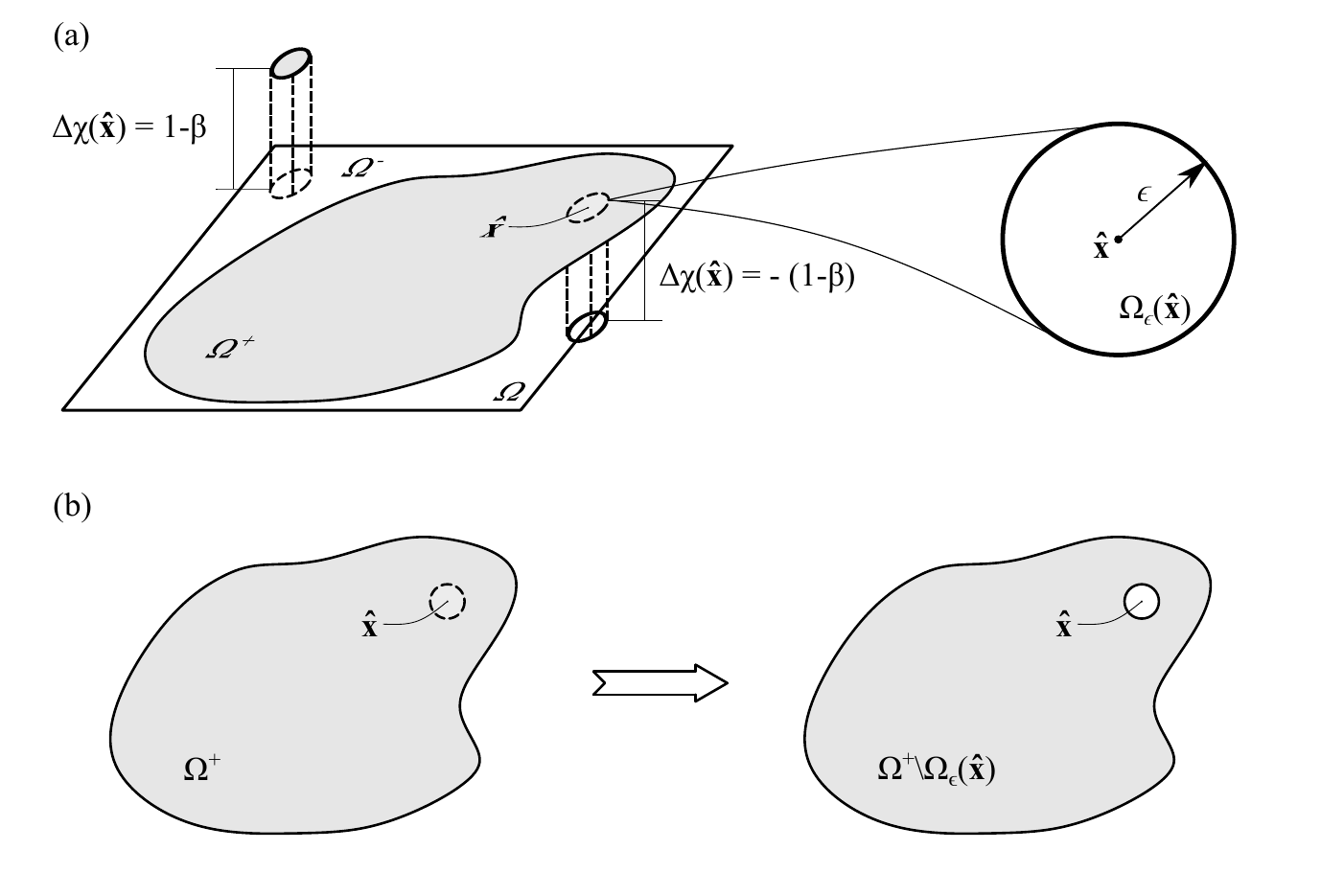}
	\caption{Topological derivative settings: (a) \textit{Relaxed} Topological Derivative (RTD), (b) \textit{Exact} Topological Derivative  (TD).}
	\label{fig_perturbation_domain}
\end{figure}

Let us now consider a specific point, $\hat{\mathbf{x}}\in\Omega$, and a sequence of balls\footnote{ ${\Omega_\epsilon}(\bf x)$ is a circle (in 2D) or a sphere in (3D).}, $\Omega_{\epsilon}(\hat{\mathbf{x}})$  of radius $ \epsilon$,
 (\textit{the perturbation domain}), surrounding $\hat{\mathbf{x}}$, i.e.
$\hat {\mathbf{x}}\in\Omega_\epsilon\subset\Omega$,   where
$\epsilon\in\mathbb{R}^+$ and $\epsilon\rightarrow 0$ is a regularization parameter (see figure \ref{fig_perturbation_domain}).
Let us finally consider the space of perturbations, due to material phase exchange \textit{only at points of the domain $\Omega_{\epsilon}(\hat{\mathbf{x}})$},
\begin{equation}
\label{eq_functional_perturbations}
\begin{split}
{\Delta\cal V}_{\hat{\mathbf{x}},\epsilon}\coloneqq\left\{\eta_{\hat{\bf x},\epsilon}\; ;\quad
\eta_{\hat{\bf x},\epsilon}(\mathbf x)=\Delta\chi({\bf x}) \quad\forall {\bf x}\in\Omega_{\epsilon}(\hat{\bf x})\; ;\quad   \eta_{\hat{\bf x},\epsilon}(\mathbf x)=0 \quad\forall\mathbf{x}\in\Omega\backslash\Omega_{\epsilon}(\hat{\bf x})\right\}
\end{split}
\end{equation}
where $\Delta\chi({\bf x)}$ stands for the increment in the value of $\chi({\bf x)}$ due to exchange of the type of material phase at point $\bf{x}$. In view of equation (\ref{eq_domain_splitting}), the value ${\Delta\chi}({\bf x})$, is:
\begin{equation}
\label{eq_alpha}
{\Delta\chi}({\bf x})=
\left\{
\begin{split}
-(&1-\beta)<0  &\textit{for}  \quad{\bf x}\in\Omega^{+}  \\ 
&1-\beta\,\,>0 &\textit{for}  \quad{\bf x}\in\Omega^{-}
\end{split} 
\right.
\end{equation}

Let us now consider the following \emph{asymptotic expansion} of the functional ${\cal J}(\chi)$ in equation (\ref{eq_functional}), in terms of the perturbation $\eta_{\hat{\bf x},\epsilon}\in{\Delta\cal V}_{\hat{\mathbf{x}},\epsilon}$ 
\begin{equation}
\label{eq_assimptotic_expansion}
{\cal J}(\chi+\eta_{\hat{\bf x},\epsilon})={\cal J}(\chi)+\dfrac{\delta {\cal J}(\chi)}{\delta \chi}(\hat{\bf x})\ \mu[\Omega_{\epsilon}(\hat{\bf x})]+{ o}(\mu[\Omega_{\epsilon}(\hat{\bf x})])
\end{equation}
where, $\mu[\Omega_{\epsilon}(\hat{\bf x})]$ is an appropriate measure\footnote{For practical purposes, it will be the volume or the surface (depending on the case) of the perturbation domain $\mu[\Omega_{\epsilon}(\hat{\bf x})]$.} of the perturbation domain $\Omega_{\epsilon}(\hat{\bf x})$, fulfilling the following conditions
\begin{equation}
\label{eq_properties}
\left\{
\begin{split}
&1) \quad\mu[\Omega_{\epsilon}(\hat{\bf x})]>0\quad&(a)\\
&2) \quad\lim\limits_{\epsilon\rightarrow{0}}\mu[\Omega_{\epsilon}(\hat{\bf x})]=0 \quad&(b)\\
&3) \quad\textit{makes $\dfrac{\delta {\cal J}(\chi)}{\delta \chi}(\hat{\bf x})$ nonzero valued and bounded}\quad&(c)
\end{split}
\right.
\end{equation}

Specific mathematical aspects of the previous derivations are out of the scope of this work and left to be proven in specialized contexts. From equation (\ref{eq_assimptotic_expansion}), and conditions in equation (\ref{eq_properties}), we define the Relaxed Topological Derivative (RTD) of the functional ${\cal J}(\chi)$ at point $\hat{\bf x}$ as 
\begin{equation}
\label{eq_RTD_definition}
\begin{split}
\dfrac{\delta {\cal J}(\chi)}{\delta \chi}(\hat{\bf x})=&
\lim\limits_{\epsilon\rightarrow{0}}\dfrac{1}{\mu[\Omega_{\epsilon}(\hat{\bf x})]}\left[{\cal J}(\chi+\eta_{\hat{\bf x},\epsilon})-{\cal J}(\chi)\right]
\end{split}
\end{equation}
In view of the previous derivations the RTD can be conceptually defined as
\begin{definition}[RTD]
	\label{def_RTD}
	The (relaxed) topological derivative (RTD) of functional ${\cal J}(\chi)$, at point $\hat{\bf x}\in\Omega$, is a finite measure of the sensitivity of ${\cal J}(\chi)$ to the change of topology  at that point, in the $\alpha$-relaxed setting described in sections \ref{sec_Relaxed_characteristic_function} and \ref{sec_relaxed_bimaterial_linear_elastic_problem}. It is computed as \textit{the limit, as $\alpha\rightarrow 0$; $\beta\rightarrow 0$, of the change of ${\cal J}(\chi)$, when the material phase type is exchanged $(\chi\leftrightarrow\chi+\Delta\chi({\hat{\bf x}}))$ at the neighborhood, $\Omega_{\epsilon}(\hat {\bf x})$} per unit of the perturbation measure,  $\mu[\Omega_{\epsilon}(\hat {\bf x})]$.
\end{definition}
	 	Notice in the previous derivations the similarity of the procedure for obtainment of the topological derivative, here acronymized as (RTD), in the \textit{relaxed bi-material setting} in equations  (\ref{eq_RTD_definition}), and that in the \textit{embedded/single-material} setting: the \textit{exact} Topological Derivative (TD) \cite{Cea2000,Sokolowski,Novotny2003,Novotny2013}. 
	The fundamental difference is that, in the second case the TD is considered in the context of  a \emph{a single (solid) phase}, occupying the domain $\Omega^{+}\subset\Omega$, the corresponding functional being ${\cal J}(\Omega^+)=\int_{\Omega^+}(\newplaceholder)({\bf x})d \Omega$. This functional is perturbed by \emph{removing} the material in an infinitesimal neighborhood, $\Omega_{\epsilon}(\hat {\bf x})$, around point $\hat{\bf x}$, so that the  \textit{perturbed solid domain} becomes $\Omega^{+}\backslash \Omega_{\epsilon}(\hat {\bf x})$ (see figure \ref{fig_perturbation_domain}).
	
	Then the topological derivative is defined as:
\begin{equation}
	\label{eq_topological_derivative}
	{\cal D}_{T}(\hat {\bf x})\coloneqq\lim\limits_{\epsilon\rightarrow 0}\dfrac{\int_{\Omega^{+}\backslash \Omega_{\epsilon}(\hat {\bf x})}(\newplaceholder)({\bf x})d \Omega-\int_{\Omega^+}\ (\newplaceholder)({\bf x})d \Omega}{\mu[\Omega_{\epsilon}(\hat{\bf x})]}
\end{equation} 

	Unlike in the RTD derivation, in the TD case the integration domains, $\Omega^{+}\backslash \Omega_{\epsilon}$ and $\Omega^{+}$, in equation (\ref{eq_topological_derivative}) are not the same, and it is not possible to establish a homeomorphism between both domains \cite{Novotny2003}. This motivates a subsequent analytical procedure that involves some additional, and problem dependent, limit and asymptotic analyzes \cite{Giusti2016}.
	
	Along this work, it will be assumed that the RTD is a convenient approximation of the TD, requesting much simpler calculations, and it will be checked it provides a \textit{sufficient approximation of the cost sensitivity} to lead to accurate solutions for a number of engineering problems. 
	
\subsection{Examples of relaxed variational topological derivatives}
In the following sections, some results for relaxed  topological derivatives, used in this work, are presented.
\subsubsection{Relaxed topological derivative of an integral over the design domain}
For the functional, ${\cal J}{\chi}$ ,  defined as in equation (\ref{eq_functional}), the asymptotic expansion in the direction of the perturbation $\eta_{\hat{\bf x},\epsilon}$ reads
\begin{equation}
\label{eq_assymptotic_expansion_F}
\begin{split}
{\cal J}(\chi+\eta_{\hat{\bf x},\epsilon})=&\int_{\Omega}F(\chi+\eta_{\hat{\bf x},\epsilon},{\bf x})d \Omega=
\int_{\Omega}\left(F(\chi,{\bf x})+\dfrac{\partial F(\chi,{\bf x})}{\partial \chi}\eta_{\hat{\bf x},\epsilon}+{ o}(\eta_{\hat{\bf x},\epsilon})\right)d \Omega=\\
=&{\cal J}(\chi)+\int_{\Omega}\dfrac{\partial F(\chi,{\bf x})}{\partial \chi}\eta_{\hat{\bf x},\epsilon}d \Omega+{ o}(\mu[\Omega_{\epsilon}(\hat{\bf x})])=\\
=&{\cal J}(\chi)+\int_{\Omega}\left(\dfrac{\partial G\left(\chi,{\bf u}_{\chi}({\bf x}),{\bf x}\right)}{\partial \chi}+
\dfrac{\partial G\left(\chi,{\bf u}_{\chi}({\bf x}),{\bf x}\right)}{\partial {\bf u}_{\chi}}\cdot\dfrac{\partial {\bf u}_{\chi} }{\partial \chi}\right)d\Omega
+{ o}(\mu[\Omega_{\epsilon}(\hat{\bf x})])
\end{split}
\end{equation}
where, when necessary, the displacement derivatives ${\partial{\bf u}_{\chi}}\slash{\partial \chi}$ are obtained from the relaxed elastic problem (the state equation) in equations (\ref{eq_bilinear_form_problem}), either directly by $\chi$-differentiation of its solution, or indirectly, by using the adjoint-problem method \cite{Cea2000}.
Then, the topological derivative of the functional ${\cal J}_{\chi}$ is obtained by replacing equation (\ref{eq_assymptotic_expansion_F}) into equation (\ref{eq_RTD_definition}), as
\begin{equation}
\label{eq_functional1}
\begin{split}
\dfrac{\delta {\cal J}(\chi)}{\delta \chi}(\hat{\bf x})=&\dfrac{\delta}{\delta \chi}\left[\int_{\Omega}F({\chi},{\bf x})d\Omega\right](\hat{\bf x})=\lim\limits_{\epsilon\rightarrow{0}}\dfrac{1}{\mu[\Omega_{\epsilon}(\hat{\bf x})]}\left[{\cal J}(\chi+\eta_{\hat{\bf x},\epsilon})-{\cal J}(\chi)\right]=\\
=&\lim\limits_{\epsilon\rightarrow{0}}\dfrac{1}{\mu[\Omega_{\epsilon}(\hat{\bf x})]}\left(
\int_{\Omega}\dfrac{\partial F({\chi},{\bf x})}{\partial{\chi}}\eta_{\hat{\bf x},\epsilon}d\Omega+{ o}(\mu[\Omega_{\epsilon}(\hat{\bf x})])\right)=\lim\limits_{\epsilon\rightarrow{0}}\dfrac{1}{\mu[\Omega_{\epsilon}(\hat{\bf x})]}
\int_{\Omega_{\epsilon}(\hat{\bf x})}{\dfrac{\partial F({\chi},{\bf x})}{\partial{\chi}}}\Delta \chi (\hat{\bf x})d\Omega=\\
=&\lim\limits_{\epsilon\rightarrow{0}}\dfrac{\vert\Omega_{\epsilon}(\hat{\bf x})\vert}{\mu[\Omega_{\epsilon}(\hat{\bf x})]}
\left[\dfrac{\partial F(\chi({\bf x}),{\bf x})}{\partial \chi}\Delta \chi({\bf x})\right]_{{\bf x}=\hat{\bf x}}
\end{split}
\end{equation}
where equations (\ref{eq_RTD_definition}) and (\ref{eq_assymptotic_expansion_F}), and the definition of the perturbation $\eta_{\hat{\bf x},\epsilon}$ in equation (\ref{eq_functional_perturbations}), have been taken into account. In order to fulfill condition in equation (\ref{eq_properties})-(c), a suitable choice for $\mu[\Omega_{\epsilon}(\hat{\bf x})]$ is 
\begin{equation}
\label{eq_choice_f_epsilon_1}
\mu[\Omega_{\epsilon}(\hat{\bf x})]=\vert\Omega_{\epsilon}(\hat{\bf x})\vert=\frac{4}{3}\pi \epsilon^3
\end{equation}
this yielding, in terms of the descriptions $F(\newplaceholder)$ and $G({\newplaceholder})$ in equation (\ref{eq_functional})
\begin{equation}
\label{eq_result_1}
\begin{split}
\dfrac{\delta {\cal J}(\chi)}{\delta \chi}(\hat{\bf x})=&\dfrac{\delta}{\delta \chi}\left[\int_{\Omega}F({\chi},{\bf x})d\Omega\right](\hat{\bf x})=
\left[\dfrac{\partial F({\chi},{\bf x})}{\partial {\chi}}\right]_{{\bf x}=\hat{\bf x}}\Delta \chi(\hat{\bf x})=\\
=&\left[\dfrac{\partial G\left(\chi,{\bf u}_{\chi},{\bf x}\right)}{\partial \chi}+\dfrac{\partial G(\chi,{\bf u}_{\chi},{\bf x})}{\partial {\bf u}_{\chi}}\cdot\dfrac{\partial{\bf u}_{\chi}}{\partial\chi}\right]_{{\bf x}=\hat{\bf x}}\Delta \chi(\hat{\bf x})
\end{split}
\end{equation}
\begin{remark}
	Notice that the linear character of the integration and differentiation operations, involved in the determination of the relaxed variational topological derivative in equations (\ref{eq_functional1}) to (\ref{eq_result_1}), confers to the RTD the same linear character, i.e.
	\begin{equation}
	\label{eq_linear_character}
	\dfrac{\delta \left({\cal J}(\chi)+\lambda{\cal K}(\chi)\right)}{\delta \chi}(\hat{\bf x})=
		\dfrac{\delta {\cal J}(\chi)}{\delta \chi}(\hat{\bf x})+\lambda
		\dfrac{\delta {\cal K}(\chi)}{\delta \chi}(\hat{\bf x})\; ;\quad{\cal J}(\chi),{\cal K}(\chi)\in{\cal V}_{\Omega}\; ;\quad\lambda\in{\mathbb R}
	\end{equation}
\end{remark}  
\subsubsection{Relaxed topological derivative of a material phase volume}
\begin{figure}[h]
	\centering
	\includegraphics[width=9cm, height=6cm]{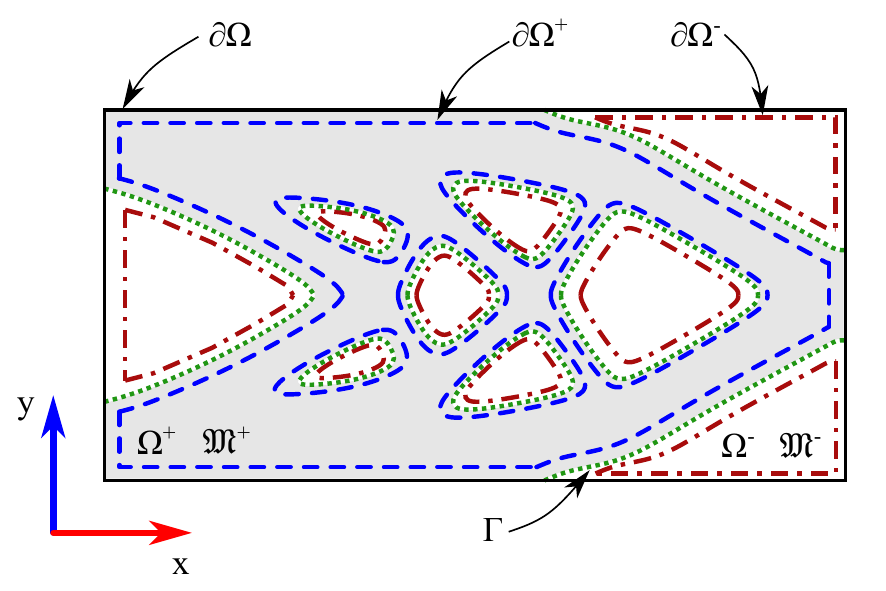}
	\caption{Design domain $\Omega$: material-phase domains ($\Omega^+$ and $\Omega^-$), material-phase boundaries ($\partial\Omega^+$ and $\partial\Omega^-$), and phase interfaces ($\Gamma$).}
	\label{fig_phase_volume}
\end{figure}

Let us consider the functional defining the volume of phase ${\mathfrak M}^{+}$ (see figure \ref{fig_phase_volume})
\begin{equation}
\label{eq_volume_phase_1}
{\cal J}(\chi)\equiv\vert\Omega^{+}(\chi)\vert=\int_{\Omega^{+}(\chi)} d\Omega\; ;\quad {\cal J(\chi)}\in {\cal V}_\Omega
\end{equation}
and the identity
\begin{equation}
\label{eq_identity_phase_volume}
\lim\limits_{\epsilon\rightarrow 0}{\cal J}(\chi+\eta_{\hat{\bf x},\epsilon})=\lim\limits_{\epsilon\rightarrow 0}\vert\Omega^{+}(\chi+\eta_{\hat{\bf x},\epsilon})\vert=\lim\limits_{\epsilon\rightarrow 0}
\left(\vert\Omega^+(\chi)\vert+\text{sgn}(\Delta\chi(\hat{\bf x}))\vert \Omega_{\epsilon}(\hat{\bf x})\vert\right)
\end{equation}
where $\text{sgn}(\Delta\chi(\hat{\bf x}))=-1\ \forall\hat{\bf x}\in\Omega^+$ and $\text{sgn}(\Delta\chi(\hat{\bf x}))=1\ \forall\hat{\bf x}\in\Omega^-$ (see equation (\ref{eq_alpha})).

The corresponding topological derivative, at point $\hat{\bf x}\in\Omega$, can be now computed from equations (\ref{eq_RTD_definition}) and (\ref{eq_identity_phase_volume}) as
\begin{equation}
\label{eq_RTD_conmputation_phase_volume}
\begin{split}
\dfrac{\delta {\cal J}(\chi)}{\delta \chi}(\hat{\bf x})=
&\lim\limits_{\epsilon\rightarrow{0}}\dfrac{1}{\mu[\Omega_{\epsilon}(\hat{\bf x})]}\left[{\cal J}(\chi+\eta_{\hat{\bf x},\epsilon})-{\cal J}(\chi)\right]=
\lim\limits_{\epsilon\rightarrow{0}}\dfrac{\vert\Omega^+(\chi+\eta_{\hat{\bf x},\epsilon})\vert-\vert\Omega^+(\chi)\vert}{\mu[\Omega_{\epsilon}(\hat{\bf x})]}\\
=&\lim\limits_{\epsilon\rightarrow{0}}\dfrac{\vert\Omega^+(\chi)\vert+\text{sgn}(\Delta\chi(\hat{\bf x}))\vert \Omega_{\epsilon}(\hat{\bf x})\vert-\vert\Omega^+(\chi)\vert}{\mu[\Omega_{\epsilon}(\hat{\bf x})]}=
\text{sgn}(\Delta\chi(\hat{\bf x}))\lim\limits_{\epsilon\rightarrow{0}}\dfrac{\vert\ \Omega_{\epsilon}(\hat{\bf x})\vert}{\mu[\Omega_{\epsilon}(\hat{\bf x})]}=\text{sgn}(\Delta\chi(\hat{\bf x}))\\
\end{split}
\end{equation}
where the choice for $\mu[\Omega_{\epsilon}(\hat{\bf x})], $ to fulfill the condition in equation (\ref{eq_properties})-(c), is 
\begin{equation}
\label{eq_choice_f_epsilon_2}
\mu[\Omega_{\epsilon}(\hat{\bf x})]=\vert\ \Omega_{\epsilon}(\hat{\bf x})\vert=\frac{4}{3} \pi \epsilon^3
\end{equation}
In summary
\begin{equation}
\label{eq_vtd_phase_hard_phase_volume}
\dfrac{\delta \vert\Omega^{+}(\chi)\vert}{\delta \chi}(\hat{\bf x})=\text{sgn}(\Delta\chi(\hat{\bf x}))=
\left\{ 
\begin{split}
&-1\quad\forall\hat{\bf x}\in\Omega^{+}\\
&+1\quad\forall\hat{\bf x}\in\Omega^{-}
\end{split}
\right.	
\end{equation}
For the RTD of the soft-phase ${\mathfrak M}^{-}$ case we consider the identity
\begin{equation}
\label{eq_identity_1}
\vert\Omega^+(\chi)\vert+\vert\Omega^-(\chi)\vert=\vert\Omega\vert
\end{equation}
which yields
\begin{equation}
\label{eq_vtd_phase_soft_phase_volume}
\dfrac{\delta \vert\Omega^{-}(\chi)\vert}{\delta \chi}(\hat{\bf x})=
-\dfrac{\delta \vert\Omega^{+}(\chi)\vert}{\delta \chi}(\hat{\bf x})=-\text{sgn}(\Delta\chi(\hat{\bf x}))=
\left\{ 
\begin{split}
&+1\quad\forall\hat{\bf x}\in\Omega^{+}\\
&-1\quad\forall\hat{\bf x}\in\Omega^{-}
\end{split}
\right.	
\end{equation}
An alternative procedure to obtain the RTD of the functional in equation (\ref{eq_volume_phase_1}), consists of rephrasing it, considering equation (\ref{eq_chi}), and applying the result in equation (\ref{eq_result_1}) (see also Table \ref{table_RTD_examples}), i.e.
\begin{equation}
\label{eq_volume_phase_2}
	\left\{
	\begin{split}
	&\vert \Omega^+(\chi)\vert=\int_{\Omega^{+}}d\Omega=\int_{\Omega}\underbrace{\dfrac{\chi-\beta}{1-\beta}}_{\displaystyle{F(\chi)}}d \Omega\\
	&F(\chi)\equiv \dfrac{\chi-\beta}{1-\beta}\Rightarrow\dfrac{\partial F(\chi)}{\partial\chi}=\dfrac{ 1}{1-\beta}\\
	&\dfrac{\delta \vert\Omega^{+}(\chi)\vert}{\delta \chi}(\hat{\bf x})=\left[\dfrac{\partial F({\chi})}{\partial {\chi}}\right]_{{\bf x}=\hat{\bf x}}
	\Delta \chi(\hat{\bf x})=\dfrac{1}{1-\beta}\underbrace{(1-\beta)(\text{sgn}(\Delta\chi(\hat{\bf x})))}_{\displaystyle{\Delta \chi(\hat{\bf x})}}=
	\text{sgn}(\Delta\chi(\hat{\bf x}))
	\end{split}
	\right.
\end{equation}
which, as expected, coincides with the result in equation (\ref{eq_vtd_phase_hard_phase_volume}). 
\subsubsection{Relaxed topological derivative of a phase perimeter}
We now consider the functional corresponding to the perimeter of the hard-phase, ${\cal P}^{+}(\chi)$, defined as (see figure \ref{fig_phase_volume})
\begin{equation}
\label{eq_perimeter}
{\cal J(\chi)}={\cal P}^{+}(\chi)=\int_{\partial\Omega^+(\chi)}d\Gamma=\vert\partial\Omega^+(\chi)\vert
\end{equation}
and the identity
\begin{equation}
\label{eq_identity_perimeter}
\lim\limits_{\epsilon\rightarrow 0}{\cal J}(\chi+\eta_{\hat{\bf x},\epsilon})=\lim\limits_{\epsilon\rightarrow 0}\vert\partial\Omega^{+}(\chi+\eta_{\hat{\bf x},\epsilon})\vert=
\lim\limits_{\epsilon\rightarrow 0}\left(\vert\partial\Omega^{+}(\chi)\vert+\vert\partial \Omega_{\epsilon}(\hat{\bf x})\vert\right)
\end{equation}
The corresponding topological derivative, at point $\hat{\bf x}\in\Omega$, can be now computed from equations (\ref{eq_RTD_definition}) and (\ref{eq_identity_perimeter}) as:
\begin{equation}
\label{eq_RTD_conmputation}
\begin{split}
\dfrac{\delta {\cal J}(\chi)}{\delta \chi}(\hat{\bf x})=
&\lim\limits_{\epsilon\rightarrow{0}}\dfrac{1}{\mu[\Omega_{\epsilon}(\hat{\bf x})]}\left[{\cal J}(\chi+\eta_{\hat{\bf x},\epsilon})-{\cal J}(\chi)\right]=
\lim\limits_{\epsilon\rightarrow{0}}\dfrac{\vert\partial\Omega^+(\chi+\eta_{\hat{\bf x},\epsilon})\vert-\vert\partial\Omega^+(\chi)\vert}{\mu[\Omega_{\epsilon}(\hat{\bf x})]}=\\
=&\lim\limits_{\epsilon\rightarrow{0}}\dfrac{\vert\partial\Omega^+(\chi)\vert+\vert\partial \Omega_{\epsilon}(\hat{\bf x})\vert-\vert\partial\Omega^+(\chi)\vert}{\mu[\Omega_{\epsilon}(\hat{\bf x})]}=
\lim\limits_{\epsilon\rightarrow{0}}\dfrac{\vert\partial \Omega_{\epsilon}(\hat{\bf x})\vert}{\mu[\Omega_{\epsilon}(\hat{\bf x})]}=1\\	
\end{split}
\end{equation}
where the choice for $\mu[\Omega_{\epsilon}(\hat{\bf x})], $ in equation (\ref{eq_properties})-(c), is 
\begin{equation}
\label{eq_choice_f_epsilon_3}
\mu[\Omega_{\epsilon}(\hat{\bf x})]=\vert\partial \Omega_{\epsilon}(\hat{\bf x})\vert=4 \pi \epsilon^2
\end{equation}
From the previous derivation it is evident that the same result would be found for the perimeter of the soft-phase domain
${\cal P}^{-}(\chi)$. Therefore, in summary,
\begin{equation}
\label{eq_phase_perimeter_derivative}
\left\{
\begin{split}
&{\cal P}^{+}(\chi)\coloneqq\vert\partial\Omega^{+}(\chi)\vert&\Rightarrow\quad&\ \dfrac{\delta{\cal P}^+(\chi)}{\delta\chi}(\hat{\mathbf{x}})=1	\\
&{\cal P}^{-}(\chi)\coloneqq\vert\partial\Omega^{-}(\chi)\vert&\Rightarrow\quad&\ \dfrac{\delta{\cal P}^-(\chi)}{\delta\chi}(\hat{\mathbf{x}})=1
\end{split}
\right.
\end{equation}
\begin{remark}[Interface measure]
	A measure frequently considered in the topological optimization literature \cite{Fernandes1999} is the \emph{measure of the phases interface} $\Gamma(\chi)$ (see figure \ref{fig_phase_volume}), which does not correspond to any of the phase perimeters, ${\cal P}^+$ and ${\cal P}^-$, in equation (\ref{eq_phase_perimeter_derivative}), but to the measure, $\vert\Gamma(\chi)\vert$,  of the geometric interface, $ \Gamma\coloneqq({\partial{\Omega}^+}\cup{\partial{\Omega}^-})$, shared by both phases (thus, excluding the boundary $\partial \Omega$, see figure \ref{fig_phase_volume}).
	
	The RTD of $\Gamma(\chi)$ can be readily obtained from the following identity
	\begin{equation}
	\label{eq_identity_perimeters}
	\underbrace{\vert\partial\Omega^{+}(\chi)\vert}_{\displaystyle{{\cal P}^+(\chi)}}+
	\underbrace{\vert\partial\Omega^{-}(\chi)\vert}_{\displaystyle{{\cal P}^-(\chi)}}= 2\vert\Gamma(\chi)\vert+\vert\partial\Omega\vert
	\end{equation}
	and the RTD of equation (\ref{eq_identity_perimeters}) yields
	\begin{equation}
	\label{eq_topological_derivative_gamma}
	\dfrac{\delta {\vert\Gamma}(\chi)\vert}{\delta \chi}(\hat{\bf x})=
	\frac{1}{2}\left( \dfrac{\delta {{\cal P}^+(\chi)}}{\delta \chi}(\hat{\bf x})+  
	\dfrac{\delta {{\cal P}^-(\chi)}}{\delta \chi}(\hat{\bf x})\right)=1
	\end{equation}
    where equations (\ref{eq_phase_perimeter_derivative}) have been taken into account.
\end{remark}
In Table \ref{table_RTD_examples}, examples of commonly used RTD derivatives are summarized.

\begin{table}\centering
\begin{tabular}{c|c|c}
	\hline 
	\rule[-2ex]{0pt}{5ex}  {Functional} ${\cal J}_{\chi}$ &  {RTD} $\left( \delta {\cal J}_{\chi} / {\delta \chi}\right)(\hat{\mathbf x}) $ &$\substack{\displaystyle{\textit{Perturbation measure}}\\ \displaystyle{\mu[\Omega_{\epsilon}(\hat{\bf x})]}}$\\ 
	\hline 
	\rule[-3ex]{0pt}{7ex}  $\int_{\Omega} F(\chi,\mathbf{x})d\Omega$  & $ \left[ \dfrac{\partial F(\chi({{\mathbf{x}}}),{{\mathbf{x}}})}{\partial \chi}\right]_{\mathbf{x}\equiv\hat{\mathbf{x}}}{\Delta\chi}({\hat{\bf x}})  $
	&$ \vert \Omega_{\epsilon}(\hat{\bf x})\vert  $\\  
	\hline 
	\rule[-2ex]{0pt}{5ex}  $\vert\Omega^{+}(\chi)\vert \coloneqq\int_{\Omega^{+}(\chi)}d\Omega$  & $\text{sgn}(\Delta\chi(\hat{\bf x}))$& $ \vert \Omega_{\epsilon}(\hat{\bf x})\vert  $\\
	\hline   
	\rule[-2ex]{0pt}{5ex}  $\vert\Omega^{-}(\chi)\vert \coloneqq\int_{\Omega^{-}(\chi)}d\Omega$  & $\text{sgn}(\Delta\chi(\hat{\bf x}))$& $ \vert \Omega_{\epsilon}(\hat{\bf x})\vert  $\\
	\hline   
	\rule[-2.5ex]{0pt}{6ex}  ${\cal P}^{+}(\chi)\coloneqq\int_{\partial\Omega^{+}(\chi)} d\Gamma$ & $1$ &$ \vert \partial\Omega_{\epsilon}(\hat{\bf x})\vert  $\\
	\hline 
	\rule[-2.5ex]{0pt}{6ex}  ${\cal P}^{-}(\chi)\coloneqq\int_{\partial\Omega^{-}(\chi)} d\Gamma$ & $1$ &$ \vert \partial\Omega_{\epsilon}(\hat{\bf x})\vert $\\
	\hline 
	\rule[-2.5ex]{0pt}{6ex}  $ \Gamma(\chi)\coloneqq\int_{({\partial{\Omega}^+}\cap{\partial{\Omega}^-})(\chi)} d\Gamma$ & $1$ & $ \vert \partial\Omega_{\epsilon}(\hat{\bf x})\vert $\\
	\hline
\end{tabular}
	\caption{Relaxed topological derivative examples. ${\Delta\chi}({\hat{\bf x}})=-(1-\beta)\ \forall\ \hat {\bf x}\in\Omega^{+}$, ${\Delta\chi}({\hat{\bf x}})=(1-\beta)\ \forall\ \hat {\bf x}\in\Omega^{-}$. } 
	\label{table_RTD_examples}
\end{table}

\section{Application to volume constrained topological optimization problems}
\label{sec_Variational_optimization_problem_equality_restricted}
Let us consider the following topological optimization problem in the design domain $\Omega$
\begin{equation} 
\label{eq_minimization_restricted}
\begin{split}
&FIND:\quad 	\chi(\mathbf{x})={\cal H}_{\beta}({\psi}\left(\mathbf{x})\right)\; ;\quad\chi:\Omega\rightarrow\{\beta,1\}\; ;\quad\psi\in H^1(\Omega)\quad\\
&FULFILLING: \\
&\hspace{1.5cm}\chi^{*}  = \underset{\chi} {\operatorname{argmin}}\quad{\cal J}\left( \chi\right)\equiv\int_{\Omega}{F(\chi,\mathbf{x})} d\Omega &\quad(a) \\
&\textit{SUBJECT TO}:  \\   
&\hspace{1.5cm}{\cal C}(\chi)\equiv \vert\Omega^+(\chi)\vert-{\bar V}=\vert\Omega\vert-\vert\Omega^-(\chi)\vert-{\bar V}=0 &\quad(b)
\end{split}
\end{equation}
where, in equation (\ref{eq_minimization_restricted})-(b), $\bar V$ is the target volume for the material \textit{hard-phase} ($ {\mathfrak M}^{+} $) and the equality $\vert\Omega^+(\chi)\vert=\vert\Omega\vert-\vert\Omega^-(\chi)\vert$ has been used.

\subsection{Penalized functional for non-smooth optimization problems. Optimality criterion}
In order to face the difficulty to impose the minimization, accounting for the constraint in equation (\ref{eq_minimization_restricted})-(b) in non-differentiable function spaces, we will resort to a sequence of penalized problems (see \cite{Bertsekas1996}), by means of the parameter $\frac{1}{k}$, penalizing the values of $\chi$ that violate the constraint ${\cal C}(\chi)=0$. Therefore, we start from the following penalized problem, in terms of the \textit{extended} functional, ${\cal J}^{ext}$,
\begin{equation} 
\label{eq_minimization_restricted_penalized}
\begin{split}
	&\left\{
		\begin{split}
		&FIND:\quad 	\chi_{k}(\mathbf{x})={\cal H}_{\beta}({\psi}_{k}\left(\mathbf{x})\right):\Omega\rightarrow\{\beta,1\} \; ;\quad\psi_{k}\in H^1(\Omega)\\
		&FULFILLING: \\
		&\hspace{1.5cm}\chi^{*}_{k}  =  \underset{\chi_{k}} {\operatorname{argmin}}\quad{\cal J}^{ext}\left( \chi_{k}\right)\equiv{\cal J}(\chi_k)+\dfrac{1}{2k}({\cal C}(\chi_k))^2  \\
		&\textit{SUBJECT TO}:  \\   
		&\hspace{1.5cm}{\cal C}(\chi_{k})\equiv \vert\Omega^+\vert(\chi_{k})-{\bar V}= 0 \\
		\end{split}
	\right.&\quad(a)\\
	&\quad\chi^*=\lim\limits_{k\rightarrow 0}\chi^{*}_{k}&\quad(b)
\end{split}
\end{equation} 
In order to solve the problem in equation (\ref{eq_minimization_restricted_penalized}) we impose the following \textit{optimality criterion}
\begin{equation}
\label{eq_condition}
\begin{split}
\dfrac{\delta {{\cal J}^{ext}}(\chi_k)}{\delta\chi_k}({\bf x})&=
\dfrac{\delta {{\cal J}}(\chi_k)}{\delta\chi}({\bf x})+\dfrac{C(\chi_k)}{k}\dfrac{\delta {\cal C}(\chi_k)}{\delta\chi_k}({\bf x})= \\
&=\left(\dfrac{\partial F \left(\chi_k,\mathbf{x}\right)}{\partial \chi_k}{\Delta\chi}({\bf x})
+\dfrac{C(\chi_k)}{k}\text{sgn}(\Delta\chi({\bf x}))\right)>{0}
\; ;\quad \forall\mathbf{x}\in\Omega
\end{split}
\end{equation}
where the topological derivation rules in equations (\ref{eq_result_1}) and (\ref{eq_volume_phase_1}) have been applied. 

\begin{remark}[Overall-increasing cost function sensitivity]
Condition in equation (\ref{eq_condition}) arises from the following argument (inspired by the one in \cite{Amstutz2006}): 
by definition, the RTD, $\left({\delta {{\cal J}^{ext}}(\chi_{k})}\slash{\delta \chi_{k}}({\bf x})\right)$, measures the sensitivity to topological changes of the functional ${{\cal J}^{ext}}(\chi_{k})$ in front of local phase exchanges, ${\Delta\chi}({\bf x})$ at point ${\bf x}$, per unit of measure of the perturbed domain (see definition \ref{def_RTD}). 
Therefore, denoting $\hat{\Omega}$ the subset of $\Omega$ where these exchanges take place, i.e.
\begin{equation}
\label{eq_Omega_1-2} 
\begin{array}{lrll}
\hat{\Omega}\subset\Omega&\rightarrow\quad &\hat{\Omega}&\coloneqq\left\{{\bf x}\in\Omega \; ;\quad \Delta\chi_{k}({\bf{x}})={\Delta\chi}\ne 0\right \}\\
\Omega\backslash\hat{\Omega}\subset\Omega&\rightarrow\quad &\Omega\backslash\hat{\Omega}&\coloneqq\left\{{\bf x}\in\Omega \; ;\quad \Delta\chi_{k}({\bf{x}})=0\right\}
\end{array}
\end{equation}
such that
\begin{equation}
\label{eq_optimality_criterion_2}
\Delta\chi_{k}(\bf{x})=
\left\{
\begin{split}
& 0\rightarrow\textit{no phase exchange at point ${\bf x}$}\\
&{\Delta\chi}\ne0\rightarrow\textit{\textit{actual phase exchange at point ${\bf x}$}}
\end{split}
\right.
\end{equation}
If condition in equation (\ref{eq_condition}) is fulfilled everywhere, for a given topology, $\chi_{k}(\bf{x})$, \textit{and for all possible phase changes of that topology}  ($\Delta\chi_{k}({\bf x});\ \forall{\bf x}\in\hat{\Omega};\ \forall{\hat \Omega}\subset{\Omega} $) then, the increment of the functional $\Delta{{\cal J}^{ext}}(\chi_{k})$ for material phase exchange at \textnormal{any}
 finite subset $\hat{\Omega}(\chi_{k})\subset\Omega$ is:
\begin{equation}
\label{eq_change_pi}
\begin{split}
\Delta{{{\cal J}^{ext}}}\left( {\chi_{k}}\right)&=
\int_{\hat{\Omega}(\chi_{k})}
\underbrace
{\dfrac{\delta {{\cal J}^{ext}}\left(\chi_{k},{\Delta\chi}\right)}{\delta\chi_{k}}({\bf x})}_{{\textstyle {> 0}}} d\Omega >0\quad \forall\hat{\Omega}(\chi _{k})\subset\Omega
\end{split}
\end{equation}
this indicating that $\chi _{k}$ is a local minimizer of ${{\cal J}^{ext}}(\chi _{k})$. This proves that equation (\ref{eq_condition}) is a \textit{sufficient condition} for minimization of the problem in equation (\ref{eq_minimization_restricted_penalized}).
\end{remark}
 \subsection{Closed-form solution}
 Replacing equation (\ref{eq_alpha}) into equation (\ref{eq_condition}) yields:  
 \begin{equation}
 \label{eq_discrimination}
 \begin{split}
&\dfrac{\partial F \left(\chi_k,\mathbf{x}\right)}{\partial \chi_k}{\Delta\chi}({\bf x})
+\dfrac{C(\chi_k)}{k}\text{sgn}(\Delta\chi({\bf x}))=\\
 &\quad\quad\quad\quad=\left\{
 \begin{split}
 -&(1-\beta)\dfrac{\partial F \left(\chi_k,\mathbf{x}\right)}{\partial \chi_k}
 -\dfrac{C(\chi_k)}{k}>0& \quad\textit{for}  &\quad{\bf x}\in\Omega^{+}(\chi_{k})  \\ 
 &(1-\beta)\dfrac{\partial F \left(\chi_k,\mathbf{x}\right)}{\partial \chi_k}
 +\dfrac{C(\chi_k)}{k}>0& \quad\textit{for}  &\quad{\bf x}\in\Omega^{-}(\chi_{k})
 \end{split} 
 \right.\\
 \end{split}
 \end{equation}
 Comparing equation (\ref{eq_discrimination}) with equations (\ref{eq_domain_splitting}) and (\ref{eq_domain_splitting_rephrased}), and since $(1-\beta)>0$ (see equation (\ref{eq_chi_definition})-(a)),  we can identify the closed-form solution of the penalized problem through
 \begin{equation}
\label{eq_solution_psi}
\left\{
\begin{split}
&\psi_k({\bf x})\coloneqq-(1-\beta)\dfrac{\partial F \left(\chi_k,\mathbf{x}\right)}{\partial \chi_k}
-\dfrac{C(\chi_k)}{k}\\
&\chi^*_{k}({\bf x})\coloneqq{\cal H}_{\beta}\left[\psi_{k}({\bf x})\right]
\end{split}
\right.
\end{equation}
where equation (\ref{eq_relaxed_chi}), has been considered. Finally, taking to the limit equation (\ref{eq_solution_psi}), as indicated in equation (\ref{eq_minimization_restricted_penalized})-(b), yields
\begin{equation}
\label{eq_solution_xi}
\chi^*{\bf(x)}=\lim\limits_{k\rightarrow 0}\chi^*_k({\bf x})
\end{equation}
  The drawback of the penalty method, considered so far, is that, for getting accurate and robust solutions in equations    (\ref{eq_solution_xi}) the penalty parameter, $k$, has to be taken very close to the limit ($k\rightarrow 0$). This issue can be circumvented by rephrasing equations (\ref{eq_minimization_restricted_penalized}) and (\ref{eq_condition}) through definition of a new variable $\lambda$ (see \cite{Bertsekas1996}),
\begin{equation}
\label{eq_lambda_definition}
\left\{
\begin{split}
&\lambda\coloneqq\lim\limits_{k\rightarrow {0}}\dfrac{{\cal C}(\chi_k)}{k}\quad\left(=\dfrac{0}{0}\right) \quad&(a)\\
&\lim\limits_{k\rightarrow {0}}{\cal C}(\chi_k)=0\quad&(b)
\end{split}
\right.
\end{equation}
where the new unknown, $\lambda:\Omega\rightarrow{\mathbb R} $, arising from the (undetermined) equation (\ref{eq_lambda_definition})-(a) is solved by imposing, in strong form, the original restriction,
${\cal C}({\chi)}=0$ in equation (\ref{eq_lambda_definition})-(b).
Taking the limit $k\rightarrow 0$ in equations (\ref{eq_minimization_restricted_penalized}) to (\ref{eq_solution_xi}), and replacing $\lim\limits_{k\rightarrow {0}}\dfrac{{\cal C}(\chi_k)}{k}$ by $\lambda$, yields the following \textit{closed-form solution}.

\begin{tcolorbox}[colback=white!50!white]
	\textbf{Box I. Volume constrained topological optimization problem. Closed form solution. }
\begin{equation}
\label{eq_rephrased_equations}
\begin{split}
&\textit{Problem:}  \\
&\left\{
\begin{split}
&\chi^{*} =\underset{\chi} {\operatorname{argmin}}\ {\cal J}(\chi)=\underset{\chi} {\operatorname{argmin}}\int_{\Omega}{F(\chi,\mathbf{x})} d\Omega\\
&s.t.\quad{\cal C}(\chi)\equiv{\vert\Omega^+(\chi)\vert}-\bar{V}=0
\end{split}
\right.   &\quad(a)\\
&\textit{Lagrangian:}\quad{\cal L}(\chi,\lambda)={\cal J}(\chi)+\lambda {\cal C}(\chi)&\quad(b)\\
&\textit{Optimality criterion:}   &\\
&\left\{
\begin{split}
&\dfrac{\delta {\cal L}(\chi,\lambda)}{\delta\chi}({\bf x})=
\dfrac{\partial F  \left(\chi,\mathbf{x}\right)}{\partial \chi}\Delta\chi({\bf x})
+\lambda\ \text{sgn}(\Delta\chi({\bf x}))>{0}\quad\forall{\bf x}\in\Omega\\
&{\cal C}(\chi)=0
\end{split}
\right.\quad\quad\quad&(c)\\
&\textit{Closed-form solution:}\\&
\begin{split}
&\psi_{\chi}({\bf x},\lambda)\coloneqq-(1-\beta)\dfrac{\partial F  \left(\chi,\mathbf{x}\right)}{\partial \chi}
-\lambda>0\quad &\text{for } {\bf x}\in\Omega^+\\
\end{split}
\\
&\Rightarrow
\left\{
\begin{split}
&\psi_{\chi}({\bf x},\lambda)\coloneqq-(1-\beta)\dfrac{\partial F  \left(\chi,\mathbf{x}\right)}{\partial \chi}
-\lambda\\
&\left\{
\begin{split}
&\chi({\bf x},\lambda)={\cal H}_{\beta}\left[ \psi_{\chi}({\bf x},\lambda) \right] \rightarrow\quad \chi_{\lambda}({\bf x})\quad&\textbf{(d-1)}\\
&{\cal C}(\chi_{\lambda},\lambda)=0&\textbf{(d-2)}
\end{split}
\right\}\rightarrow \lambda^{*}\rightarrow \chi^{*}({\bf x})=:\chi_{\lambda^{*}}({\bf x})
\end{split}
\right. &(d)\\
&\textit{Topology:}\\
&\left\{
\begin{split}
&\Omega^{+}(\chi)\coloneqq\{\mathbf{x}\in \Omega \; ;\quad \psi_{\chi}\mathbf{(x,\lambda)}>0 \}\\
&\Omega^{-}(\chi)\coloneqq\{\mathbf{x}\in \Omega \; ;\quad  \mathbf{x}\notin {\Omega^+}\}\\
&\Gamma(\chi)\ \ \ \coloneqq\{\mathbf{x}\in \Omega \; ;\quad \psi_{\chi}\mathbf{(x,\lambda)}=0 \} \\
\end{split}
\right.\\
\end{split}
\end{equation}
\end{tcolorbox}

Equation (\ref{eq_rephrased_equations})-(d-1) provides, at any  point ${\bf x}\in\Omega$ and for a given value of $\lambda$, a closed-form (fixed-point type) scalar equation\footnote{Of the type $y={\cal G}(y)$, in the argument $y\equiv\chi$.}, whose solution is $\chi_{\lambda}({\bf x})$. When this is set for all points ${\bf x}\in\Omega$, the spatial description of a topology can be parametrized in terms of the current value of $\lambda$, i.e.: $\chi_{\lambda}\equiv\chi({\bf x},\lambda) $.
Substitution into the constraint equation (\ref{eq_rephrased_equations})-(d-2) yields a scalar equation supplying the solutions $\lambda^*$,
and $\chi^*(\bf x)=\chi_{\lambda^{*}}(\bf x)$, and the solution of the original problem in equations (\ref{eq_minimization_restricted})-(a) and (\ref{eq_rephrased_equations})-(a), is retrieved.

\section{Regularization}
It is well known that the topological optimization problem may be ill-posed \cite{Sigmund1998,MartinPh.Bendsoe2003}. For instance, in case of the restriction in equation (\ref{eq_minimization_restricted_penalized})-(b),  in terms of the material \textit{hard-phase} (${\mathfrak M}^+$) volume, $\vert\Omega^{+}(\chi)\vert$, i.e.
\begin{equation} 
{\cal C}(\chi)\equiv \vert\Omega^{+}(\chi)\vert-\bar{V}=0
\end{equation}
the infimum of the problem corresponds to a topology made of infinite number of inclusions of the \textit{soft-phase} ($ {\mathfrak M}^{-} $), of infinitesimal size each, embedded into a matrix of \textit{hard-phase} ($ {\mathfrak M}^{+} $). Since, for solving the problem numerically, one uses meshes with cells (voxels or finite elements) of typical size, $h$, this determines the minimum inclusion size that can be captured by the mesh. Therefore, a \textit{local minimum} is achieved with the inclusion's size in terms of $h$. However, when finer meshes are used ($h\rightarrow0$) finer topologies (with smaller inclusion's size) are obtained, providing different and unbounded lower minima of the cost function ${\cal J}(\chi)$. No convergence is achieved in this process in terms of $h$ (lack of mesh size objectivity).

Although this issue does not appear for some other type of restrictions (like perimeter restrictions \cite{Ambrosio1993,Fernandes1999}), this fact sets serious limits to the type of applications in the topological derivative optimization based methods, unless some remedy to the \textit{ill-posed-problem/mesh-size-unobjectivity issue} is introduced. In the literature some of them can be found, typically based on introducing some modifications on the original cost function ${\cal J}$ that \textit{regularize} the problem (like the Tikhonov regularization, \cite{Yamada2010}), by using Fourier series-based regularizations \cite{White2018}, or by a combination of topological and shape derivatives \cite{Burger2004,Allaire2005a}.
\subsection{Regularization. Laplacian smoothing}
\label{sec_laplacean_smoothing}
Here we resort to a very simple and efficient regularization procedure, based on a Laplacian smoothing of the discrimination function $\psi({\bf x},\lambda)$, derived in equation (\ref{eq_rephrased_equations})-(d). 
 This function, which in principle might be discontinuous due to the characteristic function $\chi\in\{\beta,1\}$, is here \textit {regularized} through the solution of a Laplace-type smoothing equation, that returns a smooth field $\psi_{\tau}(\mathbf{x},\lambda)$ to be inserted in equation (\ref{eq_rephrased_equations})-(d):
 \begin{equation} \label{eq_regularized_psi}  
 \psi({\mathbf{x},\lambda})\leftarrow\psi_{\tau}({\mathbf{x},\lambda})
 \end{equation} 
 the smooth field $\psi_{\tau}(\mathbf{x})$ being the solution of
 \begin{equation}
 \label{eq_regularized_laplacean_smoothing}
 \left\{
 \begin{split}
 &\psi_{\tau}(\mathbf{x},\lambda)-\epsilon^2\Delta_{\bf x}\psi_{\tau}(\mathbf{x},\lambda)=\psi(\mathbf{x},\lambda)& &\quad\forall \mathbf{x}\in\Omega\\
 &\nabla_{\bf x}\psi_{\tau}(\mathbf{x},\lambda)\cdot\mathbf{n}={0}& &\quad\forall \mathbf{x}\in\partial\Omega
 \end{split}
 \right.
 \end{equation}
 where, $\Delta_{\bf x}({\bf x},\newplaceholder)$ and $\nabla_{\bf x}({\bf x},\newplaceholder)$ stand for the Laplacian and gradient operators, respectively, $\mathbf{n}$ is the outwards normal to the boundary, $\partial\Omega$, of the design domain.
Therefore, the closed-form solution of the problem in equation (\ref{eq_minimization_restricted_penalized}) reads:
\begin{equation} 
\label{eq_regularization_paramete}
\chi({\bf x})\rightarrow\textit{solution of}\ 
\left\{
\begin{split}
&\chi(\bs{x},\lambda)={\cal H}_{\beta}\left[\psi_{\tau}({\mathbf{x}})\right]\\
& {\cal C}(\chi(\bs{x},\lambda))=0
\end{split}
\right.
\quad\quad\quad\forall{\bf x}\in\Omega 
\end{equation}
where $\psi_{\tau}({\mathbf{x}})$ is the solution of equation (\ref{eq_regularized_laplacean_smoothing}), which can be solved in the context of any spatial numerical discretization method. In a finite element context, it yields:
\begin{equation}
\label{eq_discrete_laplacean_smoothing}
\left\{
\begin{split}
&\psi_\tau(\mathbf{x})=\mathbf{N}(\mathbf{x})\{{\bf\hat{\mathbf \psi}_\tau}\}
\; ;\quad \hat{\mathbf \psi}_\tau={\tilde{\mathbb G}}^{-1}{\mathbf f}\; ;\quad\tilde{\mathbb G}=\tilde{\mathbb M}+{\epsilon^2}\tilde{\mathbb K}&(a)\\
&\tilde{\mathbb M}=\int_{\Omega}{\mathbf N}^T{({\bf x})}\mathbf{N}({\bf{x}})
\; ;\quad \tilde{\mathbb K}=\int_{\Omega}\nabla {\mathbf N}{({\bf x})}^T{\nabla {\mathbf N}{({\bf x})} d\Omega};\quad
{\mathbf f}(\psi)=\int_\Omega{\mathbf{N}^T(\mathbf{x})\psi(\mathbf{x})d\Omega}&(b)\\
\end{split}
\right.
\end{equation}
where $\mathbf{N}(\bf{x})$ stands for the standard interpolation  matrix and $\hat{\mathbf \psi}_\tau$ is the vector  of  nodal values of the field $\psi_\tau(\mathbf{x})$. Notice that matrix $\tilde{\mathbb G}_\tau$, in equation (\ref{eq_discrete_laplacean_smoothing})-(a), needs to be built, and inverted, only once for ever, and that it can be used as many times as the Laplacian smoothing is required in the optimization problem. This translates into a low computational cost of the procedure.

\begin{figure}[h]
	\centering
	\includegraphics[width=12cm, height=12cm]{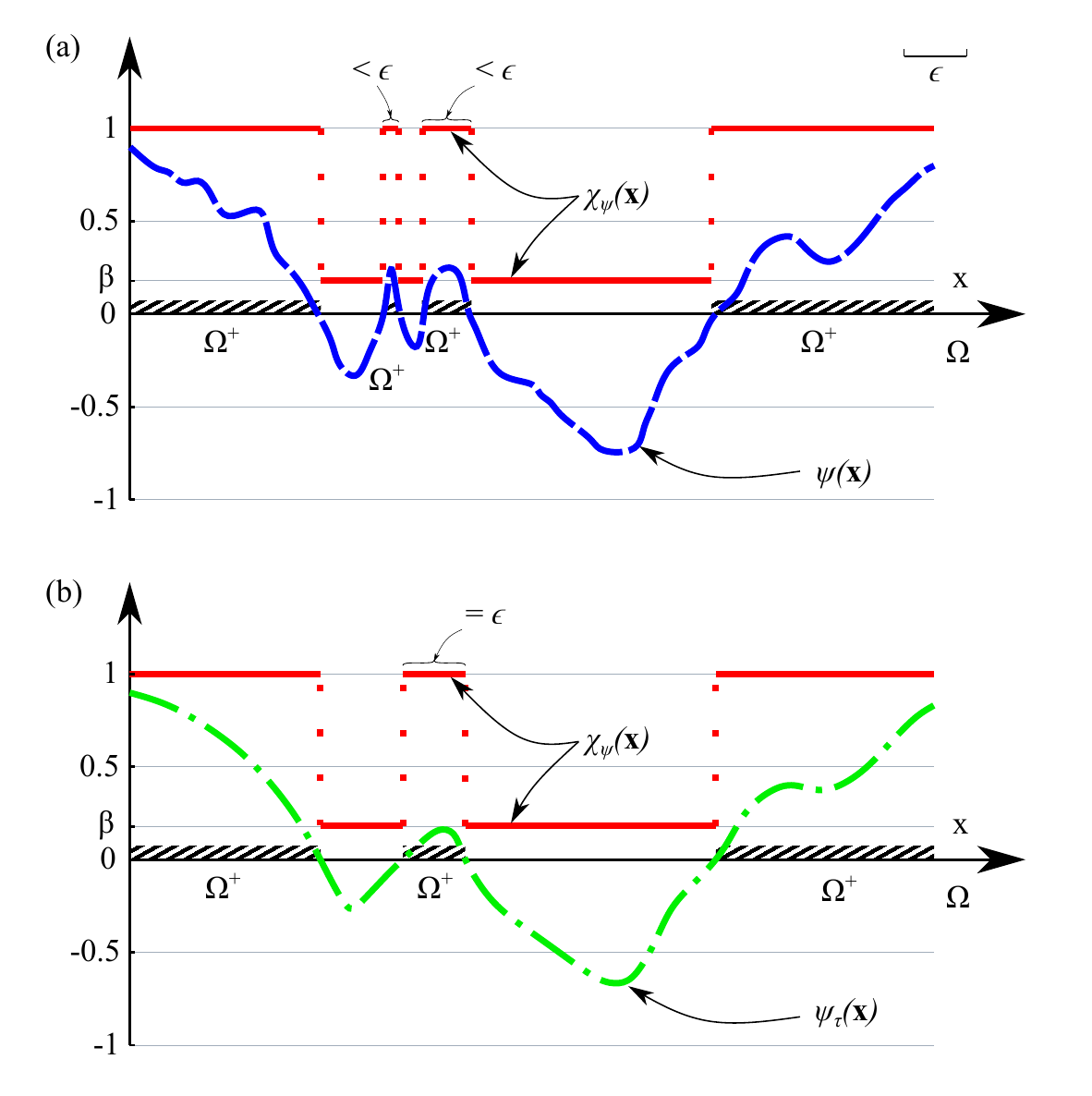}
	\caption{ Noisy modes removal and minimum phase-filament-thickness, via Laplacian smoothing of the discrimination function $\psi({\bf x})$.
	(a) Before smoothing   ;  (b) After smoothing.}
	\label{fig_laplacean_smoothing}
\end{figure}

The regularization parameter, $\epsilon\ge0$, in equations (\ref{eq_regularized_psi}) to (\ref{eq_discrete_laplacean_smoothing}), is indirectly defined through the mesh/voxel size, $h$, and the non-dimensional parameter, $\tau$, i.e.
\begin{equation} \label{eq_tau}
\epsilon=\tau h\; ;\quad \tau\in{\mathbb R^+}
\end{equation} 
where, for every mesh, $\tau$, describes the \textit{number of elements} that typically cover the filtering measure $\epsilon$.
The Laplacian smoothing procedure in equations  (\ref{eq_regularized_laplacean_smoothing}) and (\ref{eq_discrete_laplacean_smoothing})  is frequently used in digital image processing \cite{Vliet1989,patane2009computing} to remove the short wavelength spectral components (\textit{noisy components}) of the original field $\psi{(\bf x)}$, in such a way that the smoothed field, $\psi_{\tau}(\bf x)$, does not exhibit oscillation wavelengths shorter than $\epsilon$.
Therefore, the value of $\epsilon$, approximately defines a lower bound for \textit{the minimum hard-phase filaments width} (separating soft-phase inclusions) that appear in the optimized topology emerging from equations (\ref{eq_regularized_laplacean_smoothing}) (see figure \ref{fig_laplacean_smoothing}). Consequently, in equation (\ref{eq_tau}), $\tau$ defines, for a given uniform-size mesh, the number of elements covering the  width of the thinnest hard-phase filaments \cite{Guest2004}.This strategy breaks down the mesh-size dependency of the obtained topologies and the aforementioned ill-condition paradigm. Now, the resulting soft-phase (inclusions/voids) cannot be closer to each other than the distance defined by $\epsilon$, disregard the size and number of the finite elements covering this width. In some way this is equivalent to establishing a \textit{perimeter constraint}\footnote{Actually,
perimeter constraint in topology optimization is a deep question whose discussion can be
found in \cite{Amstutz2013}.} in the optimization problem to remove those shortcomings \cite{Haber1996}.  
From another point of view, the need for a regularization can be related to \textit{manufacturing constraints}, related to manufacturing issues \cite{Norato2015}, $\epsilon$ defining the minimum hard-phase ${\mathfrak M}^+$ filament width that the manufacturing technology can produce.

\section{Pseudo-time sequential approach for volume constrained optimization problem}
\label{sec_pseudo_time_approach}
 Introduction of a \textit{pseudo-time}, $t$, as an additional increasing scalar parameter in the problem, allows connecting the strategy of resolution of the topological optimization problem, with some well known algorithmic time-advancing techniques in computational mechanics. For instance, \textit {control} methods in non-linear solid mechanics problems (i.e. displacement control, arc-length methods, in which one additional algebraic equation (equivalent to the restriction in equation (\ref{eq_minimization_restricted})-(b)) is used to limit/control the evolution of the problem unknowns\footnote{The time interval $[0,T]$ is subdivided into a number of \textit{time-steps}, each one defining an independent optimization problem, with a sufficiently small length, ${\Delta t}$, so as to ensure that the  solution of the problem, at the beginning of the time step, is close enough to the solution at its end; this helping the convergence of the solving iterative process.}, in order to increase the robustness of the resolution process \cite{Crisfield1983}.

 In this context, the  pseudo-time parameter $t\in[0,T]$ is introduced in the constraint, which is now rewritten as:
\begin{equation}
\label{eq_time_parametrization}
\begin{split}
{\cal C}(\chi,t)\equiv \dfrac{\vert\Omega^+(\chi)\vert}{\vert\Omega\vert}-{\dfrac{\bar V}{\vert\Omega\vert}}=\dfrac{\vert\Omega\vert-\vert\Omega^-(\chi)\vert}{\vert\Omega\vert}-\underbrace{\dfrac{\bar V}{\vert\Omega\vert}}_{\displaystyle{1-t}}=
t-\dfrac{\vert\Omega^-(\chi)\vert}{\vert\Omega\vert}\; ;\quad t\in[0,1]
\end{split} 
\end{equation}
so that all variables become \textit{parametrized} in terms of $t$, and the original problem in equation (\ref{eq_minimization_restricted}) now reads:
\begin{equation} 
\label{eq_minimization_restricted_time}
\begin{split}
&FIND:\quad \psi:\Omega\times[0,T]\rightarrow{\mathbb R}\; ;\quad\psi(\cdot,t)\in{H}^{(1)}(\Omega)\quad  \emph{and}\\
&\chi:\Omega\times[0,1]\rightarrow\{\beta,1\}\; ;\quad\chi(\mathbf{x},t)={\cal H}_{\beta}({\psi}\left(\mathbf{x},t)\right)&\quad(a)\\
&FULFILLING: \\
&\hspace{1.5cm}\chi^{*}_{t}  = \underset{\chi} {\operatorname{argmin}}\quad{\cal J}_{t}\left( \chi\right)\equiv\int_{\Omega}{F(\chi,\mathbf{x},t)} d\Omega &\quad(b) \\
&\textit{SUBJECT TO}:  \\   
&\hspace{1.5cm}{\cal C}(\chi,t)\equiv t-\dfrac{\vert\Omega^-(\chi)\vert}{\vert\Omega\vert}=0 &\quad(c)
\end{split}
\end{equation} 

\section{Algorithmic resolution} \label{sec_Algorithmic_resolution} 
The time-discretized  version of the closed-form problem and solution in equations (\ref{eq_minimization_restricted_time}) can be now written as:
\begin{equation}
\label{eq_algebraic_problem}
\begin{array}{ll}
GIVEN:
\left\{
\begin{split}
&\mathcal{T}\coloneqq\{t_{0}=0,t_1, t_2, ....., T\le 1\}\quad(\textit{time discretization})\\
&\chi({\bf x},t_n)\equiv\chi_{n}({\bf x})\quad(\textit{topology at time $t_n$})\\
&\textit{and } \left[t_n,t_{n+1}\right]\quad\textit{(time interval)} \\
\end{split}
\right.
\\
FIND:\quad 	\chi_{n+1}:\Omega\rightarrow\{\beta,1\} \textit{, and} \ \lambda(t_{n+1})\equiv\lambda_{n+1}\in{\mathbb{R}} \\
FULFILLING: \\
\hspace{1.5cm}\begin{split}
&\psi_{\tau}(\mathbf{x},\lambda_{n+1})\coloneqq\textit{ solution of}
\left\{
	\begin{split}
	&\psi_{\tau}(\mathbf{x},\lambda_{n+1})-\epsilon^2\Delta\psi_{\tau}(\mathbf{x},\lambda_{n+1})=
	\psi(\mathbf{x},\lambda_{n+1})\quad& &\forall \mathbf{x}\in\Omega\\
	&\nabla\psi_{\tau}(\mathbf{x},\lambda_{n+1})\cdot\mathbf{n}=\mathbf{0}\quad& &\forall \mathbf{x}\in\partial\Omega\\
	&\psi(\mathbf{x},\lambda_{n+1})\coloneqq{-\left((1-\beta)\dfrac{\partial F  \left(\chi_{n+1},\mathbf{x}\right)}{\partial \chi}
		+\lambda_{n+1}\right)}\quad& 
	\end{split}
\right.
\quad&(a) \\
&\chi_{n+1}(\mathbf{x})={\cal H}_{\beta}\left[\psi_\tau(\chi_{n+1},\mathbf{x},\lambda_{n+1})\right]   \quad&(b)\\
&{\cal C}(\chi_{n+1},t_{n+1})\equiv t_{n+1}-\dfrac{\vert\Omega^-(\chi_{n+1})\vert}{\vert\Omega\vert}=0   \quad&(c)\\
\end{split}
\end{array}
\end{equation}

\begin{figure}[h]
	\centering
	\includegraphics[width=9cm, height=4.5cm]{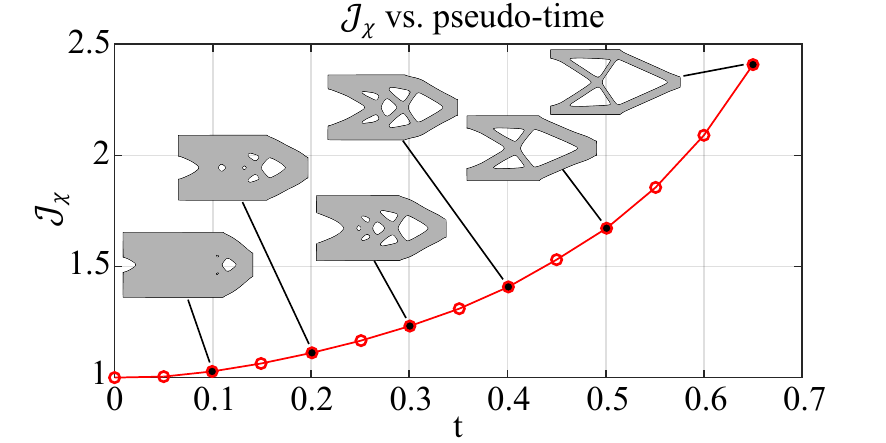} 
	\caption{Pseudo-time evolutionary analysis. Typical evolution of the cost function and topology vs. time.}
	\label{fig_time_evolution}
\end{figure}

Equations (\ref{eq_algebraic_problem}) define a discrete time-advancing problem providing the solution $(\chi_{n+1}(\mathbf{x}),\lambda_{n+1})$, at pseudo-time $t_{n+1}$, in terms of results at previous times (see figure \ref{fig_time_evolution}). Equations (\ref{eq_algebraic_problem})-(a) are associated to determination of the \textit{topology}, $\chi_{n+1}$, whereas solution of equation (\ref{eq_algebraic_problem})-(c)  provides the value of the \textit{Lagrange multiplier $\lambda_{n+1}$}.
Therefore, they can be numerically solved by means of an iterative algorithm at every time step. At time step $[t_{n},t_{n+1}]$ the iterative problem to be solved is defined by the coupled system of equations involving as unknowns \textit{the topological field}, $\chi_{n+1}(\mathbf{x})$, and the (spatially constant) \textit{Lagrange multiplier}, $\lambda_{n+1}$.
\subsection{Iterative solving strategy}
 The problem to be solved, at the representative iteration $(i+1)$ of the considered time step ${[t_n, t_{n+1}]}$, can be then sketched as: 
\begin{equation} \label{eq_coupled_algorithm}
\begin{array}{l}
\textit{GIVEN}\quad \chi_{n+1}^{(i)}:\Omega \rightarrow\left\{\beta,1\right\}\\[4pt]
\textit{SOLVE}
\left\{ 
\begin{split}
&\chi_{n+1}^{(i+1)}={\cal F}\left(\chi_{n+1}^{(i)},\lambda_{n+1}\right)\quad\forall{\bf x}\in\Omega\ &\text{(topology problem)}\quad\quad&(a)\\
&\lambda_{n+1}={\cal G}\left(\chi_{n+1}^{(i+1)},\lambda_{n+1}\right) &\text{(restriction problem)}\quad\quad&(b)
\end{split}
\right. 
\\ 
\textit{UNTIL}
\begin{split}
&\left| \left| \chi^{(i+1)}_{n+1}-\chi^{(i)}_{n+1}\right| \right|_{L^{2}(\Omega)}=
\left[\int_{\Omega}\left(\chi^{(i+1)}_{n+1}(\mathbf{x})-\chi^{(i)}_{n+1}(\mathbf{x})\right)^2 d\Omega\right]^\frac{1}{2}
\le{\text{Toler}}_\chi\\
\end{split} 
\end{array}
\end{equation}

Notice that the \textit{topology problem}, in equation (\ref{eq_coupled_algorithm})-(a), sketching equations (\ref{eq_algebraic_problem})-(a) involves as unknown the \textit{space field}, $\chi_{n+1}^{(i+1)}({\bf x})$, which is iteratively updated, in terms of its value at the previous iteration, $\chi_{n+1}^{(i)}({\bf x})$, until convergence is achieved according to the tolerance $\emph{Toler}_{\chi}$.  On the contrary, the \textit{restriction problem} in equation (\ref{eq_coupled_algorithm})-(b), corresponding to equation (\ref{eq_algebraic_problem})-(c), which involves the scalar unknown $\lambda_{n+1}$, \textit{is exactly enforced at every iteration} $(i+1)$ of the time step. This goal is achieved  via a specific method for resolution of scalar  equations e.g.: a bisection algorithm. Details on the procedure are given in next sections.

\subsubsection{Topology problem: cutting algorithm}
The problem reads as follows
\label{sec_cutting_algorithm}
\begin{equation}
 \label{eq_contour_line}
\begin{split}
&\textit{GIVEN}\quad \chi^{(i)}({\bf x}) \quad(\textit{topology at the end of iteration "i"})\\
&
\substack{\displaystyle{\textit{SOLVE}}\\ \displaystyle{\chi^{(i+1)}(\mathbf{x})}}
\left\{
\begin{split}
		&\xi_{\tau}(\chi^{(i)},\mathbf{x})\coloneqq\textit{ solution of}
		\left\{
			\begin{split}
				&\xi_{\tau}(\chi^{(i)},\mathbf{x})-\epsilon^2\Delta\xi_{\tau}(\chi^{(i)},\mathbf{x})=\xi(\chi^{(i)},\mathbf{x})\quad& &\forall \mathbf{x}\in\Omega\\
				&\nabla\xi_{\tau}(\chi^{(i)},\mathbf{x})\cdot\mathbf{n}=\mathbf{0}\quad& &\forall \mathbf{x}\in\partial\Omega\\
				&\xi(\chi^{(i)},\mathbf{x})\coloneqq
				-(1-\beta)\dfrac{\partial F  \left(\chi^{(i)},\mathbf{x}\right)}{\partial \chi^{(i)}}\quad &\\
			\end{split}
		\right.&\quad(a)
 \\
 &\psi_{\tau}(\chi^{(i)},\mathbf{x},\lambda)=\xi_\tau(\chi^{(i)},\mathbf{x})-\lambda&\quad(b)\\
  &\chi^{(i+1)}(\mathbf{x})={\cal H}_{\beta}\left[\psi_\tau(\chi^{(i)},\mathbf{x},\lambda)\right]&\quad(c)\\
 \end{split}
   \right. \\
 &\textit{UNTIL} \quad\vert\vert \chi^{(i+1)}-\chi^{(i)}\vert\vert \le{\text{Toler}}_{\chi}
 \end{split}
\end{equation}

where the Laplacian smoothing of variable $\xi_{\tau}(\chi^{(i)},\mathbf{x})$, in equation (\ref{eq_contour_line})-(a), (from now on termed the \textit{spatial energy distribution}), provides equivalent results\footnote{Since the spatially constant variable ${\lambda}$ is not affected by the smoothing procedure.} to the smoothing of variable $\psi_{\tau}(\mathbf{x},\lambda_{n+1})$ in equation (\ref{eq_algebraic_problem})-(a). 

\begin{figure}[h]
	\centering
	\includegraphics[width=17cm, height=10cm]{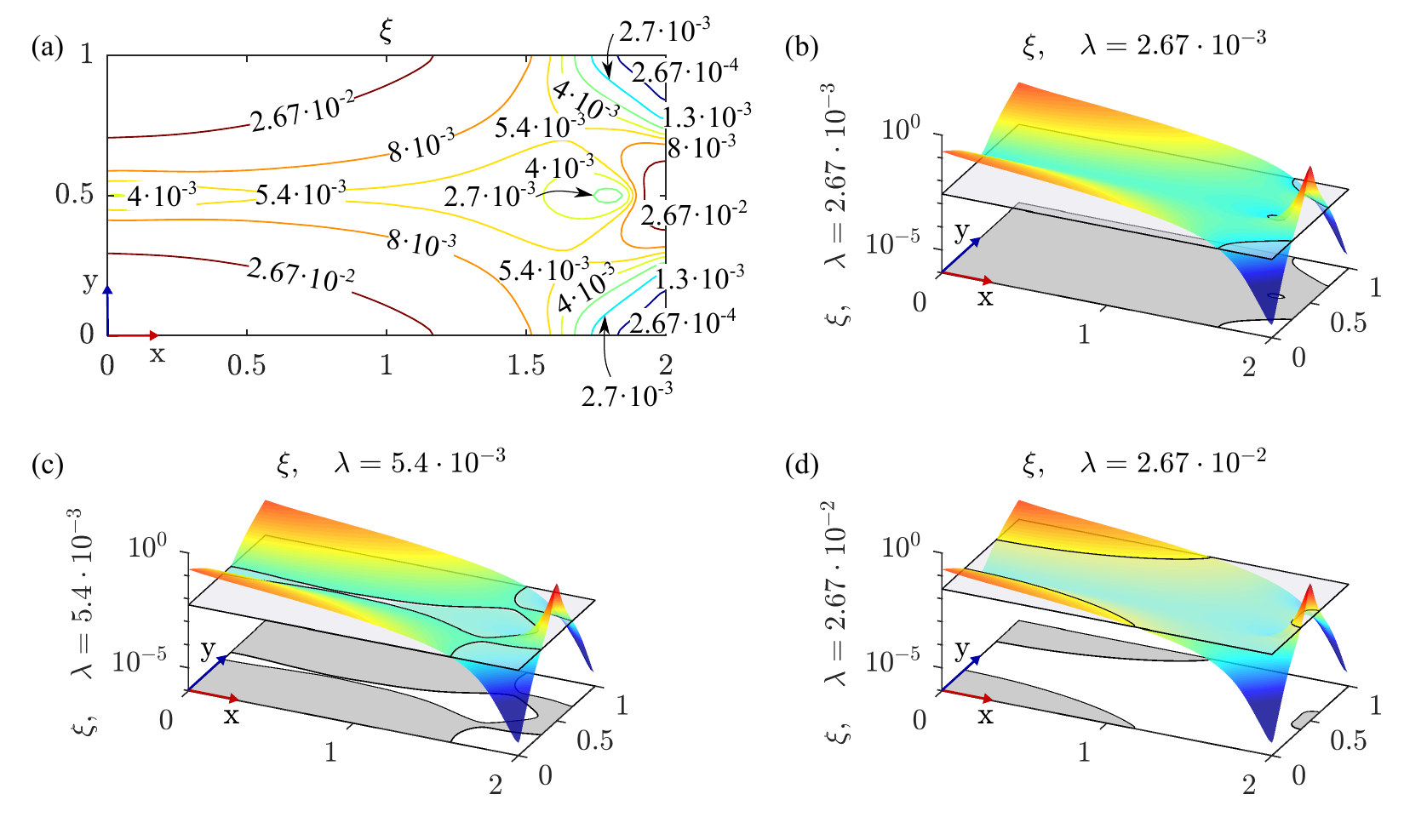}
	\caption{Isoenergy ($\lambda$ iso-level) contours of the \textit{spatial energy distribution} $\xi$: (a) contour-lines of $\xi$; (b)-(c)-(d) $\xi$ iso-levels for different values of $\lambda$.}
	\label{fig_cut_level}
\end{figure}

\begin{remark}
In view of equation (\ref{eq_contour_line})-(c), iteration $i+1$ can be regarded as performing a spatial cut, at level $\lambda$, of the spatial energy distribution, $\xi_{\tau}(\chi^{i}({\bf x}),\mathbf{x})$, to obtain the corresponding \textit{$\lambda$-contour} (line/surface), $\Gamma^{(i+1)}$, defining the topological boundaries
(see figure \ref{fig_cut_level}) i.e.
\begin{equation} \label{eq_cutoff}
{\left[\Gamma(\chi^{(i)},{\lambda})\right] }^{(i+1)}\coloneqq\left\lbrace \mathbf{x}\in{\Omega}\; ;\quad
\xi_{\tau}\left(\chi^{(i)}(\mathbf{x}),\mathbf{x}\right)=\lambda\right\rbrace  
\end{equation}
Therefore, equation (\ref{eq_cutoff}) proves that, \textit{for the optimal topology solution corresponding to any of the sequential problems in equations (\ref{eq_algebraic_problem}),  the material phase boundaries are,  iso-energetic, $\lambda$-contours  of the smoothed spatial energy distribution $\xi_{\tau}(\bf x)$}\footnote{This concept is also retrieved in the context of regularized  methods (SIMP) for the regularized domain $\Omega^{reg}$, see \cite{MartinPh.Bendsoe2003}.}.
\end{remark}
  From the result in equation (\ref{eq_cutoff}), the \textit{hard-phase} (matrix) domain, $\vert\Omega^{+}\vert(\lambda)$, and the \textit{soft-phase} (inclusions) domain, $\vert\Omega^{-}\vert(\lambda)$, are finally  obtained as (accordingly with equation (\ref{eq_domain_splitting_rephrased}))
\begin{equation} \label{eq_volume}
\begin{split}
&{\left(\Omega^{+}(\chi^{(i)},{\lambda})\right) }^{(i+1)}\coloneqq
\left\lbrace \mathbf{x}\in\Omega\;;\quad \xi_{\tau}\left(\chi^{(i)}(\mathbf{x}),\mathbf{x}\right)>\lambda \right\rbrace  \\ 
&{\left(\Omega^{-}(\chi^{(i)},{\lambda})\right) }^{(i+1)}\coloneqq
\left\lbrace \mathbf{x}\in \Omega \; ;\quad  \mathbf{x}\notin \left(\Omega^{+}\right)^{(i+1)} \right\rbrace    
\end{split}
\end{equation}
and, finally, from equations (\ref{eq_contour_line}), the updated characteristic function
\begin{equation} \label{eq_characteristic_function}
\chi^{(i+1)}(\mathbf{x})={\cal H}_{\beta}\left[\xi_\tau\left(\chi^{(i)}(\mathbf{x}),\mathbf{x}\right)-\lambda\right]
\end{equation}
can be calculated (see figure \ref{fig_characteristic_function}).
Calculations in equations (\ref{eq_contour_line}) to (\ref{eq_characteristic_function}) are repeated until convergence is achieved ($\vert\vert \chi^{(i+1)}-\chi^{(i)}\vert\vert_{L^{2}(\Omega)} \le{\text{Toler}_{\chi}}$).

\subsection{Lagrange multiplier resolution: bisection algorithm}
\label{sec_bisection_algorithm}
The algorithm consists in determining, for the \textit{(i-th) iteration characteristic function}, $\chi^{(i)}(\mathbf{x})$ and the corresponding spatial energy distribution, $\xi_{\tau}(\chi^{(i)},\mathbf{x})$, the cutting level, $\lambda$, in equation (\ref{eq_cutoff}), that fulfills the volume constraint (\ref{eq_minimization_restricted_time})-(c), i.e.  
\begin{equation} \label{eq_bisection}
\begin{split}\phi\left(\lambda\right)\equiv
{\cal C}(\chi^{(i+1)},\lambda,t)\equiv t-\dfrac{\vert\Omega^-(\chi^{(i+1)},\lambda)\vert}{\vert\Omega\vert}=0 \\
\end{split}
\end{equation}
The bisection algorithm is based on the following stages:
	\begin{itemize}
		\item[1)]Explore the appropriate range of $\lambda \in{\mathbb R}$ and find in a sequence $\{0,\lambda^{(1)},\dotsc,\lambda^{(j)},\lambda^{(j+1)}\} $ a \textit{bracket} $\{\lambda^{(j+1)}_+,\lambda^{(j)}_{-} \}$, for the sign change of function $\phi(\lambda^{j})$, in equation (\ref{eq_bisection}):
		\begin{equation} \label{eq_bracketting}
			\begin{split}	
			&\phi\left(\lambda^{(j+1)}_+\right)\equiv t-\dfrac{\vert\Omega^{-}(\chi,\lambda^{(j+1)})\vert}{\vert\Omega\vert}
			\; ;\quad \phi\left(\lambda^{(j)}_-\right)\equiv t-\dfrac{\vert\Omega^{-}(\chi,\lambda^{(j)})\vert}{\vert\Omega\vert}\\
			 &\phi\left({\lambda^{(j+1)}_+}\right)>0\; ;\quad \phi\left({\lambda^{(j)}_-}\right)<0
			\end{split}
		\end{equation}
		\item[2)]When the bracket is found, decrease the bracket's size by iteration and interpolation (\textit{regula falsi method}) \cite{Press2007} until
		\begin{equation}\label{eq_bracket}
		\begin{array}{l}
		\| \lambda^{(j+1)}_+-\lambda^{(j)}_-\|\le \text{Toler}_\lambda \Rightarrow\ \lambda\leftarrow \lambda^{(j+1)}_{+}
		\end{array}
		\end{equation}
or, equivalently,
	\begin{equation}\label{eq_constraint_tol}
	\begin{array}{l}
	\vert\phi(\lambda^{(j+1)}_+)\vert\le Toler_{\cal C}
	\end{array}
	\end{equation}

 \end{itemize} 
Once the value of $\lambda$ is determined, the topology domains $\left(\Omega^{+}\right)^{(i+1)}$ and $\left(\Omega^{-}\right)^{(i+1)}$ are computed, according to equations (\ref{eq_volume}) and the  new characteristic function, $\chi^{(i+1)}$, is obtained according to equation  (\ref{eq_characteristic_function}).
\begin{remark}A specific feature of the proposed algorithm is that, unlike in alternative options (i.e. augmented Lagrangian methods), the restriction ${\cal C}(\chi)$, in equation (\ref{eq_algebraic_problem})-(c)  is not fulfilled only at the end (convergence) of the iterative process. Instead, that constraint is exactly fulfilled (up to a small tolerance, $Toler_{\cal C}$) at every iteration $"i"$ of the algorithm. This provides, in comparison with some alternative procedures\footnote{Like \textit{augmented Lagrangian methods}, where $\lambda$ is iteratively updated until convergence to the \textit{exact} final value.}, additional control on the evolution of  topologies along iterations, which translates into global algorithmic robustness.
\end{remark}

\subsubsection{Algorithm initialization}
As commented above, a relevant issue for convergence of the topology problem algorithm in equation (\ref{eq_contour_line}), is the fact that the value of the unknown field  at the beginning of the current time interval, $\chi_{n+1}^{(0)}({\bf x})$, has to be \textit{close enough} to the final one (converged) at the end of the time step ($\chi_{n+1}^{(converged)}({\bf x})$). 

In the pseudo-time sequential approach chosen here (see sections  \ref{sec_pseudo_time_approach} and \ref{sec_Algorithmic_resolution}), this could be \textit{conceptually} achieved by choosing a sufficiently small time-step length, $\Delta t =t_{n+1}-t_{n}$, which places the iterative values into the suitable range of convergence.
However, the $\Delta t$-proportional distance, between consecutive solutions, requires that the algorithm initiates (i.e. at time $t=0\equiv t_0$), at a  \textit{solution} of the problem: i.e. fulfilling the equations of the algorithm\footnote{Here it is assumed that the algorithm is initiated at a \textit{full hard-phase} ${\mathfrak M}^+$ design ($\Omega^{+}_{t_0}=\Omega$).}.
\begin{equation}
\label{eq_initial_conditions}
t_{0}=0 \rightarrow
\left\{
	\begin{split}
	&\lambda_{t_{0}}=0&\quad(a)\\
	&{\cal C}(\chi_{t_{0}})=\dfrac{\vert\Omega^+(\chi_{t_{0}})\vert}{\vert\Omega\vert}=1\Rightarrow\vert\Omega^+(\chi_{t_{0}})\vert=
	\vert\Omega\vert\Rightarrow\chi_{t_{0}}({\bf x})=1 \quad\forall{\bf x}\in\Omega&\quad(b)\\
	&\xi_{t_{0}}({\bf x})-\lambda_{t_{0}}>0\quad \forall{\bf x}\in\Omega\Rightarrow \min_{{\bf x}\in\Omega}{\xi_{t_{0}}({\bf x)}}\ge\lambda_{t_{0}}=0&\quad(c)
	\end{split}
\right.	
\end{equation}
where equations (\ref{eq_initial_conditions})-(a) and (\ref{eq_initial_conditions})-(b)  state the initial conditions in terms of $\lambda=0$, and a \textit{full hard-phase} $\mathfrak{M}^{+}$ distribution for the initial design ($\Omega^+(\chi_{t_{0}})=\Omega\ ; \ \Omega^-(\chi_{t_{0}})=\{\emptyset\}$). However, under these initial conditions, equation (\ref{eq_initial_conditions})-(c) is not automatically fulfilled in some problems: typically those in which the  \textit{initial energy}, $\xi_{t_{0}}$, is not intrinsically positive definite (i.e. $\exists{\bf x}\in\Omega;\xi({\bf x})<0 $). In these cases, convergence in the first time interval is not achievable (or robustly achievable) disregard the considered time-step size\footnote{Even for $\Delta t=0$.} $\Delta t$.
A remedy for this problem, is to perform a \textit{constant shift}\footnote{Constant for all time-steps and iterations of the optimization algorithm.}, of value $\Delta_{shift}$, of both unknown fields: $\xi(\chi({\bf x}))$ (\textit{only at the $\Omega^+$ domain}) and $\lambda$  i.e.
\begin{equation}
\label{eq_shifted_values}
\Delta_{shift}=\min_{{\bf x}\in\Omega}{\xi_{t_0}({\bf x})}\rightarrow
\left\{
\begin{split}
&\hat{\xi}_{t_{n}}^{(i)}({\bf x})=\xi_{t_{n}}^{(i)}({\bf x})-\Delta_{shift}&\quad\forall{\bf x}\in(\Omega^{+})^{(i)}  &\quad(a)\\
&\hat{\xi}_{t_{n}}^{(i)}({\bf x})=\xi_{t_{n}}({\bf x})&\quad\forall{\bf x}\in(\Omega^{-})^{(i)}  &\quad(b)\\
&\hat{\lambda}_{t_{n}}^{(i)}=\lambda_{t_{n}}^{(i)}-\Delta_{shift}&  &\quad(c)\\
\end{split}
\right.	\quad\forall{i}\quad\forall {t_{n}}
\end{equation}	 
In addition, a \textit{normalization} of the resulting energy field, $\xi({\bf x}^{(i)})$, is done, in terms of a positive factor, $\Delta_{norm}>0$ (again constant for all iterations and time steps), in order to keep its value inside convenient ranges, i.e.
\begin{equation}
\label{eq_normalized_values}
\Delta_{norm}=\vert  \max_{{\bf x}\in\Omega}{\xi_{t_0}({\bf x})}-\min_{{\bf x}\in\Omega}{\xi_{t_0}({\bf x})}\vert\rightarrow
\left\{
\begin{split}
&\hat{\xi}_{t_{n}}^{(i)}({\bf x})=\dfrac{\xi_{t_{n}}^{(i)}({\bf x})-\Delta_{shift}}{\Delta_{norm}}  &\quad\forall{\bf x}\in(\Omega^{+})^{(i)}\   &\quad(a)\\
&\hat{\xi}_{t_{n}}^{(i)}({\bf x})=\dfrac{\xi_{t_{n}}({\bf x})}{\Delta_{norm}} &\quad\forall{\bf x}\in(\Omega^{-})^{(i)}\   &\quad(b)\\
&\hat{\lambda}_{t_{n}}^{(i)}=\dfrac{\lambda_{t_{n}}^{(i)}-\Delta_{shift}}{\Delta_{norm}} \quad& &\quad(c)\\
\end{split}
\right.	\quad\forall{i}\quad\forall {t_{n}}
\end{equation}

Therefore, the topology is updated, according to the cutting algorithm in section \ref{sec_cutting_algorithm},  in terms of the values of the corrected fields, $\hat{\xi}_{\tau,t_{n}}^{(i)}\left(\mathbf{x}\right)-\hat{\lambda}_{t_n}^{(i)}$, at the domain, $\Omega^+$, instead of the original ones, ${\xi}_{\tau,t_{n}}^{(i)}\left(\mathbf{x}\right)-{\lambda}_{t_n}^{(i)}$, (see equations (\ref{eq_contour_line}) to (\ref{eq_volume})):
\begin{equation} 
\label{eq_volume_corrected_2}
\begin{split}
&\left(\Omega^{+}_{t_n}\right)^{(i+1)}\coloneqq
 \left\lbrace \mathbf{x}\in\Omega\; ;\quad \psi_{t_n}\coloneqq\hat{\xi}_{\tau,t_{n}}^{(i)}\left(\mathbf{x}\right)-\hat{\lambda}_{t_n}^{(i)}>0 \right\rbrace &\quad(a)  \\ 
&{\left(\Omega^{-}_{t_n}\right) }^{(i+1)}\coloneqq
\left\lbrace \mathbf{x}\in \Omega \; ;\quad  \mathbf{x}\notin {\left[\Omega_{t_n}^{+}\right] }^{(i+1)} \right\rbrace &\quad(b)    
\end{split}
\end{equation}	
where $\hat{\xi}^{(i)}_{\tau,t_n}\left(\mathbf{x}\right)$ is the result of applying the Laplacian smoothing, in equations (\ref{eq_contour_line})-(a), to the normalized energy $\hat{\xi}_{t_n}^{(i)}({\bf x})$ in equations (\ref{eq_normalized_values}).	
\begin{remark}
	Notice that the shifted and normalized fields, $\hat{\xi}_{t_{n}}^{(i)}({\bf x})$ and $\hat{\lambda}_{t_{n}}^{(i)}$, in equation  (\ref {eq_normalized_values}) fulfill the following properties:
	\item 
	\begin{itemize}	
		\item[a)] By construction, at the initial time $t_{0}$, $\min\limits_{{\bf x} \in\Omega}\hat{\xi}_{t_{0}}({\bf x})=0$. Therefore, if $\Omega^{+}_{t_0}\equiv\Omega$, equations (\ref{eq_initial_conditions}) are fulfilled for $\xi_{t_0}\equiv\hat{\xi}_{t_0}$ and $\lambda_{t_0}\equiv{\hat{\lambda}}_{t_0}=0$.
		\item[b)]
		 Since the shift, $\Delta_{shift}$, is applied to both fields, ${\xi}_{t_{n}}^{(i)}({\bf x})$ and $\lambda_{t_n}^{(i)}$, their difference, $\psi^{(i)}_{t_n}\equiv{\xi}_{t_{n}}^{(i)}({\bf x})-\lambda_{t_n}^{(i)}$, and the corresponding topology, $\chi^{(i)}_{t_n}={\cal H}_{\beta}(\psi^{(i)}_{t_n})$, is not affected by that shift. Also the normalization in equation (\ref{eq_normalized_values}) does not affect neither the sign of $\psi^{(i)}_{t_n}$ nor the topology.
		\item[c)]The Laplacian smoothing, $\hat{\xi}^{(i)}_{t_n}\leftarrow\hat{\xi}^{(i)}_{\tau,t_n}$, is neither affected by the initialization operations\footnote{Except for the error associated to the (assumed small) parameter $\epsilon$ in the smoothing procedure (see equations (\ref{eq_contour_line})).}.
		 \item[d)] In consequence, if the discrimination function $\psi^{(i)}_{t_n}$, in equation (\ref{eq_volume_corrected_2})-(a), is applied in terms of the shifted and normalized entities, $\hat{\xi}_{\tau,t_{n}}^{(i)}\left(\mathbf{x}\right)$ and $\hat{\lambda}_{t_n}^{(i)}$ at $\Omega^+$ (see equation (\ref{eq_volume_corrected_2})), the topology, at convergence, is not affected. However, the robustness of the iterative process is remarkably increased.
	\end{itemize}
\end{remark}
\subsection{Finite element implementation. Numerical aspects.}
In the examples presented in section \ref{sec_representative_numerical_simulations}, the design space, $\Omega$, has been discretized in uniform structured meshes (labeled $Mi, i=1,2..$) made of regular hexahedra, i.e. elements, $e$, occupying the domain $\Omega^{(e)}\subset\Omega$, with typical size $h^{(e)}$. These finite element meshes have been used both for solving the state bi-material elastic problem, in equation (\ref{eq_elastic_problem_2}), and for obtaining the geometrical data (typically the volume  of the soft-phase $\Omega^-$) necessary for the cutting and bisection algorithm in sections \ref{sec_cutting_algorithm}  and \ref{sec_bisection_algorithm}. In general, the integration (Gauss) points of regular elements have been used as \textit{sampling points}, ${\bf x}_i$, to evaluate the necessary data i.e. \textit{characteristic function} $\chi({\bf x}_i)$, \textit{discrimination function} $\psi(\chi,{\bf x}_i)$, etc.

\subsubsection{Topology properties determination. Modified marching cubes strategy}
Determination of the volumes, $\Omega^+$ or $\Omega^-$, in the cutting algorithm (equation (\ref{eq_volume}), see also figure \ref{fig_cut_level}) has been performed via the classic strategy of \textit{marching cubes} \cite{Lorensen1987}. In appendix \ref{app_marching_cubes}, additional details on the method are provided.

\subsubsection{Bi-material elements. Three field mixed displacement-strain-stress formulation}
A specific issue arises in the treatment of the finite element formulation of \textit{bi-material elements}, i.e. those elements containing both the hard-phase, $\mathfrak {M}^+$, and soft-phase, $\mathfrak {M}^-$. They are identified during the marching cubes strategy and  the subset of bi-material elements $\Omega^{(+,-)}\subset\Omega$ is determined, as well as the portions $\Omega^{(e,+)}\subset\Omega^{(e)}$ and $\Omega^{(e,-)}\subset\Omega^{(e)}$ ($\Omega^{(e,+)}\cup\Omega^{(e,-)}=\Omega^{(e)}$) containing hard and soft material phases in every element $(e)$. 
\begin{equation}
	\left\{
	\begin{split}
	&\Omega^{(e,+)}\coloneqq\{{\bf x}\in\Omega^{(e)}\; ;\quad{\bf x}\in\Omega^{+}\}\\
	&\Omega^{(e,-)}\coloneqq\{{\bf x}\in\Omega^{(e)}\; ;\quad{\bf x}\in\Omega^{-}\}\\
	&\Omega^{(+,-)}\coloneqq\left\{\cup_{e=1}^{e=n_{elem}}\left(\Omega^{(e,+)}\cup\Omega^{(e,-)}\right)\right\}
	\end{split}
	\right.
\end{equation}
Details on the derivation of this formulation are given in Appendix \ref{app_mixed_finite_element_formulation}.

\section{Structural compliance problems}
\label{sec_structural_compliance_problems}  
Let us now specify the equations for the  \textit{minimal compliance} topological optimization in a linear elastic problem, e.g.:
\begin{equation}  \label{eq_cost_function_compliance}
\begin{split}
	&\chi(\mathbf{x},t)=
	\underset{\chi}
	{\operatorname{argmin}}\ {\cal J}(\mathbf{u}_{\chi}(\mathbf{x},t))=
	\underset{\chi}
	{\operatorname{argmin}}\  l(\mathbf{u}_{\chi}(\mathbf{x},t))\quad&(a)  \\
	&\textit{subject to:} \quad
	{\mathcal C}(\chi,t)\coloneqq t-\dfrac{\vert\Omega^-\vert(\chi)}{\vert\Omega\vert}=0\; ;\quad t\in[0,1]   &(b) 
\end{split}
\end{equation}
The problem above, belongs to the class  of problems considered in equation (\ref{eq_minimization_restricted_time}) with
\begin{equation}
\label{eq_compliance_cost_function}
\begin{split}
{\cal J}({\bf u}_\chi) = l(\mathbf{u_{\chi}})&=\int_{\Omega} \mathbf{b}\cdot\mathbf{u_{\chi}}d\Omega+
\int_{\partial_{\sigma}\Omega}
{\mathbf{t}}^{*}\cdot{\mathbf u_{\chi}}d\Gamma=a_{\chi}(\mathbf{u}_{\chi},\mathbf{\bf u}_\chi)=\\
&=\int_{\Omega}\bm{\varepsilon}(\mathbf{u}_{\chi}):{\mathbb C}_{\chi}:\bm{\varepsilon}(\mathbf{u}_{\chi}) d\Omega=\int_\Omega2{\cal U_\chi}d\Omega
\end{split}
\end{equation}
where equations (\ref{eq_bilinear_form_problem})-(d)-(e) have been considered for ${\bf w}\equiv{\bf u}$, and ${\cal U}_\chi$ can be identified as the \emph{actual elastic energy density } (${\cal U}_{\chi}=\frac{1}{2}\bm{\varepsilon}:{\mathbb C}_{\chi}:\bm{\varepsilon}$).
Comparing equations (\ref{eq_compliance_cost_function}) and (\ref{eq_minimization_restricted}) we can identify
\begin{equation}
\label{eq_identification1}
F(\chi,{\bf x})\equiv\bm{\varepsilon}\left(\mathbf{u}_{\chi}({\bf x})\right):{\mathbb C}_{\chi}:\bm{\varepsilon}\left(\mathbf{u}_{\chi}({\bf x})\right)=2{\cal U}_\chi({\bf x})
\end{equation}
The corresponding finite element discretization counterpart of the problem in equation (\ref{eq_cost_function_compliance}) reads
\begin{equation}\label{eq_discrete_form}
\begin{split}
&\chi(\mathbf{x},t)=
\underset{\chi}
{\operatorname{argmin}}\ {\cal J}^{(h)}(\mathbf{d}_{\chi}(t))=
\underset{\chi}
{\operatorname{argmin}}\  \mathbf{f}^{T}\mathbf{d}_{\chi}(t)\quad&(a)  \\
&\textit{subject to:} \quad
{\mathcal C}(\chi,t)\coloneqq t-\dfrac{\vert\Omega^-\vert(\chi)}{\vert\Omega\vert}=0\; ;\quad t\in[0,1]   &(b) 
\end{split}
\end{equation}
where $h$ stands for the typical size of the finite element mesh, and the term $\mathbf{f}^{T}\mathbf{d}_{\chi}(t)$, in equation (\ref{eq_discrete_form})-(a) corresponds to the structural compliance value.
\subsection{Cost function topological sensitivity}
The \textit{adjoint equation method}  \cite{Cea2000} for sensitivity analysis is mimicked here for the case of RTD. The goal is to compute the variational topological derivative (RTD) of the cost-function\footnote{Omitting time dependencies.}, ${\cal J}^{(h)}(\mathbf{d}_{\chi})$, in equation (\ref{eq_discrete_form})-(a), without explicitly computing the sensitivity of the nodal displacement field ($\partial {\bf d}_{\chi} \slash \partial \chi$).
The sensitivity of ${\cal J}^{(h)}(\chi)$ can be, then, obtained through the following steps:
\begin{enumerate}
	\item{} Rephrase the cost function accounting for the state equation (\ref{eq_equilibrium})-(a) 	
	\begin{equation} \label{eq_rephrased}
	{\cal J}^{(h)}(\chi)=\mathbf{f}^{T}\mathbf{d}_{\chi}=
	\mathbf{f}^{T}\mathbf{d}_{\chi}-\mathbf{w}^T
	\underbrace{\left( {\mathbb K}_{\chi}\mathbf{d}_{\chi}-\mathbf{f}\right)}_{\textstyle{=\mathbf{0}}} 
	\end{equation}
	where $\mathbf{w}$ is a vector to be chosen.
	\item{} Compute the RTD of the equation (\ref{eq_rephrased})
	\begin{equation}\label{eq_diferentiation}
	\dfrac{\delta{\cal J}^{(h)}(\chi)}{\delta\chi}(\hat{\mathbf{x}})=
	(\mathbf{f}^{T}-\mathbf{w}^T{\mathbb K}_{\chi})\dfrac{\delta\mathbf{d}_{\chi}}{\delta{\chi}}({\hat{\mathbf{\mathbf{x}}}})-
	\mathbf{w}^T
	\dfrac{\delta {\mathbb K}_{\chi}}{\delta\chi}({\hat{\mathbf{\mathbf{x}}}})\mathbf{d}_{\chi}
	\end{equation}
	\item{} Choose $\mathbf{w}\equiv\mathbf{d}_{\chi}$ and replace it into equation (\ref{eq_diferentiation})
	\begin{equation}
	\dfrac{\delta{\cal J}^{(h)}(\chi)}{\delta\chi}(\hat{\mathbf{x}})=
	\underbrace{(\mathbf{f}^{T}-\mathbf{d}_{\chi}^T{\mathbb K}_{\chi})}_{\textstyle{=\mathbf{0}}}\dfrac{\delta\mathbf{d}_{\chi}}{\delta{\chi}}({\hat{\mathbf{\mathbf{x}}}})-
	\mathbf{d}_{\chi}^T\dfrac{\delta {\mathbb K}_{\chi}}{\delta\chi}({\hat{\mathbf{\mathbf{x}}}})\mathbf{d}_{\chi}=
	-\left[ \mathbf{d}_{\chi}^T\dfrac{\delta {\mathbb K}_{\chi}}{\delta\chi}({{\mathbf{\mathbf{x}}}})\mathbf{d}_{\chi}\right]
	_{\mathbf{x}\equiv\hat{\mathbf{x}}} 
	\end{equation}	
\end{enumerate}
where equations (\ref{eq_equilibrium}) and the symmetry of $\mathbb{K}_{\chi}$ ($\mathbb{K}_{\chi}=\mathbb{K}_{\chi}^T)$ have been considered, yielding the final result
\begin{equation}\label{eq_final_derivative}
\begin{split}
\dfrac{\delta{\cal J}^{(h)}(\chi)}{\delta\chi}(\hat{\mathbf{x}})
&=-\underbrace{\mathbf{d}^T_{\chi}\mathbf{B}^T(\hat{\mathbf{x}})}_{\textstyle {\varepsilon^T_{\chi}(\hat{\mathbf{x}})}}
\dfrac{\partial {\mathbb{D}_{\chi}}}{\partial\chi}({{\mathbf{\hat{\mathbf{x}}}}})
\underbrace{\mathbf{B}(\hat{\mathbf{x}})\mathbf{d}_{\chi}}_{\textstyle {\bm{\varepsilon}_{\chi}(\hat{\mathbf{x}})}}\Delta\chi(\hat{\bf x})
=-\left[ \bm{\varepsilon}^T_{\chi}(\mathbf{x})\dfrac{\partial \mathbb{D}_{\chi}}{\partial{\chi}}\bm{\varepsilon}_{\chi}(\mathbf{x})\right]
_{\mathbf{x}\equiv\hat{\mathbf{x}}}\Delta\chi(\hat{\bf x})=\\
& =-\left[ m{\chi}^{m-1}\bm{\varepsilon}^T_{\chi}(\mathbf{x}) \mathbb{D}\bm{\varepsilon}_{\chi}(\mathbf{x})\right]_{\mathbf{x}\equiv\hat{\mathbf{x}}}{{\Delta\chi}({\hat{\bf x}}) }  
\end{split}
\end{equation}
where the result, in matrix form, in equation (\ref{eq_contrasted_constitutive_tensor}) has been considered. The result in equation (\ref{eq_final_derivative}) can be then written as
\begin{equation}
\label{eq_compliance_topological_derivative}
\begin{split}
&\left\{
\begin{split}
\dfrac{\delta \left({\bf f}^T{\bf d}_{\chi}\right)}{\delta \chi}=\dfrac{\delta{{\cal J}^{(h)}(\mathbf{u}_{\chi})}}{\delta\chi}(\hat{\mathbf{x}})&=
-\left[ m{\chi}^{m-1}\bm{\varepsilon}^T(\mathbf{d}_{\chi}){\mathbb D}:\bm{\varepsilon}(\mathbf{d}_{\chi})\right]_{\mathbf{x}=\hat{\mathbf{x}}}
{{\Delta\chi}({\hat{\bf x}}) }\\
&=-2m \left(\chi(\hat{\mathbf{x}})\right) ^{m-1}{\overline{\cal U}}(\hat{\mathbf{x}}){{\Delta\chi}({\hat{\bf x}}) } \\
\end{split}
\right. \quad&(a)\\
&\overline{\cal U}(\hat{\mathbf{x}})=\dfrac{1}{2}\left(\bm{\varepsilon}^T(\mathbf{d}_{\chi}){\mathbb D}\bm{\varepsilon}(\mathbf{d}_{\chi})\right)(\hat{\mathbf{x}})\quad&(b)
\end{split}
\end{equation}
where $\overline{\cal U}_{\chi}(\hat{\mathbf{x}})$ is \textit{the nominal elastic energy density}\footnote{Notice the difference of the \emph{nominal elastic energy density} $\overline{\cal U}(\hat{\mathbf{x}})=\dfrac{1}{2}\left(\bm{\varepsilon}^T(\mathbf{d}_{\chi}){\mathbb D}\bm{\varepsilon}(\mathbf{d}_{\chi})\right)(\hat{\mathbf{x}})$ (in terms of $\mathbb D$), in equation (\ref{eq_compliance_topological_derivative})-(b), and the \textit{actual elastic energy density}, ${\cal U}(\hat{\mathbf{x}})=\dfrac{1}{2}\left(\bm{\varepsilon}^T(\mathbf{d}_{\chi}){\mathbb D}_{\chi}\bm{\varepsilon}(\mathbf{d}_{\chi})\right)(\hat{\mathbf{x}})$ (in terms of ${\mathbb D}_\chi=\chi^m{\mathbb D}$).}. The result in equation (\ref{eq_compliance_topological_derivative}) is close to the one obtained for the sensitivity in the (regularized/smoothed) method SIMP \cite{MartinPh.Bendsoe2003}, except for the \textit{material exchange term} ${{\Delta\chi}({\hat{\bf x}})}$ (see equation (\ref{eq_alpha})).
 
Now the original compliance functional ${\cal J}^{(h)}$, in equation (\ref{eq_discrete_form})-(a)  can be extended to account for the restriction in equation (\ref{eq_discrete_form})-(b). Following the methodology in section \ref{sec_Variational_optimization_problem_equality_restricted}, and accounting for the RTD results in Table (\ref{table_RTD_examples}) one arrives to
\begin{tcolorbox}[colback=white!50!white]
	\textbf{Box II. Topological optimization for mean compliance problems}
\begin{equation}
\label{eq_compliance}
\begin{split}
&\textit{Problem:}  \\
&\left\{
\begin{split}
&\chi^{*} =
\underset{\chi}
{\operatorname{argmin}}\  {\cal J}^{(h)}{(\chi)}\coloneqq {\bf f}^T{\bf d}_{\chi}\\
&s.t.\quad{\cal C}(\chi)\equiv{\vert\Omega^+(\chi)\vert}-\bar{V}=0
\end{split}
\right.   &\quad(a)\\
&\textit{Lagrangian:}\quad{\cal L}(\chi,\lambda)={\cal J}^{(h)}(\chi)+\lambda {\cal C}(\chi)&\quad(b)\\
&\textit{Optimality criterion:}   &\\
&\left\{
\begin{split}
&\dfrac{\delta {\cal L}(\chi,\lambda)}{\delta\chi}({\bf x})=
-2m\chi^{m-1}({\mathbf{x}})\overline{{\cal U}}({\mathbf{x}}){\Delta \chi}({\bf x})
+\lambda\ \text{sgn}(\Delta\chi({\bf x}))\ >{0}\quad\forall{\bf x}\in\Omega\\
&{\cal C}(\chi)=0
\end{split}
\right.\quad\quad\quad&(c)\\
&\textit{Closed-form solution:}\\
&
\left\{
\begin{split}
&\psi_{\chi}({\bf x},\lambda)\coloneqq2m(1-\beta) \chi^{m-1}({\mathbf{x}})\overline{{\cal U}}({\mathbf{x}})
-\lambda\; ;\quad \overline{\cal U}(\hat{\mathbf{x}})=\dfrac{1}{2}\left(\bm{\varepsilon}^T(\mathbf{d}_{\chi}){\mathbb D}\bm{\varepsilon}(\mathbf{d}_{\chi})\right)(\hat{\mathbf{x}})\ge 0\\
&\left\{
\begin{split}
&\chi({\bf x},\lambda)={\cal H}_{\beta}\left[ \psi_{\chi}({\bf x},\lambda) \right] \rightarrow\quad \chi_{\lambda}({\bf x})\quad&\textbf{(d-1)}\\
&{\cal C}(\chi_{\lambda},\lambda)=0&\textbf{(d-2)}
\end{split}
\right\}\rightarrow \lambda^{*}\rightarrow \chi^{*}({\bf x})=:\chi_{\lambda^{*}}({\bf x})
\end{split}
\right. &(d)\\
&\textit{Topology:}\\
&\left\{
\begin{split}
&\Omega^{+}(\chi)\coloneqq\{\mathbf{x}\in \Omega \; ;\quad \psi_\chi\mathbf{(x,\lambda)}>0 \}\\
&\Omega^{-}(\chi)\coloneqq\{\mathbf{x}\in \Omega \; ;\quad  \mathbf{x}\notin {\Omega^+}\}\\
&\Gamma(\chi)\ \ \ \coloneqq\{\mathbf{x}\in \Omega \; ;\quad \psi_\chi\mathbf{(x,\lambda)}=0\} \\
\end{split}
\right.\\
\end{split}
\end{equation}
\end{tcolorbox}
\section{Compliant mechanisms}
\label{sec_compliant_mechanism}
\begin{figure}[h]
	\centering
	\includegraphics[width=17cm, height=11cm]{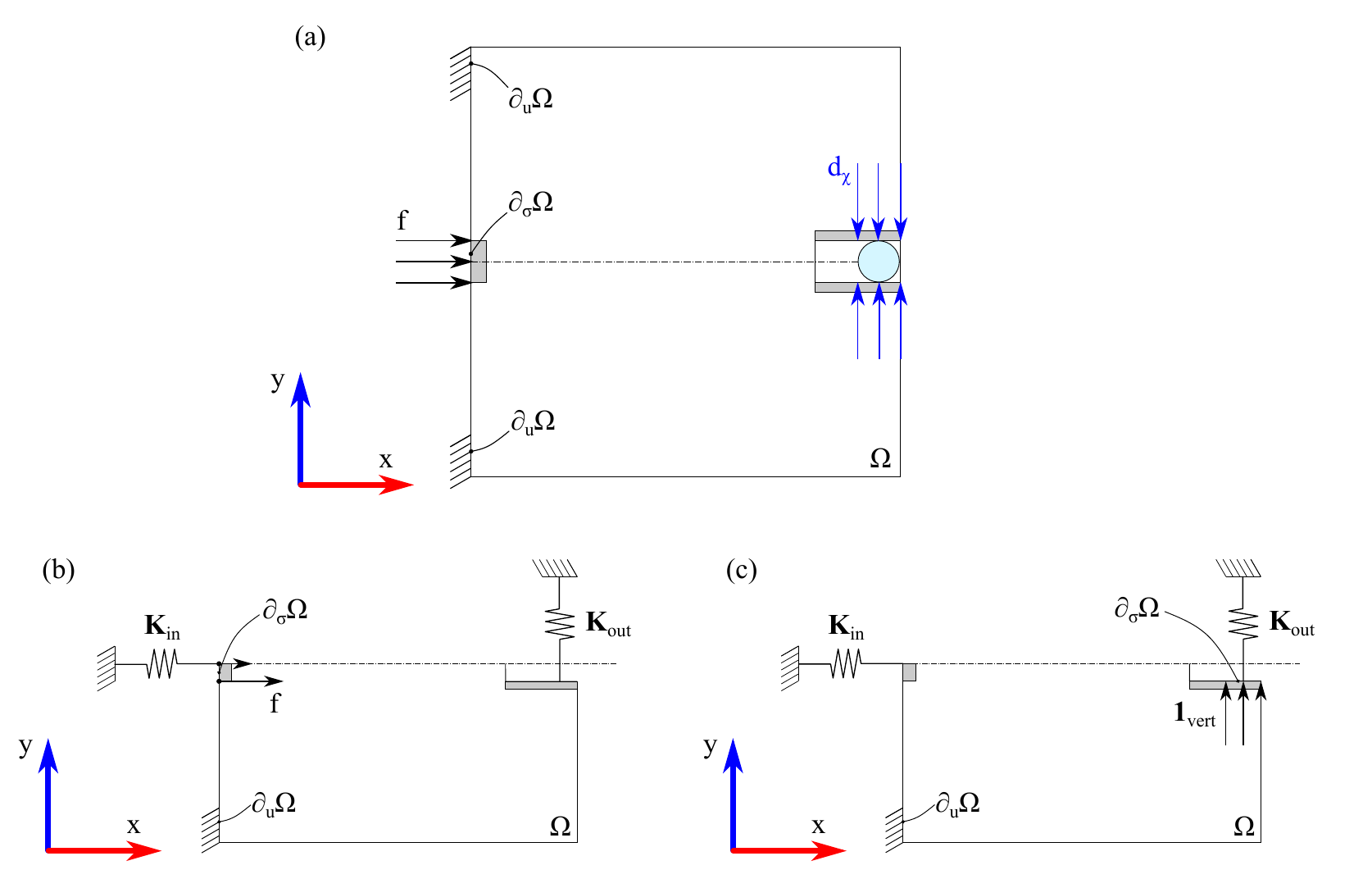}
	\caption{Compliant mechanism: (a) problem setting, (b) system (1) (half domain), (c) system (2) (half domain).}
	\label{fig_compliant_mechanism}
\end{figure}

We consider now the optimal design of the compliant mechanism sketched in figure \ref{fig_compliant_mechanism}.
The goal is to design the topology of a gripper, so as to maximize the compressive displacement of a spring placed at \textit{port (out)}, with stiffness ${\bf K}_{out}$ (standing for the stiffness of the gripped object), under the action of a force acting on \textit{port (in)}, ${\bf f}$, also applied by means of spring of stiffness ${\bf K}_{in}$ (see figure \ref{fig_compliant_mechanism}) \cite{Sigmund1997,nishiwaki1998topology,frecker1997topological}. Therefore the structural system is constituted by the gripper and the two springs (at ports \textit{in} and  \textit{out}).
In the context of a finite element discretization, like in section \ref{sec_finite_element_discretization}, the problem is expressed as
\begin{equation}
\begin{split}
\label{eq_compliant_mechanism}
&\min_{\chi}{\cal J}^{(h)}({\bf d}_\chi)=
-{\bf d}_{\chi}^T{\bf 1}_{\text{vert}} \quad&(a)\\
&\textit{subject to}:\quad{\mathcal C}(\chi,t)\coloneqq t-\dfrac{\vert\Omega^-\vert(\chi)}{\vert\Omega\vert}=0\; ;\quad t\in[0,1] \quad&(b)
\end{split}\\
\end{equation}
where, in equation (\ref{eq_compliant_mechanism})-(a), ${\bf 1}_{\text{vert}}$ stands for that nodal vector force corresponding to uniform vertical components (\textit{in the spring stretching sense}) acting on the gripper jaws\footnote{Therefore, its components are zero everywhere excepting at the gripper jaws nodes.}.
 
The \textit{original elastic system (1)} is supplemented with an \textit{auxiliary elastic system (2)}, subject only to the forces ${\bf 1}_{\text{vert}}$ (see figure \ref{fig_compliant_mechanism}). Both of them are ruled by the elastic problem in equations (\ref{eq_equilibrium}), with the same stiffness matrix ${\mathbb K}_{\chi}$ and \textit{different actions}, with solutions ${\bf d}_{\chi}^{(1)}$ and  ${\bf d}_{\chi}^{(2)}$, respectively, i.e.:
\begin{equation}
\label{eq_split_system}
\left\{
\begin{split}
&{\mathbb K}\ {\bf d}_{\chi}^{(1)}={\bf f}\; ;\quad{\bf d}_{\chi}={\bf d}_{\chi}^{(1)} &(a)\\
&{\mathbb K}\ {\bf d}_{\chi}^{(2)}={\bf 1}_{\text{vert}}\quad&(b)
\end{split}
\right.
\end{equation}
Standard algebraic manipulations allow replacing the cost function in equation (\ref{eq_compliant_mechanism})-(a) by
\begin{equation}
\label{eq_cost_function_compliant_mechanism}
\begin{split}
{\cal J}^{(h)}({\bf u}_\chi)
&\equiv -\int_{\Omega}\bm{\varepsilon}\left({\bf d}_{\chi}^{(1)}\right):{\mathbb C}_{\chi}:\bm{\varepsilon}\left({\bf d}_{\chi}^{(2)}\right) d\Omega=-\int_\Omega2{\cal U_\chi}({\bf x})d\Omega
\end{split}
\end{equation}
where, again, ${\cal U_\chi}$ is defined as a \textit{pseudo elastic energy} of the combined system. 
\begin{equation}
\left\{
\begin{split}
&{\cal U}_{\chi}({\bf x})=\frac{1}{2}\bm{\varepsilon}^T\left({\bf d}_{\chi}^{(1)}({\mathbf{x}})\right){\mathbb D}_{\chi}\bm{\varepsilon}\left({\bf d}_{\chi}^{(2)}({\mathbf{x}})\right)\equiv \frac{1}{2}\left(\bm{\varepsilon}_{\chi}^{(1)}({\mathbf{x}})\right)^T{\mathbb D}_{\chi}\bm{\varepsilon}_{\chi}^{(2)}({\mathbf{x}})\\
&\bm{\varepsilon}\left({\bf d}_{\chi}^{(1)}({\mathbf{x}})\right)\equiv\bm{\varepsilon}_{\chi}^{(1)}({\mathbf{x}})=\mathbf{B}({\mathbf{x}})\mathbf{d}_{\chi}^{(1)}
\; ;\quad \bm{\varepsilon}\left({\bf d}_{\chi}^{(2)}({\mathbf{x}})\right)\equiv\bm{\varepsilon}_{\chi}^{(2)}({\mathbf{x}})=\mathbf{B}({\mathbf{x}})\mathbf{d}_{\chi}^{(2)}
\end{split}
\right.
\end{equation}

\subsection{Topological sensitivity}
The sensitivity of ${\cal J}^{(h)}(\chi)$ is then obtained through the following steps
\begin{enumerate}
	\item{} Rephrase the cost function accounting for the state equation (\ref{eq_split_system})-(a) (adjoint equation method)	
	\begin{equation} \label{eq_rephrased_1}
	\begin{split}
	{\cal J}^{(h)}(\chi)=&-{\bf 1}^{T}_{\text{vert}}{\bf d}_{\chi}=-{\bf 1}^{T}_{\text{vert}}{\bf d}_{\chi}^{(1)}\\
	=&-{\bf 1}_{\text{vert}}^T{\bf d}^{(1)}_{\chi}+
	{\bf w}^T
	\underbrace{\left( {\mathbb K}_{\chi}\mathbf{d}_{\chi}^{(1)}-\mathbf{f}\right)}_{\textstyle{=\mathbf{0}}} 
	\end{split}
	\end{equation}
	where $\mathbf{w}$ is a vector to be chosen.
	\item{} Compute the RTD of the equation (\ref{eq_rephrased_1})
	\begin{equation}\label{eq_diferentiation_1}
	\begin{split}
	\dfrac{\delta{\cal J}^{(h)}(\chi)}{\delta\chi}(\hat{\mathbf{x}})=&
	-({\bf 1}_{\text{vert}}^T-\mathbf{w}^T{\mathbb K}_{\chi})\dfrac{\delta{\bf d}_{\chi}^{(1)}}{\delta{\chi}}({\hat{\mathbf{\mathbf{x}}}})+\mathbf{w}^T
	\dfrac{\delta {\mathbb K}_{\chi}}{\delta\chi}({\hat{\mathbf{\mathbf{x}}}})\mathbf{d}_{\chi}^{(1)}
	\end{split}
	\end{equation}
	\item{} Choose ${\bf{w}}\equiv{\bf d}_{\chi}^{(2)}$  and replace it into equation (\ref{eq_diferentiation_1})
	\begin{equation}\label{eq_diferentiation_2}
	\begin{split}
	\dfrac{\delta{\cal J}^{(h)}(\chi)}{\delta\chi}(\hat{\mathbf{x}})=&
	-\underbrace{\left({\bf 1}^T_{\text{vert}}-\left({\bf d}_{\chi}^{(2)}\right)^T{\mathbb K}_{\chi}\right)}_{\displaystyle {\bf =0}}
	\dfrac{\delta{\bf d}_{\chi}^{(1)}}{\delta{\chi}}({\hat{\mathbf{\mathbf{x}}}})
	+&\left({\bf d}_{\chi}^{(2)}\right)^T
	\dfrac{\delta {\mathbb K}_{\chi}}{\delta\chi}({\hat{\mathbf{\mathbf{x}}}})\mathbf{d}_{\chi}^{(1)}\\
	=&\left({\bf d}_{\chi}^{(1)}\right)^T\dfrac{\delta {\mathbb K}_{\chi}}{\delta\chi}({\hat{\mathbf{\mathbf{x}}}})\mathbf{d}_{\chi}^{(2)}
	\end{split}
	\end{equation}	
\end{enumerate}
where equation (\ref{eq_split_system})-(b) and the symmetry of $\mathbb{K}$ ($\mathbb{K}=\mathbb{K}^T)$ have been considered, yielding the final result
\begin{equation}\label{eq_final_derivative_1}
\begin{split}
\dfrac{\delta\left(-{\bf 1}^{T}_{\text{vert}}{\bf d}_{\chi}\right)}{\delta\chi}(\hat{\mathbf{x}})&=\dfrac{\delta{\cal J}^{(h)}(\chi)}{\delta\chi}(\hat{\mathbf{x}})
=\underbrace{\left({\bf d}_{\chi}^{(1)}\right)^T\mathbf{B}^T(\hat{\mathbf{x}})}_{\textstyle {\left(\varepsilon_{\chi}^{(1)}\right)^T}(\hat{\mathbf{x}})}
\dfrac{\delta {\mathbb{D}_{\chi}}}{\delta\chi}({{\mathbf{\hat{\mathbf{x}}}}})
\underbrace{\mathbf{B}(\hat{\mathbf{x}}){\bf d}_{\chi}^{(2)}}_{\textstyle {\bm{\varepsilon}_{\chi}^{(2)}(\hat{\mathbf{x}})}}
=\left[ \left(\bm{\varepsilon}^{(1)}_{\chi}\right)^T\dfrac{\delta \mathbb{D}_{\chi}}{\delta{\chi}}\bm{\varepsilon}^{(2)}_{\chi}\right]
_{\mathbf{x}\equiv\hat{\mathbf{x}}}=\\
&=\left[ m{\chi}^{m-1}\left(\bm{\varepsilon}^{(1)}_{\chi}\right)^T \mathbb{D}\bm{\varepsilon}_{\chi}^{(2)}\right]_{\mathbf{x}\equiv\hat{\mathbf{x}}}{{\Delta\chi}({\hat{\bf x}}) } =
2m\left({\chi}({\hat{\bf x}})\right)^{m-1}\overline{\cal U}(\hat{\bf x}){\Delta\chi}({\hat{\bf x}}) 
\end{split}
\end{equation}
where equations (\ref{eq_equilibrium})-(b) (${\mathbb D}_{\chi}=\chi^m{\mathbb D})$ has been considered. From equation (\ref{eq_final_derivative_1}) the \textit{nominal pseudo elastic strain energy density} of the combined system, $\overline{\cal U}({\bf x})${\footnote{Notice that, unlike in equation (\ref{eq_compliance_topological_derivative})-(b), for the mean structural compliance problem, now ${\overline{\cal U}(\hat{\bf x})}$ may be positive or negative. This makes a substantial difference in the sign of the RTD in equation (\ref{eq_final_derivative_1}).}, is defined as
\begin{equation}
\overline{\cal U}(\hat{\bf x})\coloneqq\frac{1}{2}\left(\bm{\varepsilon}^{(1)}_{\chi}(\hat{\bf x})\right)^T \mathbb{D}\ \bm{\varepsilon}_{\chi}^{(2)}(\hat{\bf x})
\end{equation}
\subsection{Closed-form solution}
Following the methodology in section \ref{sec_Variational_optimization_problem_equality_restricted}, and accounting for the RTD results in table \ref{table_RTD_examples}, one arrives to
\begin{tcolorbox}[colback=white!50!white]
\textbf{Box III.  Topological optimization of compliant mechanisms}	
\begin{equation}
\label{eq_compliance_mechanism}
\begin{split}
&\textit{Problem:}  \\
&\left\{
\begin{split}
&\min_{\chi}{\cal J}^{(h)}({\chi})=
-{\bf d}_{\chi}^T{\bf 1}_{\text{vert}} \quad&\textbf{(a-1)}\\
&s.t.\quad{\cal C}(\chi)\equiv{\vert\Omega^+(\chi)\vert}-\bar{V}=0\quad&\textbf{(a-2)}
\end{split}
\right.   &\quad(a)\\
&\textit{Lagrangian:}\quad{\cal L}(\chi,\lambda)={\cal J}^{(h)}(\chi)+\lambda {\cal C}(\chi)&\quad(b)\\
&\textit{Optimality criterion:}   &\\
&\left\{
\begin{split}
&\dfrac{\delta {\cal L}(\chi,\lambda)}{\delta\chi}({\bf x})=
2m{\chi}^{m-1}({\mathbf{x}})\overline{\cal U}({\bf x}){\Delta\chi}({{\bf x}})
+\lambda\ \text{sign}(\Delta\chi({\bf x}))\ \quad\forall{\bf x}\in\Omega\\
&{\cal C}(\chi)=0
\end{split}
\right.\quad\quad\quad&(c)\\
&\textit{Closed-form solution:}\\
&\left\{
\begin{split}
&\psi_{\chi}({\bf x},\lambda)\coloneqq-(1-\beta)2m \chi^{m-1}({\mathbf{x}})\overline{{\cal U}}({\mathbf{x}})
-\lambda\; ;\quad \overline{\cal U}({\bf x})\coloneqq\dfrac{1}{2}\left(\bm{\varepsilon}^{(1)}_{\chi}({\bf x})\right)^T \mathbb{D}\ \bm{\varepsilon}_{\chi}^{(2)}({\bf x})\\
&\left\{
\begin{split}
&\chi({\bf x},\lambda)={\cal H}_{\beta}\left[ \psi_{\chi}({\bf x},\lambda) \right] \rightarrow\quad \chi_{\lambda}({\bf x})\quad&\textbf{(d-1)}\\
&{\cal C}(\chi_{\lambda},\lambda)=0&\textbf{(d-2)}
\end{split}
\right\}\rightarrow \lambda^{*}\rightarrow \chi^{*}({\bf x})=:\chi_{\lambda^{*}}({\bf x})
\end{split}
\right. &(d)\\
&\textit{Topology:}\\
&\left\{
\begin{split}
&\Omega^{+}(\chi)\coloneqq\{\mathbf{x}\in \Omega \; ;\quad \psi_\chi\mathbf{(x,\lambda)}>0 \}\\
&\Omega^{-}(\chi)\coloneqq\{\mathbf{x}\in \Omega \; ;\quad  \mathbf{x}\notin {\Omega^+}\}\\
&\Gamma(\chi)\ \ \ \coloneqq\{\mathbf{x}\in \Omega \; ;\quad \psi_\chi\mathbf{(x,\lambda)}=0\} \\
\end{split}
\right.\\
\end{split}
\end{equation}
\end{tcolorbox}

\section{Representative numerical simulations}
\label{sec_representative_numerical_simulations}
In the following a set of 3D numerical simulations are presented, in order to display the performance of the proposed approach. Unless it is differently indicated, the following material properties for the isotropic elastic material are considered:
\begin{equation}
\label{eq_elastic_material_properties}
\begin{split}
&\text{Young modulus}\  E=210 GPa \; ;\quad \text{Poisson ratio}\  \nu=0.3 \; ;\quad \text{Stiffness contrast factor} \ \alpha=1.0\cdot10^{-6}\\ 
&\text{Stiffness contrast exponent}\  \ m=5 \quad  \text{(see equation (\ref{eq_contrasted_constitutive_tensor}))}\Rightarrow\text{Relaxation factor}
 \ \beta=\alpha^{\frac{1}{m}}=6.3\cdot10^{-2} 
\end{split}
\end{equation}
The following additional algorithmic data used for running the proposed algorithm is
\begin{equation}
\label{eq_algorithmic data}
\begin{split}
&\text{Tolerance error for topology}\  Tol_{\chi}=10^{-1} \\ 
&\text{Tolerance error for Lagrange multiplier}\  Tol_{\lambda}=10^{-1}\\
&\text{Tolerance error for bisection}\  Tol_{\cal C}=10^{-5}
\end{split}
\end{equation}

\subsection{Mean compliance optimization. Cantilever beam.}
\label{sec_compliance_cantilever_beam}
The 3D design domain, in figure \ref{fig_Cantilever_domain},  is used. A first mesh, M1, (made of 432000 bi-linear hexahedral finite elements), displayed in figure \ref{fig_Cantilever_domain}-(b), is used to mesh one-half of the design domain $\Omega$ (see figure \ref{fig_Cantilever_domain}-(a)). The mean compliance design of the structure is done for the distributed loading forces at the right, lower, boundary of the design domain (see figure \ref{fig_Cantilever_domain}-(d)).  

In figure \ref{fig_Cantilever_final_topology_M1}-(a), the pseudo-time, ($t=\frac{\vert\Omega^{-}\vert}{\vert\Omega\vert}$),
 evolution of the analysis is shown, both in terms of the cost-function  and the topology. In figure \ref{fig_Cantilever_final_topology_M1}-(b), the final design, when the hard-phase (solid material) equals $8\%$ of the initial one ($100\%$), is depicted. Figures \ref{fig_Cantilever_final_topology_M1}-(c) and \ref{fig_Cantilever_final_topology_M1}-(d) illustrate the convergence rate of the cost function, $\cal{J}_\chi$, for the first ten steps. As it can be observed, the objective function tends to a local optimal value in a few iterations for each step.

\begin{figure}[h]
	\centering
	\includegraphics[width=13cm, height=6cm]{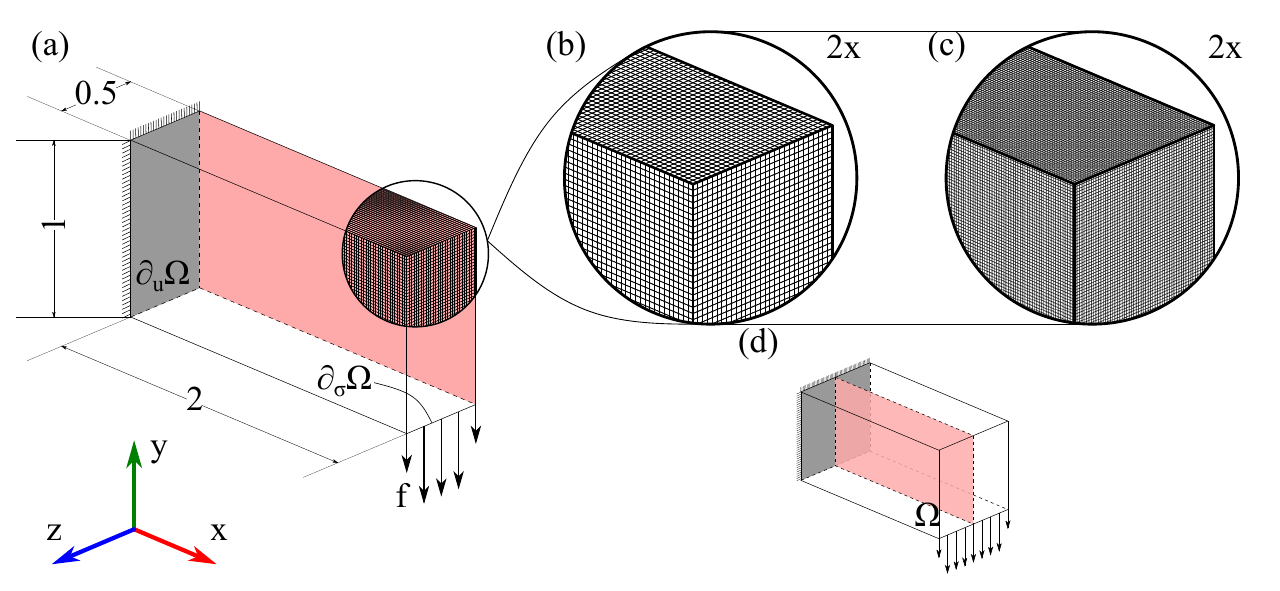}	
	\caption{Minimum compliance cantilever beam: (a) Meshed domain, (b) Detailed mesh M1, (c) Detailed mesh M2,  (d) Symmetrized design domain  $\Omega$.} 
	\label{fig_Cantilever_domain}
\end{figure}

\begin{figure}[H]
	\centering
	\includegraphics[width=17cm, height=11cm]{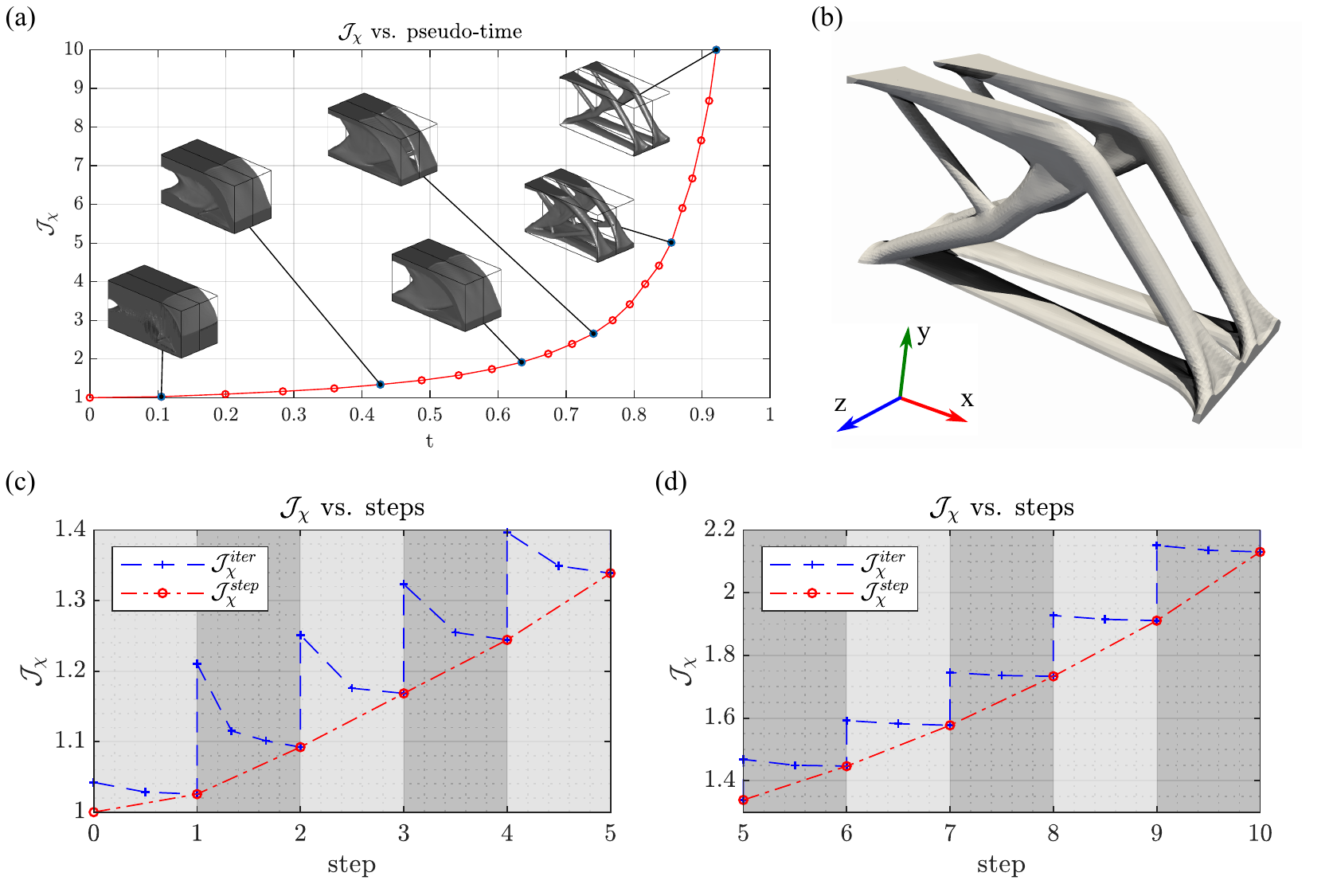}
	\caption{Cantilever beam.  Mean structural compliance optimization for Mesh M1: 432.000 elements.\\ $\tau=1 \quad(\epsilon=\tau\cdot h=1.67\cdot10^{-2})$ (a) Cost function and topology evolution\hspace{0.0cm}, (b) Topology for $t=\frac{\vert\Omega^-\vert}{\vert\Omega\vert}=0.92$,  $\frac{\vert\Omega^+\vert}{\vert\Omega\vert}=0.08$, (c--d) In-step convergence of the cost function for steps 1 to 10. Vertical stripes correspond to steps, and crosses inside them correspond to step-iterations.}	
	\label{fig_Cantilever_final_topology_M1}	
\end{figure}
   
\subsubsection{Mesh size objectivity}

The effects of the regularization parameter, $\epsilon$,  in the Laplacian smoothing equations (\ref{eq_regularized_psi}) to (\ref{eq_tau}), are examined, now for two different meshes in the design domain $\Omega$. Mesh M1 (120x60x30=216.000 elements, with size $h^{(1)}$), and mesh M2 (obtained by doubling the number of elements in every direction $\{x,y,z\}$, this leading to a total of 1.728.000 elements $\Rightarrow$  $h^{(2)}=h^{(1)}/2$, see figure \ref{fig_Cantilever_domain}-(c)).

\begin{figure}[H]
	\centering
	\includegraphics[width=17cm, height=16.5cm]{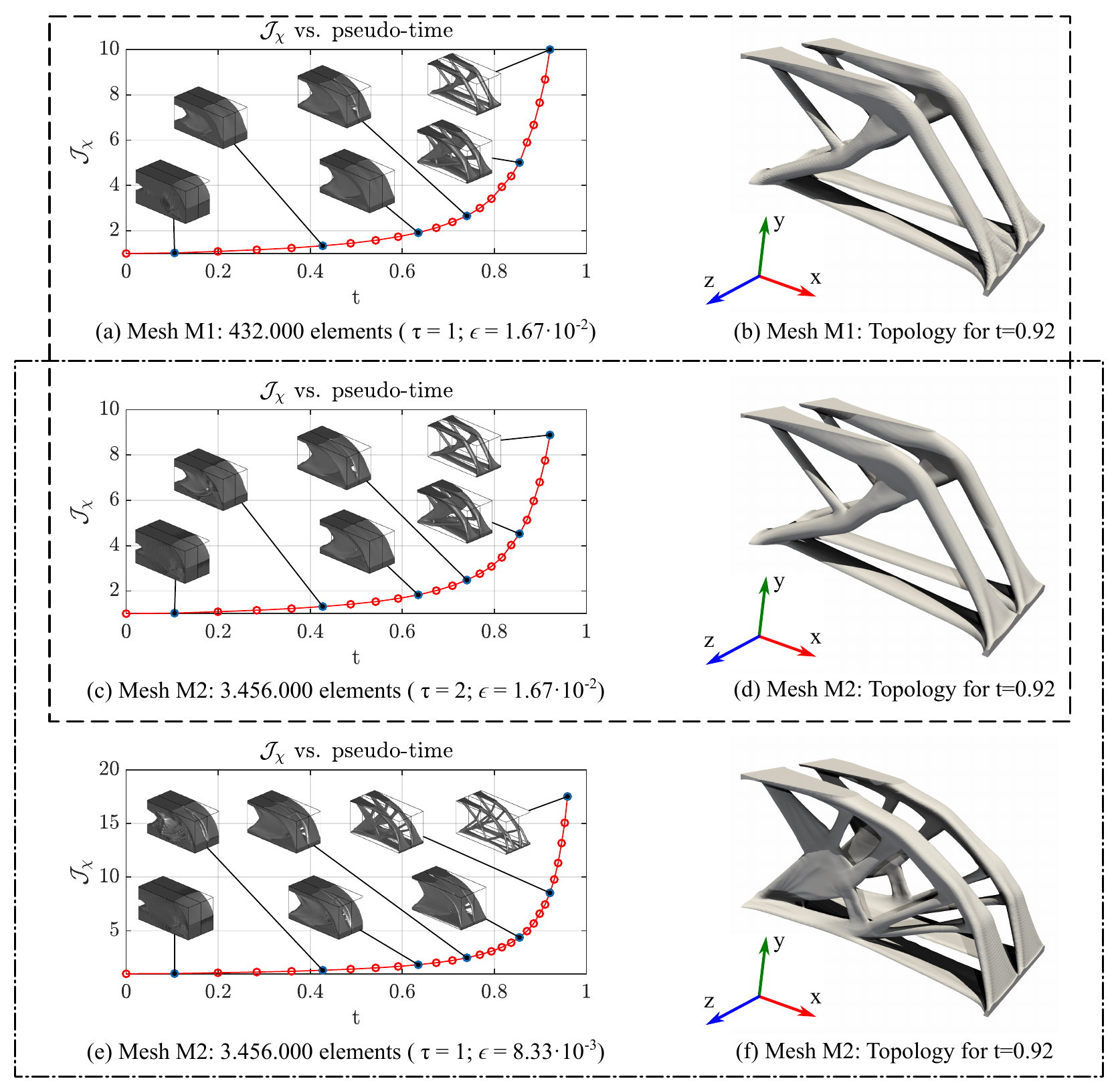}
	\caption{Cantilever beam.  Mesh-size objectivity results $\quad(\epsilon=\tau\cdot h)$. (a) to (d): results for meshes M1 and M2 and the \textit{same regularization parameter} ($\epsilon=1.67\cdot10^{-2}$), 
	(c) to (f): results for the same mesh, M2, and different regularization values ($\epsilon=1.67\cdot10^{-2}$ and $\epsilon=8.33\cdot10^{-3}$).}
	\label{fig_cantilever_beam}	
\end{figure}

Mesh size objectivity is assessed in figures \ref{fig_cantilever_beam}-(a)-(b)-(c)-(d), when the \textit{value of the regularization parameter } $\epsilon=1.67\cdot10^{-2}$ {is kept constant}  for the original, M1, and direction-doubled, M2, meshes, giving rise to very similar cost-function and topological design evolution. If, instead, \textit{no action is taken to keep this parameter constant in mesh} M2 ($\epsilon^{(2)}=8.33\cdot10^{-3}$ $=1/2\epsilon^{(1)}$, \textit{thus keeping $\tau$=constant in equation} $\epsilon^{(i)} =\tau \  h^{(i)}$), the results, in figures \ref{fig_cantilever_beam}-(e)-(f), substantially change with respect to the ones in the figure above, and \textit{they become strongly dependent on the finite element mesh size} $h^{(i)}$.  

\subsubsection{2-1/2D extruded design}
Following the steps in \cite{Yamada2010} (where a \textit{Tikhonov-like} regularization was used) we show here that, changing the tensor order of the Laplacian smoothing regularization parameter $\epsilon$, from scalar to orthotropic second order tensor or, equivalently, the parameter $\tau=\dfrac{\epsilon}{h}$ in equations (\ref{eq_regularized_psi}) to (\ref{eq_tau}), one can obtain \textit{2-1/2D extruded designs} in a selected direction.

For instance, let us replace the scalar value $\epsilon$, in these equations, by an orthotropic second order tensor, $\bm{\epsilon}$, defined through,
\begin{equation}
\label{eq_orthotropic_epsilon}
\bm{\epsilon}\coloneqq
\begin{bmatrix}
\epsilon_x& 0 & 0 \\ 
0& \epsilon_y&  0\\ 
0&  0& \epsilon_z
\end{bmatrix}=
\begin{bmatrix}
1& 0 & 0 \\ 
0& 1&  0\\ 
0&  0& \eta
\end{bmatrix}
\tau h=
\underbrace{
	\begin{bmatrix}
	\tau_x& 0 & 0 \\ 
	0& \tau_x&  0\\ 
	0&  0& \tau_z
	\end{bmatrix}
}_{\displaystyle{\bm \tau}}
h={\bm \tau}h
\; ;\quad \eta\gg1 
\end{equation}
where the second order tensor ${\bm\epsilon}={\bm \tau} h$ ($\tau_z=\eta\tau \gg\tau_x=\tau_y$)	is an orthotropic tensor defining the \textit{minimum filament width size} (see section \ref{sec_laplacean_smoothing}) in the directions $\{x,y,z\}$. The value $\eta\gg1$ sets the length of the  \textit{minimum solid filament width in the $z$-direction} (=filtered wavelength in the $z$-direction) much larger than in the other two orthogonal directions, this resulting in $z$-extruded designs (see figure \ref{fig_extruded_cantilever_design}). 
This action was performed in the mean compliance optimization of the cantilever beam, in section \ref{sec_compliance_cantilever_beam}, with the only modification of considering the orthotropic character of the filtering tensor ${\bm \epsilon}$ in equation (\ref{eq_orthotropic_epsilon}). The corresponding results are shown in figure \ref{fig_extruded_cantilever_design}.  

\begin{figure}[H]
	\centering
	\includegraphics[width=14cm, height=6cm]{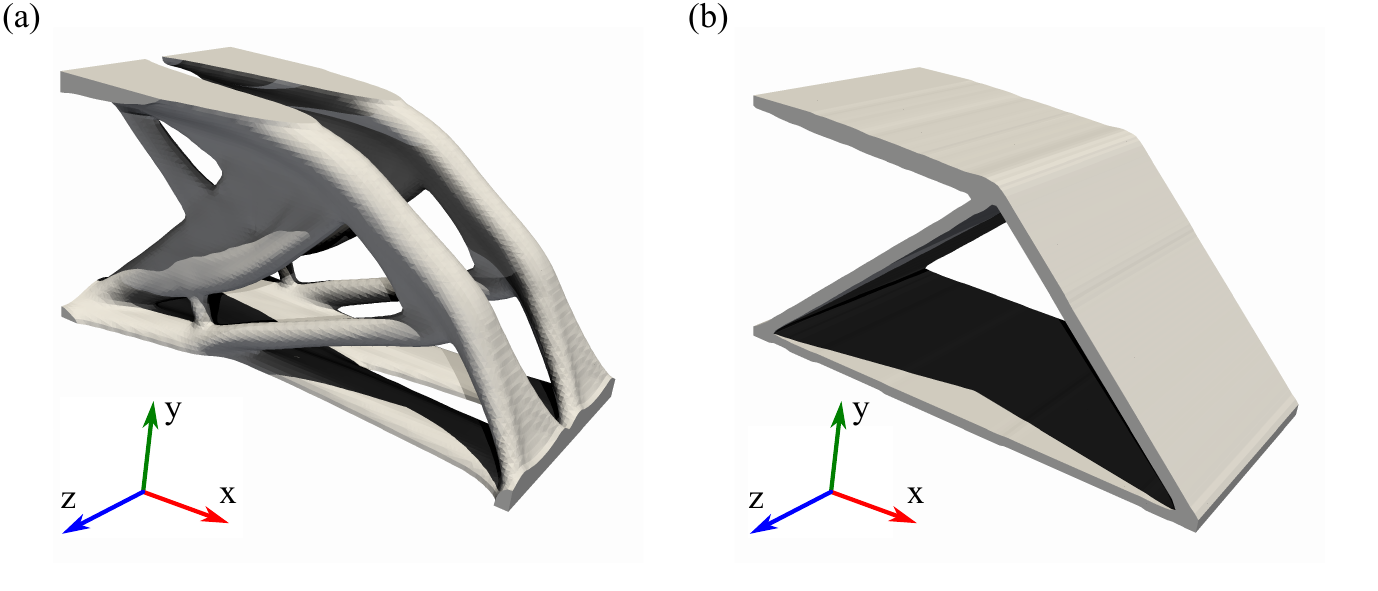}
	\caption{Cantilever beam.  Comparison of full 3D and $2\frac{1}{2}D$ (extruded) optimal topologies for t=0.85 and mesh $M1$.\\ (a) full 3D optimal topology ($\tau =1;\epsilon=1.67\cdot10^{-2}$) ; (b) Extruded optimal topology ($\tau=1;\epsilon_x=\epsilon_y=1.67\cdot10^{-2} ; \epsilon_z= 1.67\cdot10^{3}$).}
	\label{fig_extruded_cantilever_design}		
\end{figure}

\subsection{Mean compliance optimization. 3D bridge design}
The problem of minimum compliance design of a bridge is tackled here. The design domain, $\Omega$, is depicted in figure \ref{fig_Bridge_domain}. For symmetry reasons, only one fourth of the domain is discretized in a structured mesh of 240x204x40, which leads to 1.084.800 hexahedra (see figure \ref{fig_Bridge_domain}-(a)). A uniform load on the bridge deck, $\partial_{\sigma}\Omega$, is prescribed, as well as, the position for the eventual  support at boundary,  $\partial_{u}\Omega$.
\begin{figure}[htb]
	\centering
	\includegraphics[width=12cm, height=7cm]{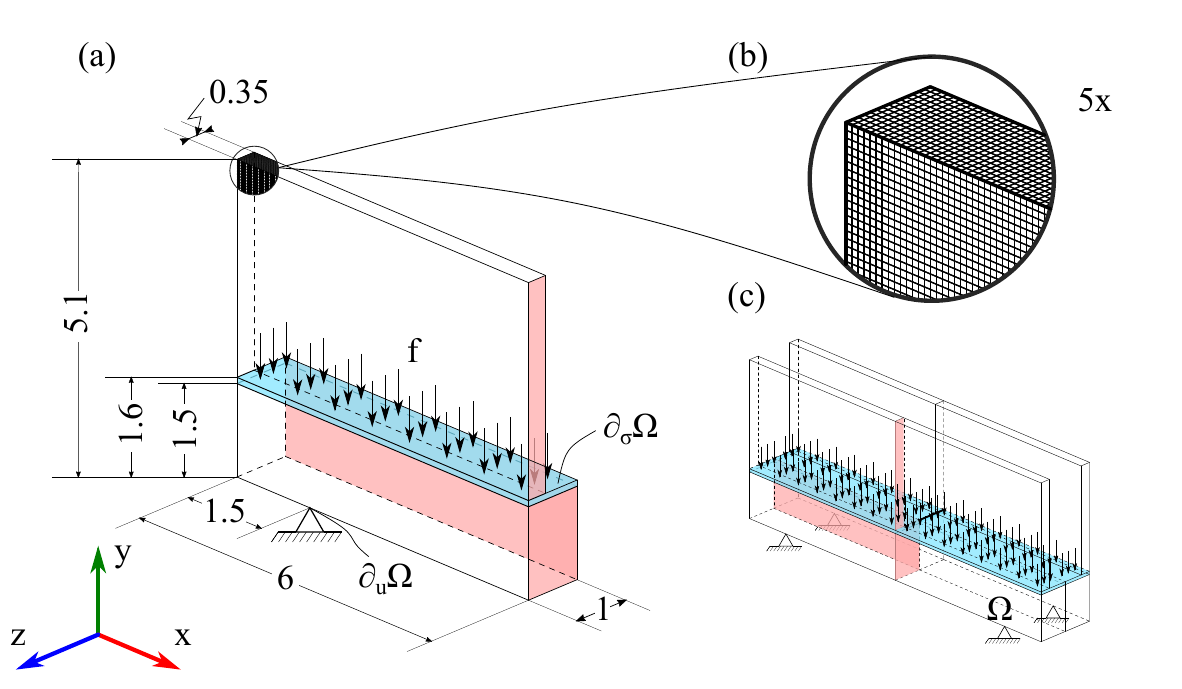}
	\caption{Minimum compliance bridge design: (a) Meshed domain, (b) Detailed mesh, (c) Symmetrized design domain  $\Omega$.}
	\label{fig_Bridge_domain}		
\end{figure}
\begin{figure}[H]
	\centering
	\includegraphics[width=14cm, height=15cm]{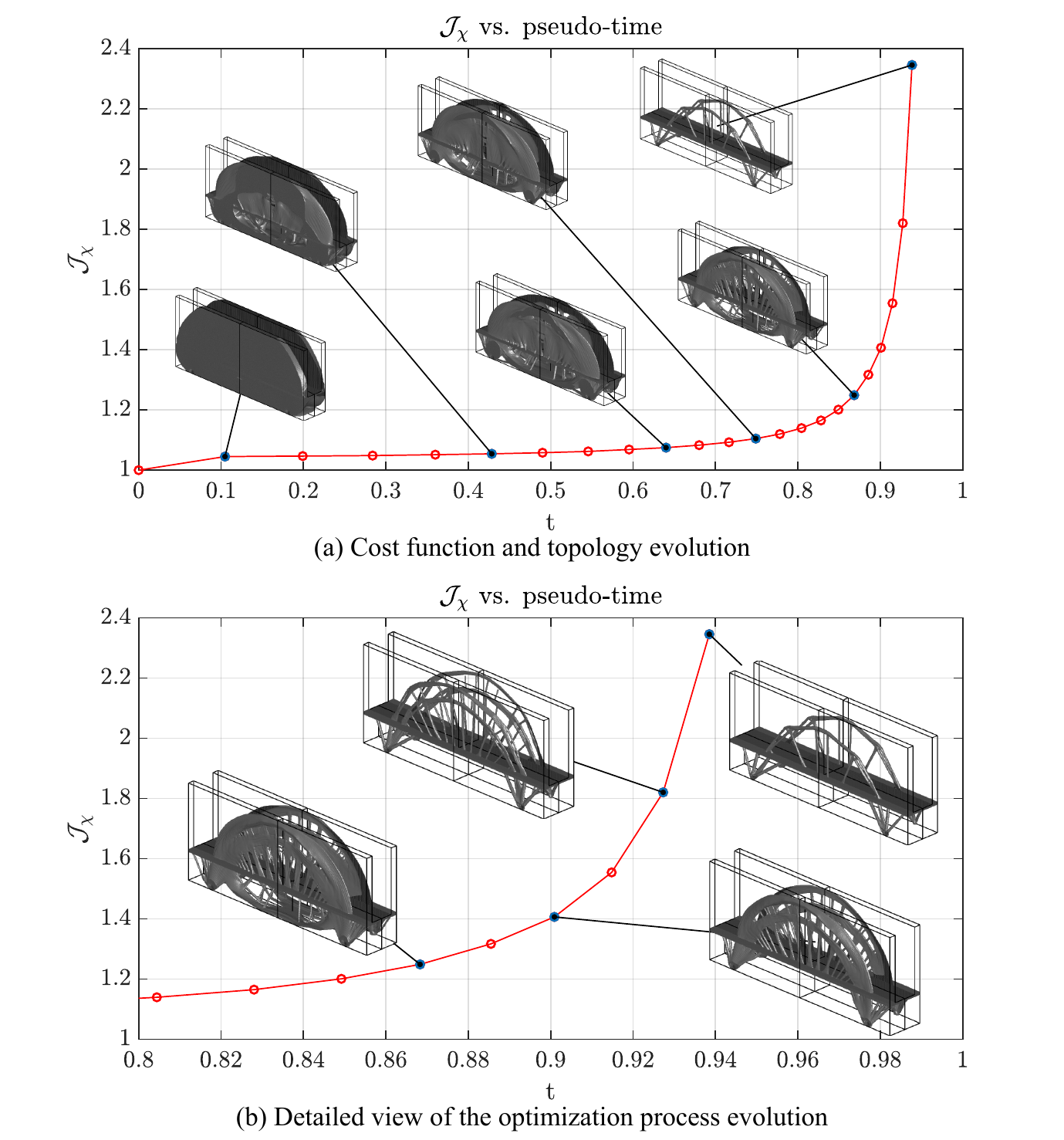}
	\caption{Bridge. Mean structural compliance optimization. $\tau=0.316 \quad(\epsilon=\tau\cdot h=7.9\cdot10^{-3})$}
	\label{fig_bridge_optimization_evolution}		
\end{figure}

Then, the pseudo-time evolving optimization process, starting from a design at $t=0$ made of  \textit{hard-phase} material over all $\Omega$, evolves along the increasing pseudo time $t=\frac{\vert\Omega^-\vert}{\vert\Omega\vert}$. In figure \ref{fig_bridge_optimization_evolution}, the cost function and the topology evolution are shown. Finally, in figure \ref{fig_Bridge_final_topology_M1} the final topology, for 
$\vert\Omega^+\vert=0.06\vert\Omega\vert$, is displayed in detail. Amazingly, subtle  structural design details, for the  structural family of \textit{arch bridges}, well known by structural designers, are captured by the numerically obtained optimized solution.

\begin{figure}[H]
	\centering
	\includegraphics[width=14cm, height=6.5cm]{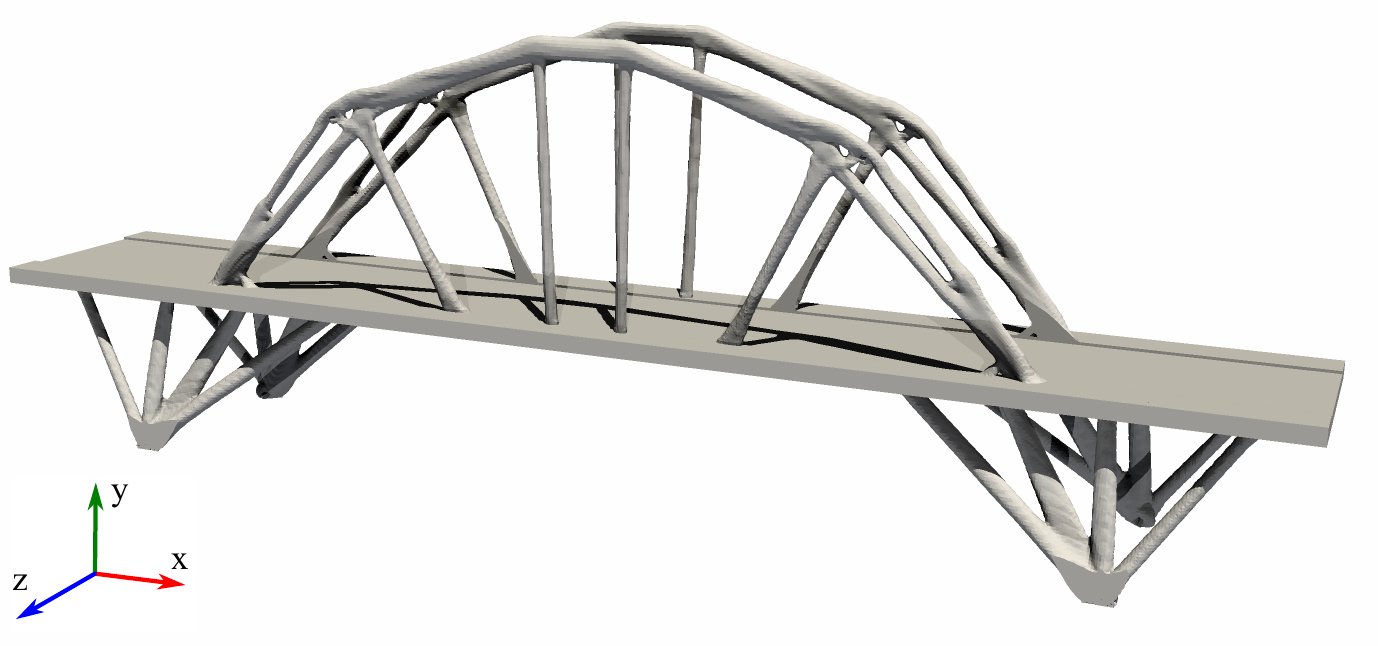}
	\caption{Bridge. Mean structural compliance optimization. Topology for t=0.94 ($\vert\Omega^+\vert=6\%\vert\Omega\vert$).}
	\label{fig_Bridge_final_topology_M1}
\end{figure}

\subsection{Compliant mechanisms optimization.}
The computational design of compliant mechanisms is another example of structural optimization problem for which the approach considered in this work is evaluated.
The RTD and closed form solution derived in section \ref{sec_compliant_mechanism} (see equation (\ref{eq_compliance_mechanism}) and figure \ref{fig_compliant_mechanism}) is applied here to the design of a compliant 3D gripper. 

\begin{figure}[H]
	\centering
	\includegraphics[width=9cm, height=6cm]{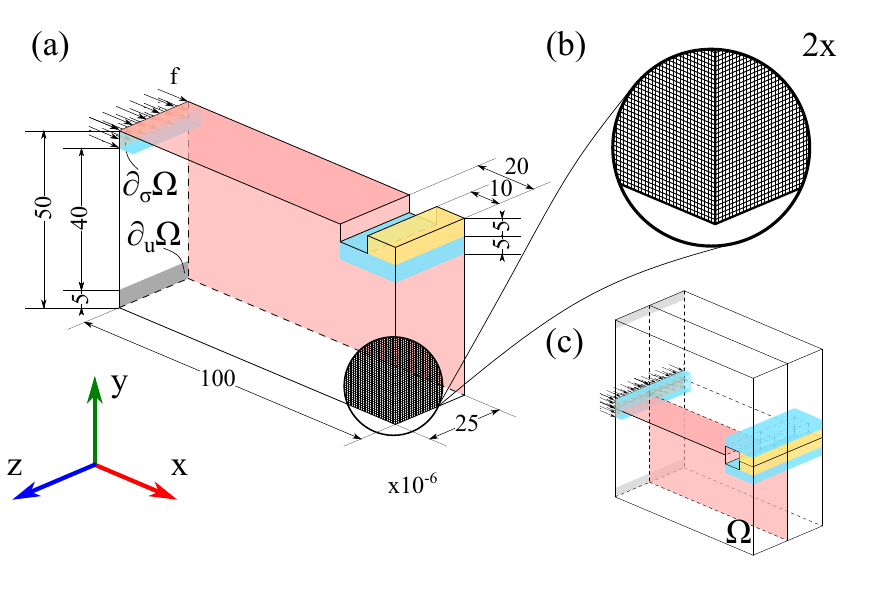}
	\caption{Compliant mechanism (gripper). (a) Design domain $\Omega$, (b) detailed mesh, (c) symmetrized design domain.}	
	\label{fig_gripper_domain}	
\end{figure}

The problem is inspired in that in \cite{Yamada2010} using a {Tikhonov-regularized level set method}, and the mechanical properties considered here are\footnote{At the output port, the existence of a \textit{gripped material mass} has been considered. Therefore, it has been equipped with elastic material properties, softer than the design solid material.},  
\begin{equation}
\label{eq_elastic_material_properties_gripper}
\begin{split}
&\text{Young modulus}\  E=210 GPa \; ;\quad \text{Poisson ratio}\  \nu=0.31 \; ;\quad \text{Stiffness contrast factor} \ \alpha=1.0\cdot10^{-2}\\ 
&\text{Stiffness contrast exponent}\  \ m=3 \quad  \text{(see equation (\ref{eq_contrasted_constitutive_tensor}))}\\
&\text{${\bf K}_{in}=3.19\cdot10^{14} \dfrac{N}{m^3} $}  \ ;  \quad \text{${ E}_{out}=10 GPa$}  \ ;  \quad \nu_{out}=0  \ ;  \quad \text{${\bf f}_{in}=3.2\cdot10^{13} \dfrac{N}{m^2}$}  \ ;  \quad \text{${\bf f}_{out}=3.2\cdot10^{10} \dfrac{N}{m^2}$}
\end{split}
\end{equation}
Due to symmetry considerations only one fourth of the design domain, $\Omega$, is meshed with a structured hexahedra with 160x80x40, which leads to 506.880 elements (see figure \ref{fig_gripper_domain}).

In figure \ref{fig_gripper_domain_sys}, the state problem (system (1)) and the adjoint state problem (system (2)) are also presented. 
\begin{figure}[H]
	\centering
	\includegraphics[width=12cm, height=5cm]{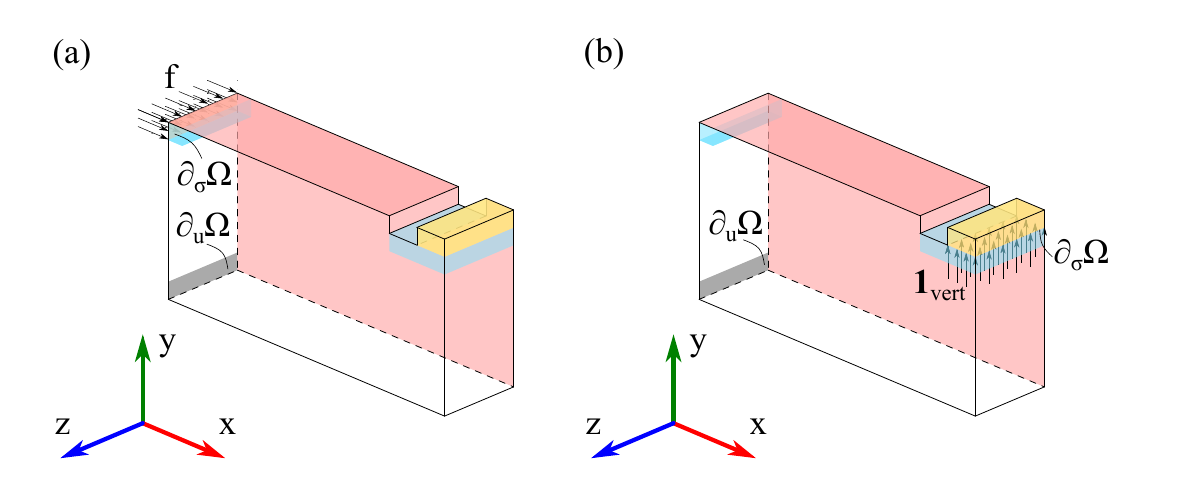}
	\caption{Compliant mechanism (gripper): (a) system (1) (1/4th-domain), (b) system (2) (1/4th-domain)).}
	\label{fig_gripper_domain_sys}
\end{figure}
 In figure \ref{fig_griper_optimization}-(a), results concerning the cost-function (opening displacement of the gripper jaws is minimized, therefore the closing motion is maximized) are presented, as well as the topology evolution. It is worth noting here that, unlike in the mean structural compliance case, the cost function is not monotonously increasing, and a \textit{minimum of the minima} compliant design can be identified for $t=\frac{\vert\Omega^-\vert}{\vert\Omega\vert}\approx0.48$. In addition, one can identify a sudden change of the topology, around $t=0.3$, translating into a change of trend of the cost function to, subsequently, fall into a new local minimum.
 
  Also notice that the chosen value for $\tau$ ($\tau=0.5$) corresponds to $\epsilon=\frac{1}{2}h$, i.e. one half of the element size, in order to allow the formation of hinges (thin solid-phase filaments) typical of this problem in some stages of the optimization process (see figure \ref{fig_griper_optimization}-(b)-(c)-(d)).

\begin{figure}[H]
	\centering
	\includegraphics[width=17cm, height=12cm]{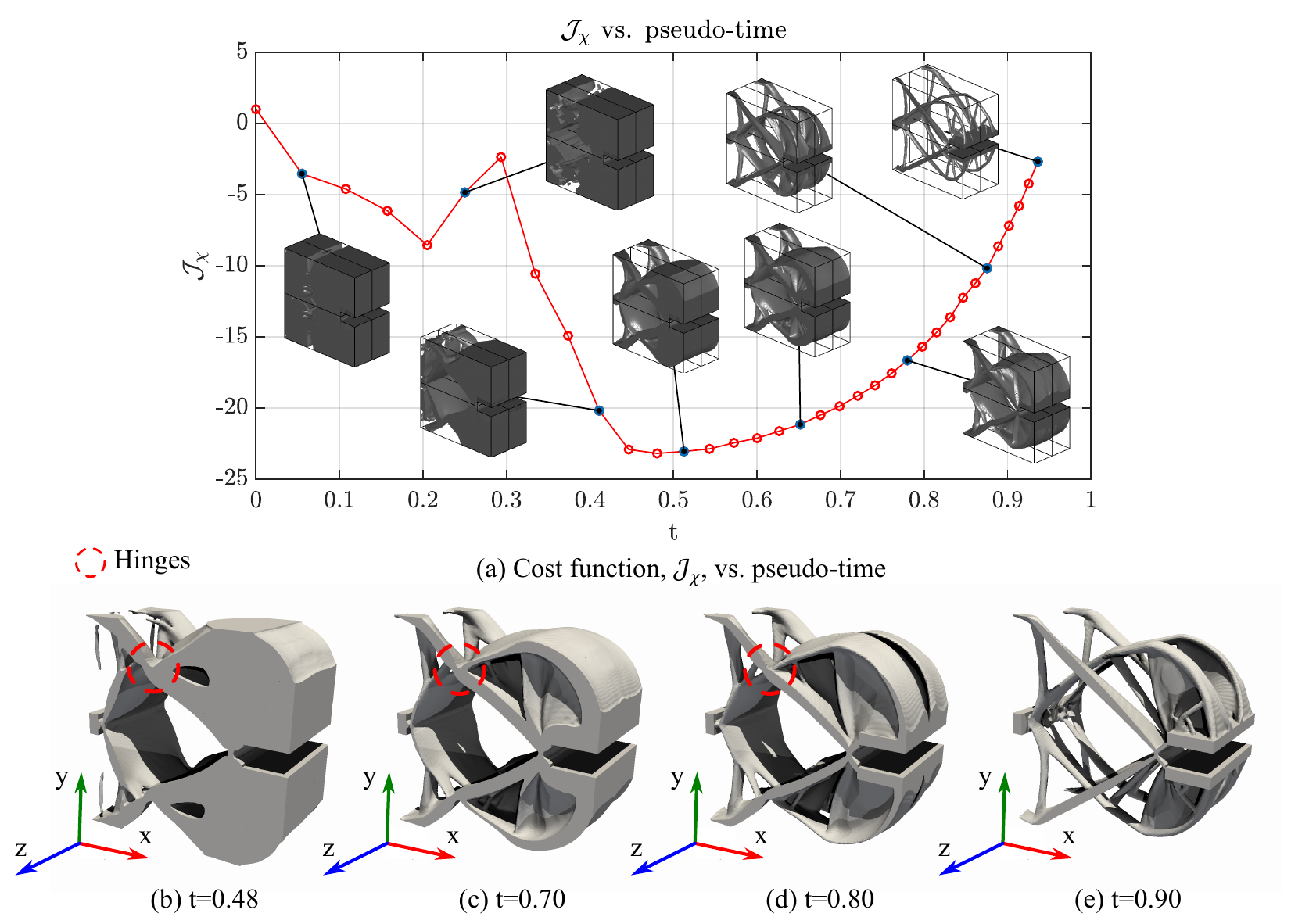}
	\caption{3D-Gripper optimization: (a) cost function evolution, (b) to (e) optimal topologies at different pseudo-times. $\quad(\tau=0.5;\;\epsilon=\tau\cdot h=3.125\cdot10^{-7})$. Notice, in figures (b) to (d) the appearance of hinges, in terms of thin short-bars, to provide additional compliance to the design.} 
	\label{fig_griper_optimization}
\end{figure}
 
\subsection{Computational assessment. Variational closed-form solution vs. level-set method}
To assess the performance of the variational closed-form approach used in this work, in this section we compare the results for the cantilever beam, obtained in section \ref{sec_compliance_cantilever_beam} with the so far described setting, with those using a level-set method driven by the relaxed topological derivative (RTD) in equation (\ref{eq_compliance})-(c). Details on the level-set numerical algorithm and its properties are given in Appendix \ref{eq_levelset_for_bi_material}.

In order to have a fair comparison, in both methods (\textit{variational closed-form solution} and \textit{level set method}) \textit{the same sequence of time steps, $ t_{i}$, has been imposed} along the pseudo-time interval $[0,T]\equiv[0,1]$, i.e. 22 exponentially spaced time steps from $t=0$ to $t=0.92$  as indicated in equation (\ref{eq_minimization_restricted_time}). The pseudo-time, $t$, evolution is, then, defined according to the following exponential law 
\begin{equation}
\label{eq_time_evolution_cantilever}
\begin{split}
	 t_i = t_0+\frac{T-t_0}{1-e^{K}}\left(1-e^{K(i/n)}\right) \quad ; \quad i\in\{0,1,\dotsc,n\} \\
\end{split}
\end{equation}
where $n=40$ denotes the number of steps required to reach the time $t_n=T=1$, and the  constant $K$ is set to obtain the desired exponential function ($K=-4.5$).

Then, both algorithms are run until convergence in terms of the topology (with the same tolerance $Tol_{\chi}=10^{-1}$). Their relative computational cost, is evaluated  in terms of the number of iterations required to converge for each of the methods\footnote{In fact, the solution of the state (equilibrium) problem dominates the computational cost in both methods. Therefore, the computational cost is closely proportional to the number of required iterations to converge.}.

\begin{figure}[H]
	\centering
	\includegraphics[width=14cm, height=16cm]{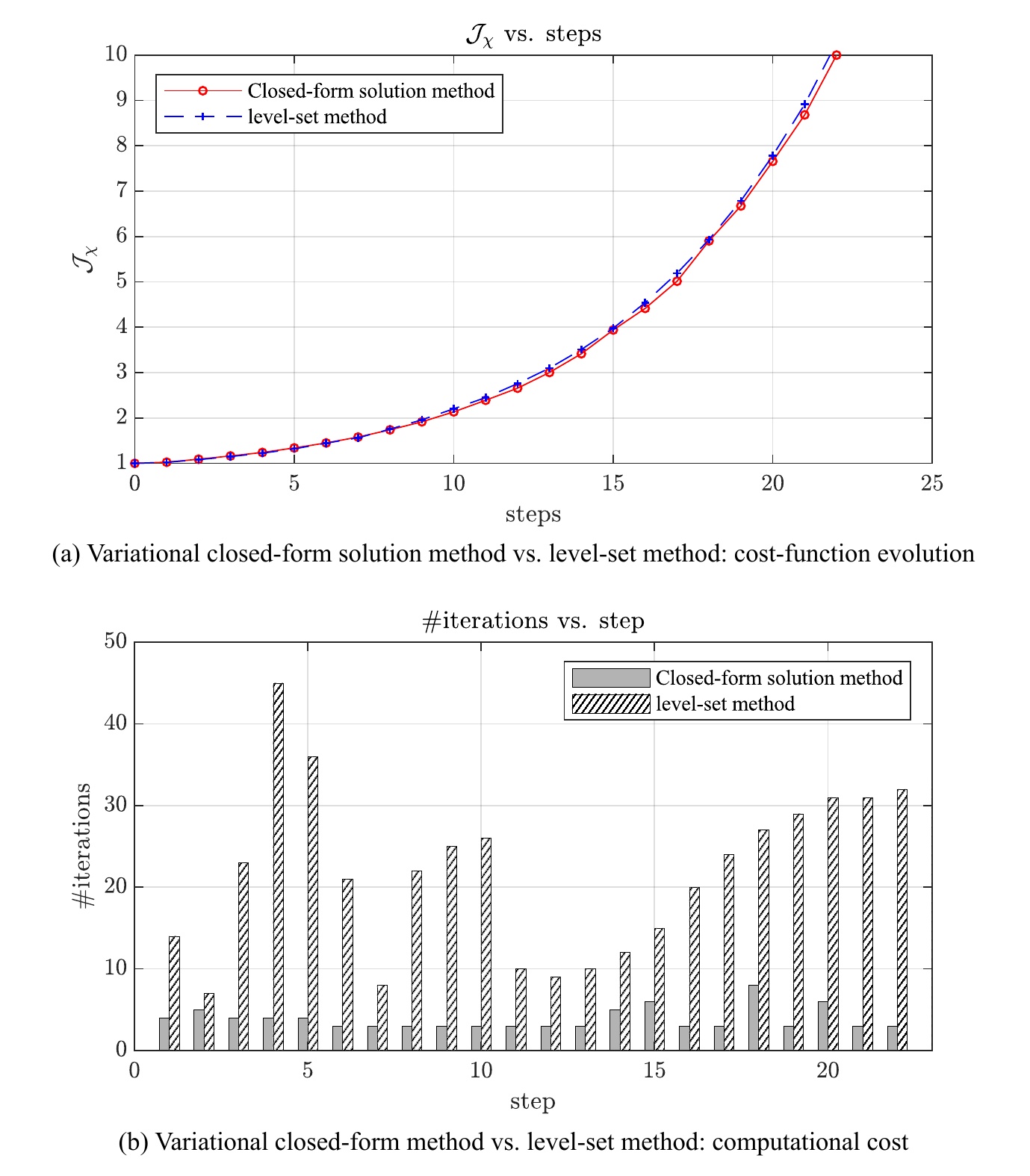}
	\caption{Cantilever beam. Variational closed-form solution vs. level set method.\\ $\quad(\varepsilon=\tau\cdot h=1.67\cdot10^{-2})$}
	\label{fig_comparison_ST_HJ}		
\end{figure}
In figure \ref{fig_comparison_ST_HJ}-(a), the cost function evolution is displayed, showing a very similar quantitative performance of both methods. As for the computational cost (number of required iterations) the variational closed-form solution is about 5 times (in average) cheaper to obtain than in the level-set method (see figure \ref{fig_comparison_ST_HJ}-(b)). The fact that the volume restriction is enforced at every iteration, in the variational closed from method, plays, in opinion of the authors, a relevant role in this cost reduction. On the other hand, the number of requested iterations tends to be almost constant, along time intervals, in the variational closed-form solution, whereas, in the level-set method it exhibits a large increase in the required number of iterations for high values of the pseudo-time (i.e. small volumes of the solid-phase $\vert\Omega^+\vert$). 

 Similar trends, to those specifically emphasized here, has been observed in other optimization problems solved using both methods. In summary, lower computational cost, in front of level-set methods, while providing similar quantitative results, seems to be an argument in favor of the approach in the present work.

\section{Concluding remarks}
Along this work a variational approach to relaxed topological optimization has been explored, and assessed through its application to a number of structural problems. The conclusions to be highlighted by the authors, about this work, are the following:
\begin{itemize}
	\item[$\bullet$] Though \textit{the relaxed character of the optimization setting} is not specific of the present approach, the formulation of a \textit{relaxed topological derivative} (RTD), clarifies the issue of using a topological sensitivity \textit{consistent with the relaxed setting in which the optimization is performed}. This is in contrast with the frequent use, in relaxed optimization settings, of the \textit{exact topological derivative} (TD),  derived in an \textit{immersed setting} at the cost of heavy, and problem dependent, mathematical derivations. Although, from the mathematical point of view, it could be argued that the TD represents the exact sensitivity with respect to singular domain perturbations whereas the RTD is just an approximation of it, the results obtained using the later, for practical applications as the ones in this work, are manifestly similar  to those obtained with the TD. In consequence,  the RTD can be considered a powerful and, at the same time very simple tool, for topological optimization in engineering applications.   
	\item[$\bullet$] The difficulties inherent to the discrete (discontinuous) character of the design variable (the characteristic function $\chi({\bf x})$), and the resulting non-smoothness of the cost function, have traditionally suggested detouring  the formulation from the original variational scenario; either by resorting to regularization of the characteristic function (like in SIMP) or by totally abandoning  the variational formulation when facing the impossibility to obtain the stationarity conditions based on differentiation of that cost function (like in level-set methods).
	Surprisingly enough, the difficulties arising from that non-smooth character of the problem can be easily overcome. Indeed, the analysis of the sensitivity of the cost function, in front of topology perturbations, via the aforementioned expressions of the RTD, yields an \textit{inequality-type optimality criterion: the overall-increasing cost function sensitivity}. Amazingly, the emerging inequality, in terms of the, here termed, \textit{discrimination function} $\psi({\bf x})$, leads straightforwardly to \textit{closed-form solutions of the topological optimization problem in the form of simple algebraic non-linear equations}. These equations, involving the characteristic function, $\chi({\bf x})$, can be solved, in turn, via simple non-linear algorithms. This is a very specific result of the proposed approach, which suggests possible extensions of the technique to further methods and applications. In the cases studied here, typically the mean structural compliance and compliant mechanisms problems, the resulting optimality solutions are interpreted as fixed-point equations, to be solved by combination of a cutting algorithm, of energy-like surfaces, and the exact imposition of the volume restriction (via bisection methods).
	\item[$\bullet$]Since the presented approach is not  regularized in terms of the characteristic function, the obtained topological designs are completely \textit{black and white designs}, and there is no need of subsequent filtering techniques to eliminate gray-scales. 
	\item[$\bullet$]Besides, the characteristic function, $\chi({\bf x})$, is never regularized, even for the treatment of bi-material elements. Instead, three-field \textit{mixed formulations} (${\bm \sigma}-{\bm \varepsilon}-{\bf u})$ \cite{Zienkiewicz2005} in hexahedral finite elements, detailed in appendix \ref{app_mixed_finite_element_formulation}, are used for accurately capturing the discontinuous stress field in those bi-material elements. Albeit, the LBB stability condition  is not fulfilled in those elements, numerical instabilities do not appear, thus not leading to the classical \textit{checkerboards}, due to their isolate position, in a mesh of \textit{regular elements}, which avoids propagation of the possible instability modes.
	 Also the marching cubes strategy, with the specific variant of \textit{marching tetrahedra method} (see appendix \ref{app_marching_cubes}) has provided the expected robustness and accuracy in the determination of volumes and surfaces on the finite element mesh.  
	\item[$\bullet$] An \textit{evolving pseudo-time framework} can be  naturally inserted in the considered setting. Albeit this is not an exclusive feature of the proposed approach, and it can be used, for instance, in combination  with the level set methods, it is worth mentioning some of the benefits provided by this framework, i.e. the fact that a number of \textit{intermediate optimal designs for a set of values of the volume restriction} are obtained "for free" and, in some problems, like compliant mechanisms design, information of the \textit{minimum of the minima} designs in the explored range can be obtained.
	\item[$\bullet$] The approach considered here has proven to provide results, in the considered structural examples, very similar to the ones obtained with other methods, the main difference being that the computational cost is smaller (about five times in a benchmark test checked in the work) when compared with the level-set method, using the same code, the same RTD and the same computational setting. 
	\item[$\bullet$] The explored approach allows envisaging new computational strategies for topological optimization, which could overcome some of the classical drawbacks of relaxed optimization settings, e.g. the \textit{necessity of using high dense meshes in the design domain involving millions of elements} and the application to new topological design problems (e.g. multiscale topological design \cite{Kato2013,Ferrer2018,Podesta2018,Coelho2018}, immersed boundary methods combined with XFEM techniques \cite{Sharma2017}, convected methods  \cite{Yaji2016}, etc.). This is an issue to be explored in a subsequent research.
\end{itemize}

\section*{Acknowledgements}
This research has received funding from the European Research Council (ERC) under the European Union’s Horizon 2020 research and innovation programme (Proof of Concept Grant agreement n 779611) through the project
“Computational catalog of multiscale materials: a plugin library for industrial finite element codes” (CATALOG). The authors also acknowledge the funding received from the Spanish Ministry of Science, Innovation and Universities through the research grant DPI2017-85521-P for the project “Computational design of Acoustic and Mechanical Metamaterials” (METAMAT).

\begin{appendices}
	\section{Modified marching cubes strategy}
	\label{app_marching_cubes}
	Different strategies can be found in the literature to extract the zero-level surface, $\Gamma\coloneqq\{{\bf x}\in \Omega^{(e)}\ ;\ \psi({\bf  x})=0\}$, from a scalar field, $\psi({\bf  x})$. Among them, the marching cubes (MC) approach, see \cite{Lorensen1987}, has been the most popular technique used for volume visualization. However, the classical MC-method implementation requires the analysis of 15 different cases in every cubic cell, and the corresponding rotations and symmetries. In spite of this, it does not guarantee the topology correctness of the isosurface of a trilinear function.
	 
	For this reason, a modified algorithm, which takes into account 33 topologically different configurations to correctly treat  topological ambiguities, both on the faces and inside the cell, was subsequently developed in \cite{chernyaev1995marching}. Nevertheless, this algorithm is heavy to code and difficult to verify its faultlessness. 
	
	In this work, an alternative \textit{marching tetrahedra method} (MT) has been implemented, which overcomes the aforementioned obstacles while avoiding ambiguities.
	The proposed strategy is based on an adaptive cube tessellation, see \cite{Velasco2008}, which ensures topologically correct isosurfaces in case of ambiguities. Furthermore, this tessellation preserves the symmetries in the domain. The method is briefly sketched next. 
	\begin{figure}[H]
		\centering
		\includegraphics[width=16cm, height=9cm]{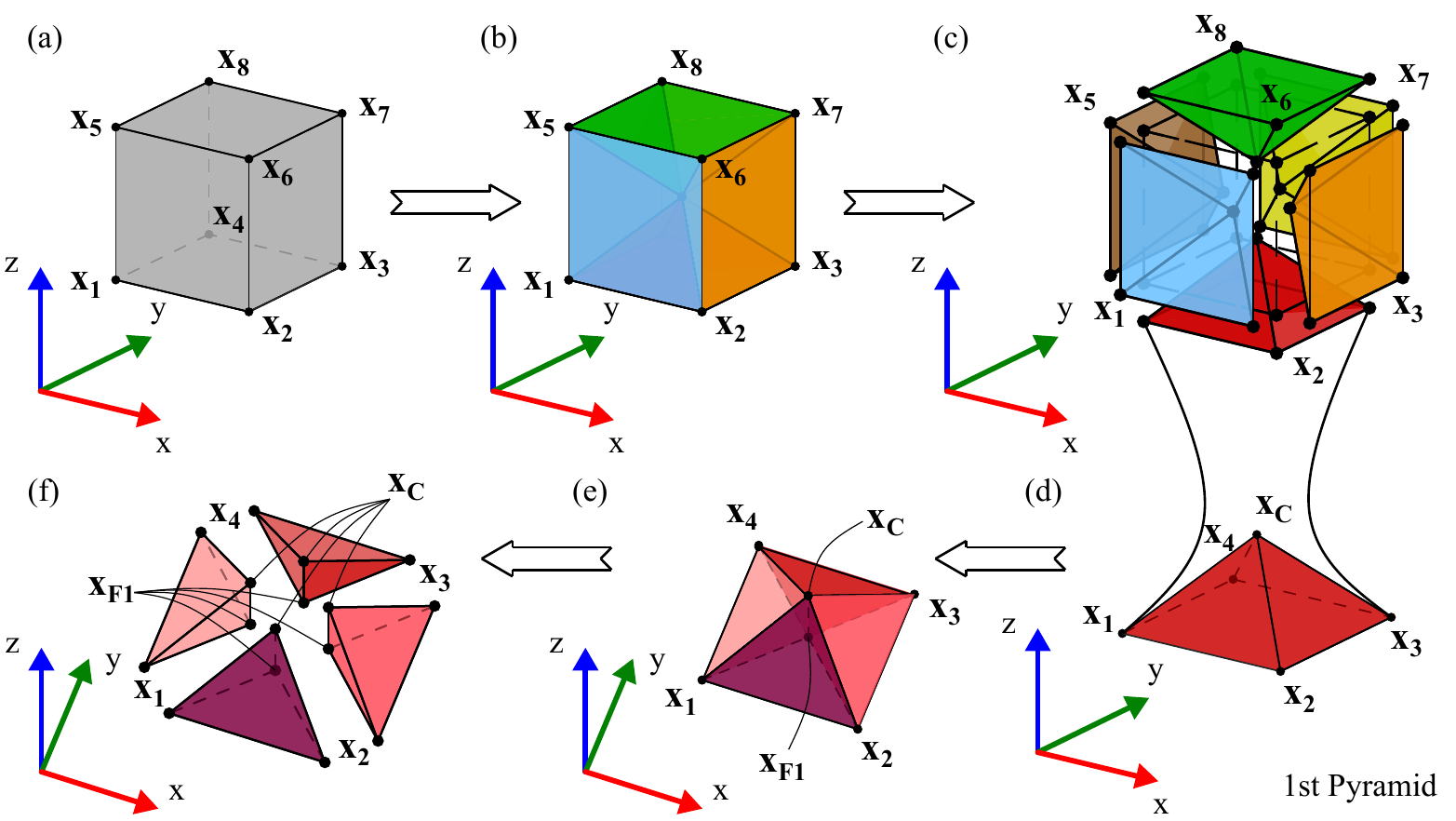}
		\caption{Marching cubes method: tessellation process.}
		\label{fig_annex3}
	\end{figure}
	
	The cubic cell is divided into six pyramids, a central extra-node, ${\bf x}_{C}$, is added to the cell and the value $\psi_C\equiv\psi({\bf x}_{C})$ is computed by interpolation from the regular nodes. In turn, each pyramid is split into four tetrahedra, including an extra-node, ${\bf x}_{Fi}$,  on each face, and the value $\psi_{Fi}\equiv\psi({\bf x}_{Fi})$ is computed, by interpolation, from its value at the vertices, see figure \ref{fig_annex3}. Therefore, seven extra values must be computed based on the trilinear approximation (green and red dots in figure \ref{fig_annex4_v4}-(b)).
	\begin{figure}[H]
		\centering
		\includegraphics[width=16cm, height=5cm]{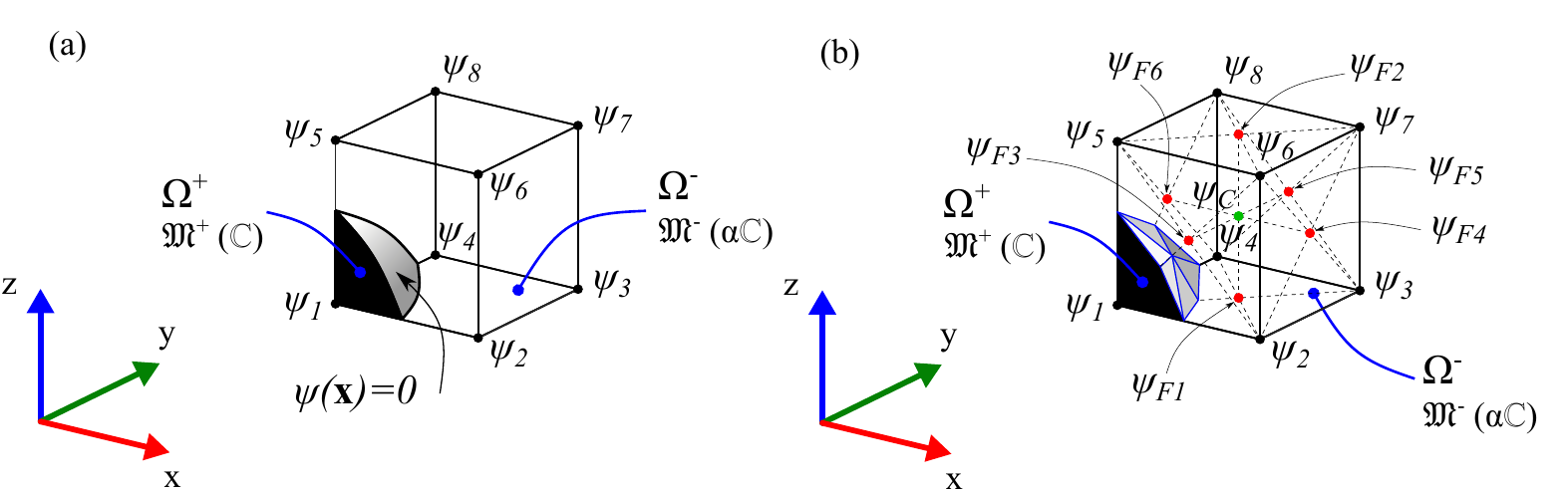}
		\caption{Marching cubes method: nodal values interpolation.}
		\label{fig_annex4_v4}
	\end{figure}
	
	Then, each tetrahedron is treated as usually in marching tetrahedra, where triangular surfaces and tetrahedral volumes are computed. This  procedure readily allows the geometrical computation of the perimeter and volume of the cell.
	
	\section{Mixed finite element formulation for bi-material finite elements}
	\label{app_mixed_finite_element_formulation} 
	Let us consider the subset of bi-material elements $\Omega^{(+,-)}\subset\Omega$  as well as the portions $\Omega^{(e,+)}\subset\Omega^{(e)}$ and $\Omega^{(e,-)}\subset\Omega^{(e)}$ ($\Omega^{(e,+)}\cup\Omega^{(e,-)}=\Omega^{(e)}$) containing hard and soft material phases, in every element $e$. This subset of elements is treated considering a specific \textit{discontinuous elemental material approach} based on a \textit{three field mixed} stress-strain-displacement (${\bm \sigma_{\chi}}-{\bm \varepsilon_{\chi}}-{\bf d}_{\chi}$) finite element formulation \cite{Zienkiewicz2005}. The elemental stresses, ${\bm \sigma}_{\chi}^{(e)}{(\bf x)}$, strains, ${\bm \varepsilon}_{\chi}^{(e)}{(\bf x)}$, and displacements, ${\bf u}_{\chi}^{(e)}{(\bf x)}$, are independently interpolated inside the element $e$. For the case of hexahedral elements, we will consider element-wise constant stresses and strains. Therefore, following the notation in equation (\ref{eq_equilibrium}), they read
	\begin{equation}
	\left\{
	\begin{split}
	&{\bf u}_{\chi}^{(e)}({\bf x})=[{\bf N}_{u}^{(e)}({\bf x})]{\bf d}^{(e)}_\chi \; ;\quad[{\bf N}_{u}^{(e)}({\bf{x}})]=[{\bf N}^{(e)}_{1}({\bf{x}}),...,{\bf N}^{(e)}_{8}({\bf{x}})]\; ;\quad{\bf d}_\chi^{(e)}=[{\bf d}_{1}^{(e)}, ...,{\bf d}_{8}^{(e)}]^T
	\quad&(a)\\
	&{\bm \varepsilon}_{\chi}^{(e)}({\bf x})={\bm \varepsilon}^{(e)}=constant \; \quad \forall {\bf x}\in\Omega^{(e)}\quad&(b)\\
	&{\bm \sigma}_{\chi}^{(e)}({\bf x})={\bm \sigma}^{(e)}=constant \quad \forall {\bf x}\in\Omega^{(e)}\quad&(c)
	\end{split}
	\right.
	\end{equation}
	where ${\bf N}_{u}^{(e)}({\bf x})$ is the elemental displacement bilinear
	interpolation matrix, and ${\bf d}_\chi^{(e)}$  the corresponding nodal vectors. The nodal displacement interpolation matrices ${\bf N}_{i}^{(e)}({\bf x})\ i\in\{1,...,8\}$ have a standard nodal support, and the nodal displacements ${\bf d}_{i}^{(e)}\  i\in\{1,...,8\}$ correspond to the displacements at the element nodes. As for the strain and stress interpolation, due to their element-wise discontinuous character, they can be associated to one \textit{fictitious} internal element node in the center of every element  $\Omega^{(e)}$, with coordinates ${\bf x}_{C}$, (see figure \ref{fig_Mixed_element}), where the (constant) elemental strain vector, ${\bm \varepsilon}^{(e)}$, is defined.
	The elemental strain, ${\bm \varepsilon}^{(e)}$, is, then, determined via the following additional variational equation (uncoupled for every element in $\Omega^{(+,-)}$)
	\begin{equation}
	\label{eq_variational_equation}
	\left\{
	\begin{split}
	&GIVEN\\
	&{\cal V}_{\varepsilon}^{(e)}\coloneqq\left\{{\bm \eta}_{\varepsilon}^{(e)}\in{\mathbb R}^6 \right\}\; ;\quad {\cal V}_{\sigma}^{(e)}\coloneqq\left\{{\bm \eta}_{\sigma}^{(e)}\in{\mathbb R}^6 \right\} \\
	&FIND: {{\bm \varepsilon}}^{(e)}\in{\cal V}_{\varepsilon}^{(e)}\  \textit {and} \ {{\bm \sigma}}^{(e)}\in{\cal V}_{\sigma}^{(e)}\\ 
	&FULFILLING:\\
	&\hspace{1.2cm}\int_{\Omega^{(e)}}\left({\bm \eta}_{\varepsilon}^{(e)}\right)^T\left({{\bm \varepsilon}}^{(e)}-{\bm \epsilon}_{\chi}^{(e)}({\bf x}) \right)d\Omega=\left({\bm \eta}_{\varepsilon}^{(e)}\right)^T\int_{\Omega{(e)}}\left({{\bm \varepsilon}}^{(e)}-{\bm \epsilon}_{\chi}^{(e)}({\bf x}) \right)d\Omega=0
	& &\quad\forall{\bm \eta}^{(e)}_{\varepsilon}\in {\cal V}_{\varepsilon}^{(e)}\\
	&\hspace{1.2cm}\int_{\Omega^{(e)}}\left({\bm \eta}_{\sigma}^{(e)}\right)^T\left({{\bm \sigma}}^{(e)}-{\mathbb D}^{(e)}({\bf x}){{\bm \varepsilon}}^{(e)} \right)d\Omega=\left({\bm \eta}_{\sigma}^{(e)}\right)^T\int_{\Omega{(e)}}\left({{\bm \sigma}}^{(e)}-{\mathbb D}^{(e)}({\bf x}){{\bm \varepsilon}}^{(e)}\right)d\Omega=0
	& &\quad\forall{\bm \eta}^{(e)}_{\sigma}\in {\cal V}_{\sigma}^{(e)}\\
	&\hspace{1.2cm}{\bm \epsilon}_{\chi}^{(e)}({\bf x})={\bf B}^{(e)}({\bf x}){\bf d}_\chi^{(e)}
	\end{split}
	\right.
	\end{equation}
	where $\bf B({\bf x})^{(e)}$ and ${\mathbb D}^{(e)}({\bf x})$ stand, respectively, for the elemental deformation matrix and the elastic constitutive matrix in equation (\ref{eq_equilibrium}).
	Equation (\ref{eq_variational_equation}) can be solved for ${{\bm \varepsilon}}^{(e)}$ and ${{\bm \sigma}}^{(e)}$ yielding 
	\begin{equation}
	\label{eq_b_matrix}
	\left\{	
	\begin{split}
	&{\bm \varepsilon}^{(e)}=\dfrac{1}{\vert{\Omega^{(e)}}\vert}\int_{\Omega^{(e)}}{\bm \epsilon}_{\chi}^{(e)}({\bf x})d\Omega=
	\int_{\Omega^{(e)}}{\bf B}^{(e)}({\bf x}){\bf d}_\chi^{(e)}d\Omega=\underbrace{\int_{\Omega^{(e)}}{\bf B}^{(e)}({\bf x})d\Omega}_
	{\displaystyle{\overline{{\bf B}^{(e)}}({\bf x})}}\ {\bf d}_\chi^{(e)}=\overline{{\bf B}^{(e)}}({\bf x})\ {\bf d}_\chi^{(e)}\quad&(a)\\
	&\overline{{\bf B}^{(e)}}({\bf x})\simeq{\bf B}^{(e)}({\bf x}_{C})\quad&(b)
	\end{split}
	\right.	
	\end{equation}
	where $\overline{(\newplaceholder)}$ stands for the \textit{mean value} of ${(\newplaceholder)({\bf x})}$ in $\Omega^{(e)}$, and equation (\ref{eq_b_matrix})-(b) stems from the bi-linear character of the element. Now, considering the stress problem in equation (\ref{eq_variational_equation}), and solving for ${\bm \sigma}^{(e)}$	yields
	\begin{equation}
	\left\{
	\begin{split}
	&{\bm \sigma}^{(e)}=\dfrac{1}{\vert{\Omega^{(e)}}\vert}\int_{\Omega^{(e)}}{\mathbb D}({\bf x}){{\bm \varepsilon}}^{(e)}d\Omega=
	\dfrac{1}{\vert{\Omega^{(e)}}\vert}
	\underbrace{\int_{\Omega^{(e)}}{\mathbb D}({\bf x})d\Omega}_{\displaystyle{\overline{{\mathbb D}({\bf x})}}} \ {{\bm \varepsilon}}^{(e)}=
	\overline{{\mathbb D}({\bf x})}\ {{\bm \varepsilon}}^{(e)}\quad&(a)\\
	&\overline{{\mathbb D}({\bf x})}=\dfrac{\vert\Omega^{(e,+)}\vert}{\vert\Omega^{(e)}\vert}{\mathbb D}_{\chi}^{+}+\dfrac{\vert\Omega^{(e,-)}\vert}{\vert\Omega^{(e)}\vert}{\mathbb D}_{\chi}^{-}\quad=\left(\dfrac{\vert\Omega^{(e,+)}\vert}{\vert\Omega^{(e)}\vert}+\alpha\dfrac{\vert\Omega^{(e,-)}\vert}{\vert\Omega^{(e)}\vert}\right){\mathbb D}&(b)
	\end{split}
	\right.
	\end{equation}
	where the values of the elastic constitutive tensor in $\Omega^+$ (${\mathbb D}^+_{\chi}$) and in $\Omega^-$ (${\mathbb D}^-_{\chi}$) in equations (\ref{eq_contrasted_constitutive_tensor}) and (\ref{eq_equilibrium}) have been considered.
	\begin{figure}
		\centering
		\includegraphics[width=15cm, height=4cm]{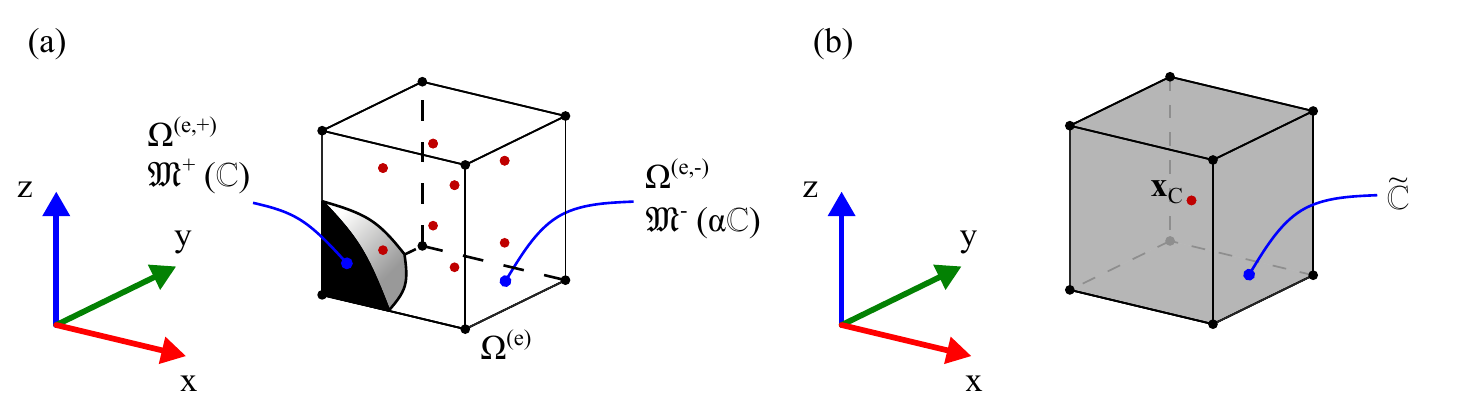}
		\caption{Mixed finite element representation: (a) Hexahedra element with 8 integration points  (b) Mixed element with 1 pseudo-integration point.}
		\label{fig_Mixed_element}
	\end{figure}
	
	\section{A level set algorithm for relaxed bi-material topological optimization}
    \label{eq_levelset_for_bi_material}		
	\subsection{Level set algorithm}
	\label{ap_level_set_algorithm}
	Let us consider the topological optimization problem in equation (\ref{eq_rephrased_equations})-(a)
	for the design domain $\Omega$, 
	and the \textit{hard-material phase}, $\Omega^+_t$ and \textit{soft-material phase}, $\Omega^-_t$, corresponding to a pseudo-time, $t$, in turn associated to an iteration procedure. The interface between the two phases is denoted by $\Gamma_t$, and ${\bf n}$ denotes the unit normal to $\Gamma_t$ pointing to $\Omega^+$ (see figure \ref{fig_Domain_HJ}).
	\begin{figure}[H]
	\centering
	\includegraphics[width=9cm, height=6cm]{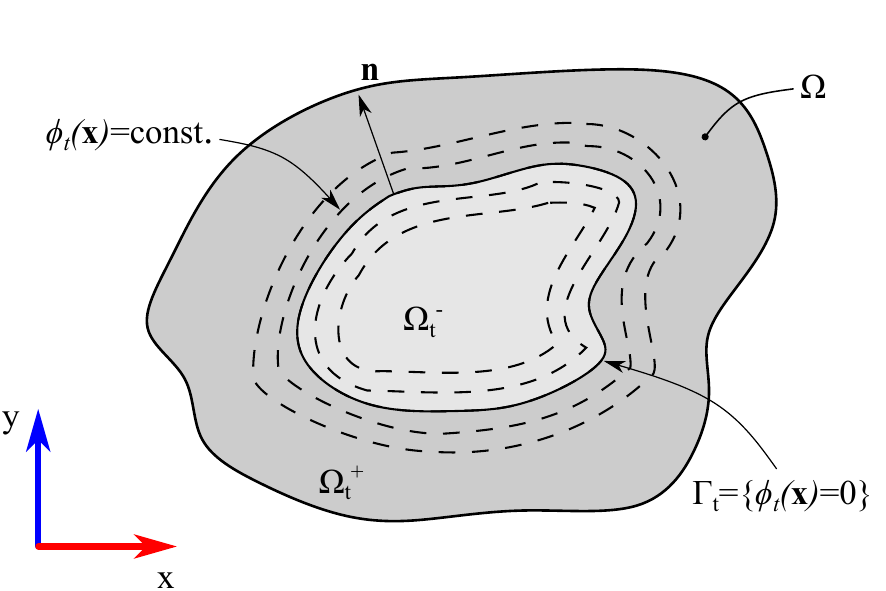}
	\caption{Level-set problem definition. The zero level-set ($\phi_t({\bf x})=0$) defines, at time $t$, the position of the material phase interface, $\Gamma_t$, in the design domain, $\Omega$, and, therefore, the phase domains $\Omega_t^+$ and $\Omega_t^-$.}
	\label{fig_Domain_HJ}
\end{figure}	
	Let $\phi_t\in{\cal V}_{\phi}$ be the level set function, considered as the actual unknown of the problem, through which the characteristic function, $\chi_t({\bf x})$, is defined (see equations (\ref{eq_chi_definition}) to (\ref{eq_heaviside})). Let us also consider $\phi_t({\bf x})$ defined through a time-evolving  (Hamilton-Jacobi) equation \cite{Allaire2005a} i.e.
	\begin{equation}
	\label{eq_hamilton_jacobi_1}
	\begin{split}
	&{\cal V}_{\phi}\in H^1(\Omega)\coloneqq\left\{\phi:\Omega\times[0,T]\rightarrow{\mathbb R} \right\}\\
	&\left\{
	\begin{split}
	&\dot{\phi_t}({\bf x},\lambda)\equiv\dot{\phi}({\bf x},\lambda_t)=-k\dfrac{1}{\Delta \chi_t({\bf x})}\dfrac{\delta {\cal L}({\chi}_t,\lambda_t)}{\delta {\chi}_t}({\bf x}) \; ;\quad k>0\; ;\quad-1\le\phi_t\le 1\quad\; ;\quad \forall {\bf x}\in\Omega&(a)\\
	&{\cal L}_t(\phi,\lambda)\equiv{\cal L}({\chi}(\phi_t),\lambda_t)={\cal J}({\chi}(\phi_t))+\lambda_t{\cal C}(\chi(\phi_t))\quad&(b)\\
	&\chi_t({\bf x})={\cal H}_{\beta}(\phi_t({\bf x}))\quad&(c)\\
	\end{split}
	\right.
	\end{split}
	\end{equation}
	where the relaxed topological derivative of the Lagrange functional ${\cal L}_t(\phi,\lambda)$ in equation (\ref{eq_hamilton_jacobi_1}) is given by (see equations (\ref{eq_compliance_mechanism})-(a)-(b)-(c))
	\begin{equation}
	\label{eq_level_set_equations}
	\left\{
	\begin{split} 
	&\dfrac{\delta {\cal L}_t(\chi,\lambda)}{\delta\chi}({\bf x})=
	2m{\chi}_t^{m-1}({\mathbf{x}})\overline{\cal U}_t({\bf x}){\Delta\chi_t}({{\bf x}})
	+\lambda_t\ \text{sign}(\Delta\chi_t({\bf x}))\ \quad\forall{\bf x}\in\Omega\quad&(a)\\
	&\overline{\cal U}_t({\bf x})\coloneqq\dfrac{1}{2}\left(\bm{\varepsilon}^{(1)}_{\chi}({\bf x},t)\right)^T \mathbb{D}\ \bm{\varepsilon}_{\chi}^{(2)}({\bf x},t)\quad&(b)
	\end{split}
	\right.
	\end{equation}
	The Lagrange multiplier, $\lambda_t$, is determined via the augmented Lagrange method (see reference \cite{Simo1992}) i.e
	\begin{equation}
	\label{eq_solve_lambda}
	\left\{
	\begin{split}
	&\lambda_t\leftarrow\lambda_t+\rho\ {\cal C}(\chi(\phi_t))\quad&(a)\\
	&{\cal C}(\chi(\phi_t))\equiv t-\dfrac{\vert\Omega^-(\chi(\phi_t))\vert}{\vert\Omega\vert}\quad&(b)\\
	\end{split}
	\right.
	\end{equation}
	where $\rho\in{\mathbb R}^+$ is a suitable penalty value.
	
	The discrete time-evolution form of equations (\ref{eq_hamilton_jacobi_1})-(a) and (\ref{eq_solve_lambda})-(a) is obtained after time integration, giving rise to the following iterative update, from iteration $i$ to $i+1$, accounting for equation (\ref{eq_hamilton_jacobi_1})-(c),
	\begin{equation}
	\label{eq_iterative_process}
	\left\{
	\begin{split}
	&\phi^{(i+1)}({\bf x})=\phi^{(i)}({\bf x})-k\dfrac{1}{\Delta \chi^{(i)}({\bf x})}\dfrac{\delta {\cal L}({\chi}^{(i)},\lambda^{(i)})}{\delta {\chi}^{(i)}}({\bf x})\Rightarrow \chi^{(i+1)}={\cal H}_{\beta}(\phi^{(i+1)}({\bf x}))\quad\forall{\bf x}\in\Omega\\
	&\lambda^{(i+1)}=\lambda^{(i)}+\rho\ {\cal C}(\chi(\phi^{(i)}))
	\end{split}
	\right.
	\end{equation}
	The iterative update in equation (\ref{eq_iterative_process}) is stopped when the steady state solution of the problem is achieved, determined in terms of some given tolerances on the variation of the unknowns  $\chi(\phi({\bf x}))$  and $\lambda$, i.e.
	\begin{equation}
	\begin{split}
	&\vert\vert \chi^{(i+1)}-\chi^{(i)}\vert\vert_{L^{2}(\Omega)} \le{\text{Toler}_{\chi}}\\
	&\vert\lambda^{(i+1)}-\lambda^{(i)}\vert\le{\text{Toler}_{\lambda}}
	\end{split}
	\end{equation}
	The corresponding characteristic function, $\chi (i+1)(\bf x)$ is then determined through equation (\ref{eq_iterative_process}).
	\subsubsection{Descending character of the algorithm}
	The algorithm in equations (\ref{eq_hamilton_jacobi_1}) to (\ref{eq_iterative_process}) leads to a continuous descent, along pseudo-time or iterations, of the Lagrange functional to be minimized. This  warranties that, when a steady state is reached, the solution is a local minimum of the problem. The proof is readily achieved through the following sequence of propositions. Let us consider the time-evolutionary problem in equation (\ref{eq_hamilton_jacobi_1})
	\begin{equation}
	\label{eq_hamilton_jacobi}
	\begin{split}
	&{\cal V}_{\phi}\coloneqq\left\{\phi\;/\;\phi:\Omega\times[0,T]\rightarrow{\mathbb R} \right\}\\
		&\left\{
		\begin{split}
		&\dot{\phi_t}({\bf x})=-k\dfrac{1}{\Delta \chi({\bf x})}\dfrac{\delta {\cal L}({\chi},\lambda)}{\delta {\chi}}({\bf x}) \; ;\quad k\in{\mathbb R}^+\; ;\quad-1\le\phi_t\le 1\quad&(a)\\
		&{\cal L}_t(\phi,\lambda)\equiv{\cal L}({\chi}(\phi_t),\lambda)={\cal J}({\chi}(\phi_t))+\lambda{\cal C}(\chi(\phi_t))\quad&(b)
	 	\end{split}
		\right.
	\end{split}
	\end{equation}
	\begin{lemma}
		Given the functional
		\begin{equation}
		\label{eq_int1}
		{\cal L}(\chi)=\int_{\Omega}{ G}({\bf u}({\bf x}),{\bf x})d \Omega\equiv\int_{\Omega}F(\chi,{\bf x})d \Omega 
		\end{equation}
		and considering the \textit{function derivative} of ${\cal L}(\chi)$ as a, parameter-$\chi$-dependent, integral, the following equality is fulfilled
		\begin{equation}
		\label{eq_result_equation}
		\dfrac{\partial{\cal L}(\chi)}{\partial \chi}=\int_\Omega\dfrac{1}{\Delta \chi ({\bf x})} \dfrac{\delta{\cal L}(\chi)}{\delta \chi}({\bf x}) d\Omega
		\end{equation}
	\end{lemma}
	\begin{proof}
		The \textit{function derivative} of ${\cal L}(\chi)$, in equation (\ref{eq_int1}), with respect to $\chi$ reads
		\begin{equation}
		\label{eq_result_equation_1}
		\begin{split}
		\dfrac{\partial{\cal L}(\chi)}{\partial \chi}=\int_\Omega \dfrac{\partial F(\chi,{\bf x})}{\partial \chi} d\Omega
		\end{split}
		\end{equation}
		and, equation (\ref{eq_result_1}) (replacing ${\cal J}$ by ${\cal L}$), yields
		\begin{equation}
		\label{eq_result_equation_2}
		\left\{
		\begin{split}
		&\left[\dfrac{\partial { F}({\chi},{\bf x})}{\partial {\chi}}\right]_{{\bf x}=\hat{\bf x}}\Delta \chi(\hat{\bf x})=\dfrac{\delta}{\delta \chi}\left[\int_{\Omega}{F}({\chi},{\bf x})d\Omega\right](\hat{\bf x})\quad \forall{\hat{\bf x}}\in\Omega\Rightarrow\quad&(a)\\
		& \Rightarrow\dfrac{\partial F({\chi},\hat{\bf x})}{\partial {\chi}}=\dfrac{1}{\Delta \chi (\hat{\bf x})}\dfrac{\delta {\cal L}(\chi)}{\delta\chi}(\hat{\bf x})\quad \forall{\hat{\bf x}}\in\Omega\quad&(b)
		\end{split}
		\right.
		\end{equation}
		Substitution of the result in equation (\ref{eq_result_equation_2})-(b) into equation (\ref{eq_result_equation_1}), yields the proposition in equation (\ref{eq_result_equation}).
	\end{proof}
	\begin{lemma}
		\begin{equation}
		\label{eq_result_equation_3}
		\dfrac{\partial \chi(\phi_t({\bf x}))}{\partial \phi_t}\equiv \dfrac{\partial \chi(\phi_t)}{\partial \phi_t}({\bf x})=(1-\beta)\vert\vert{\bm \nabla}\phi_t({\bf x})\vert\vert^{-1}\delta_{\Gamma_t}({\bf x})
		\end{equation}
		where $\delta_{\Gamma_t}({\bf x})$ stands for the line/surface Dirac's-delta function, shifted to $\Gamma_t$ (see figure \ref{fig_Domain_HJ}) thus fulfilling
		\begin{equation}
		\label{eq_delta_Gamma}	
		\int_{\Omega}\delta_{\Gamma_t}({\bf x})\varphi({\bf x})d\Omega=\int_{\Gamma_t}\varphi({\bf x})d\Gamma
		\end{equation}
		for any sufficiently regular $\varphi({\bf x})$.
	\end{lemma}
	\begin{proof}
	\begin{equation}
	\begin{split}
	&\chi(\phi_t({\bf x}))={\cal H}_{\beta}(\phi_t({\bf x}))\Rightarrow
	\left\{
		\begin{split}
		&{\bm \nabla}\chi(\phi_t({\bf x}))={\bm \nabla}{\cal H}_{\beta}(\phi_t({\bf x}))=
		(1-\beta)\delta_{\Gamma_t}({\bf x})\otimes {\bf n}({\bf x})\; ;\quad {\bf n}({\bf x})=\dfrac{{\bm \nabla} \phi_t({\bf x})}
		{\vert\vert {\bm \nabla} \phi_t({\bf x})  \vert\vert}\\
		&{\bm \nabla}\chi(\phi_t({\bf x}))=\dfrac{\partial \chi(\phi_t({\bf x}))}{\partial \phi_t}{\bm \nabla}{\phi_t({\bf x})}=
	\dfrac{\partial \chi(\phi_t({\bf x}))}{\partial \phi_t}{\vert\vert {\bm \nabla} \phi_t({\bf x})  \vert\vert}\otimes {\bf n}({\bf x})
		\end{split}
	\right.\quad\Rightarrow\\
	 &\Rightarrow\dfrac{\partial \chi(\phi_t({\bf x}))}{\partial \phi_t}{\vert\vert {\bm \nabla} \phi_t({\bf x})  \vert\vert}=
	 	(1-\beta)\delta_{\Gamma_t}({\bf x})\Rightarrow
	 	\dfrac{\partial \chi(\phi_t({\bf x}))}{\partial \phi_t}=(1-\beta)\vert\vert{\bm \nabla}\phi_t({\bf x})\vert\vert^{-1}\delta_{\Gamma_t}({\bf x})
	 \end{split}
	\end{equation}
\end{proof}

\begin{theorem}
Time evolution of the Lagrangian functional ${\cal L}_t(\phi,\lambda)\equiv{\cal L}({\chi}(\phi_t),\lambda)$, in equation (\ref{eq_hamilton_jacobi}), is always negative i.e. $\dot{\cal L}({\chi}(\phi_t),\lambda)<0$. 
\end{theorem}

\begin{proof}
Time differentiation of ${\cal L}(\phi_t,\lambda)$, in equation (\ref{eq_hamilton_jacobi})-(b), yields 
\begin{equation}
\label{eq_theorem_proof}
\begin{split}
\dfrac{\partial{\cal L}\left(\chi(\phi({\bf x},t)),\lambda\right)}{\partial t}&\equiv\dot{\cal L}({\chi}(\phi_t({\bf x})),\lambda)=
\int_\Omega\dfrac{\partial \left({\cal J}({\chi}(\phi_t))+\lambda{\cal C}(\chi(\phi_t))\right)}{\partial\chi}
\dfrac{\partial \chi(\phi_t)}{\partial \phi_t}\dot{\phi_t}({\bf x})d\Omega=\\
	&=\int_\Omega(1-\beta)\vert\vert{\bm \nabla}\phi_t({\bf x})\vert\vert^{-1}\delta_{\Gamma_t}({\bf x})
\dfrac{\partial {({\cal J}({\chi}(\phi_t))+\lambda{\cal C}(\chi(\phi_t)))}}{\partial\chi}\dot{\phi_t}({\bf x})d\Omega=\\
&=\int_\Omega(1-\beta)\vert\vert{\bm \nabla}\phi_t({\bf x})\vert\vert^{-1}\dfrac{1}{\Delta \chi({\bf x})}
\dfrac{\delta {{\cal L}(\chi(\phi_t),\lambda)}}{\delta\chi}({\bf x})\delta_{\Gamma_t}({\bf x})\dot{\phi_t}({\bf x})d\Omega=\\
&=\int_{\Gamma_t}(1-\beta)\vert\vert{\bm \nabla}\phi_t({\bf x})\vert\vert^{-1}
\dfrac{1}{\Delta \chi({\bf x})}
\dfrac{\delta {{\cal L}(\chi(\phi_t),\lambda)}}{\delta\chi}({\bf x})\dot{\phi_t}({\bf x})d \Gamma
\end{split}
\end{equation}
where results in equations (\ref{eq_result_equation}), (\ref{eq_result_equation_3}) and (\ref{eq_delta_Gamma}) have been used. Replacement of equation (\ref{eq_hamilton_jacobi}) into equation (\ref{eq_theorem_proof}) yields
\begin{equation}
\begin{split}
&\left\{
\begin{split}
&\dot{\cal L}({\chi}(\phi_t),\lambda)=\int_{\Gamma_t}(1-\beta)\vert\vert{\bm \nabla}\phi_t({\bf x})\vert\vert^{-1}
\dfrac{1}{\Delta \chi({\bf x})}
\dfrac{\delta {{\cal L}(\chi(\phi_t),\lambda)}}{\delta\chi}({\bf x})\ \dot{\phi_t}({\bf x})d \Gamma\\
&\dot{\phi_t}({\bf x})=-k\dfrac{1}{\Delta \chi({\bf x})}\dfrac{\delta {\cal L}({\chi},\lambda)}{\delta {\chi}}({\bf x}) \; ;\quad k\in{\mathbb R}^+
\end{split}
\right.\\
&\Rightarrow\dot{\cal L}({\chi}(\phi_t),\lambda)=-k\int_{\Gamma_t}
\underbrace{(1-\beta)\vert\vert{\bm \nabla}\phi_t({\bf x})\vert\vert^{-1}}_{\displaystyle >0}
\underbrace{\left(\dfrac{1}{\Delta \chi({\bf x})}\dfrac{\delta {\cal L}({\chi}(\phi_t),\lambda)}{\delta {\chi}}({\bf x})\right)^2}_
{\displaystyle{>0}} d \Gamma< 0\quad\forall t
\end{split}
\end{equation}
\end{proof}
\begin{corollary}
	\textit{Therefore, at the stationarity of the characteristic function ($\dot{\chi}(\phi_t({\bf x}))=0 \ \forall {\bf x}\in\Omega$) the achieved,  topology $\chi({\bf x},t)$, defines a (local) minimum of ${\cal L}_t(\phi,\lambda)\equiv{\cal L}({\chi}(\phi_t),\lambda)$ in equation (\ref{eq_hamilton_jacobi}).}
\end{corollary}
\end{appendices}

\section*{References}
\bibliographystyle{abbrvnat}
\bibliography{TopOpt}

\end{document}